\documentclass[a4paper,11pt]{article}
\pdfoutput=1 

\usepackage{jheppub} 

\usepackage[T1]{fontenc} 
\usepackage{graphicx,xcolor}
\usepackage{multicol}
\usepackage{csquotes}
\usepackage[normalem]{ulem}
\usepackage{verbatim}
\usepackage{etoolbox}
\usepackage{booktabs}

\newcommand{\be}{\begin{eqnarray}}
	\newcommand{\ee}{\end{eqnarray}}

\title{Linear causality and stability constraints on relativistic second-order magnetohydrodynamics}

\author[a]{Yiwei Qiu}
\author[b,a]{, Duan She}
\author[a]{, Defu Hou}

\affiliation[a]{Institute of Particle Physics and Key Laboratory of Quark and Lepton Physics (MOE), Central China Normal University, Wuhan 430079, China}
\affiliation[b]{Institute of Physics, Henan Academy of Sciences, Zhengzhou 450046, China}

\emailAdd{qiuyw@mails.ccnu.edu.cn}
\emailAdd{sheduan@hnas.ac.cn}
\emailAdd{houdf@mail.ccnu.edu.cn}

\abstract{
	In this work, we construct a theoretical framework for relativistic second-order magnetohydrodynamics based on entropy current analysis. The formalism consistently incorporates finite relaxation dynamics of the dissipative fluxes, providing a transient framework suitable for analyzing linear stability and short-wavelength causality. Utilizing linear mode analysis, we investigate the constraints imposed by causality and stability on this anisotropic system. By linearizing the theory around a homogeneous equilibrium state, we identify the magnetosonic, Alfv\'en, and charge-diffusion sectors, {together with their coupling terms}. For each sector, we derive asymptotic dispersion relations in both the long-wavelength (small-$k$) and short-wavelength (large-$k$) regimes, validating them against exact numerical roots. Our numerical analysis confirms the accuracy of these asymptotic solutions and uncovers a nontrivial angular dependence, especially near special propagation directions where the ordinary momentum expansion becomes less reliable. By evaluating the large-$k$ behavior of the propagating branches alongside the damping properties of non-hydrodynamic modes, we delineate the corresponding causality constraints. We find that the admissible linear causal domain is governed by the interplay between {propagation direction,} anisotropic transport coefficients, and relaxation times, with the resulting bounds being intrinsically mode-dependent. These findings provide a systematic theoretical basis for investigating { linear} stability and short-wavelength causality in relativistic magnetohydrodynamics beyond the first-order approximation.}
\makeatletter

\newcommand{\setnewrefcolor}[1]{%
	\color{black}%
	\ifstrequal{#1}{BenenowskiPoovuttikul2019}{}{}%
	\ifstrequal{#1}{PoovuttikulRajagopal2021}{}{}%
	\ifstrequal{#1}{ArmasCamilloni2022}{}{}%
	\ifstrequal{#1}{VardhanGrozdanovLeutheusserLiu2024}{}{}%
	\ifstrequal{#1}{VardhanGrozdanovLeutheusserLiu2025}{}{}%
	\ifstrequal{#1}{LandryLiu2026}{}{}%
}

\let\original@bibitem\@bibitem
\let\original@lbibitem\@lbibitem

\def\@bibitem#1{%
	\setnewrefcolor{#1}%
	\original@bibitem{#1}%
}

\def\@lbibitem[#1]#2{%
	\setnewrefcolor{#2}%
	\original@lbibitem[#1]{#2}%
}

\AtEndEnvironment{thebibliography}{\color{black}}

\makeatother
\begin{document}
	
	\maketitle
	
	\section{Introduction}
	\label{section1}
	Electromagnetic fields and relativistic fluid motion coexist in some of the most extreme environments in the universe, spanning from the quark-gluon plasma (QGP) generated in relativistic heavy-ion collisions to magnetars and compact-object mergers. In non-central heavy-ion collisions, event-by-event simulations indicate the generation of ultra-strong magnetic fields during the early pre-equilibrium stages of the reaction. This has motivated an extensive research program focused on electromagnetic transport, anomalous phenomena, and collective dynamics in hot QCD matter~\cite{BzdakSkokov2012,DengHuang2012,Tuchin2013,RoyPu2015,LiShengWang2016,Huang2016,HattoriHuang2017,MiranskyShovkovy2015,Fukushima2019,GursoyKharzeevRajagopal2014}. On the experimental front, the medium's response to transient vortical and electromagnetic backgrounds has been explored through diverse observables. The global polarization of $\Lambda$ and $\bar{\Lambda}$ hyperons, initially reported by the STAR Collaboration and subsequently scrutinized by ALICE, demonstrates that spin observables are highly sensitive to the rotational and electromagnetic environment of the fireball~\cite{STARLambda2017,ALICEGlobalPol2020,ALICEBeamPol2022}. More direct evidence of electromagnetic influence emerges from charge-dependent directed-flow measurements at RHIC and the LHC, which provide increasingly quantitative constraints on the role of early-time electromagnetic fields in subsequent collective evolution~\cite{ALICEDirectedFlow2020,STAREMField2024,STARCharge2009,STARCharge2010,ALICEChargeSeparation2013,CMSChargeDependent2017,CMSCME2018,CMSChiralMagneticWave2019,ALICEXeXeCME2024}. Furthermore, these strong magnetic fields serve not only as probes of the QGP's electromagnetic properties but also as a unique window into macroscopic quantum anomalous phenomena. While a definitive experimental confirmation of the chiral magnetic effect (CME) remains a subject of intense investigation and debate~\cite{Kharzeev2006,KharzeevMcLerranWarringa2008,FukushimaKharzeevWarringa2008,KharzeevLiaoVoloshinWang2016,Liao2015,ZhaoWang2019}, its discovery would signify local CP symmetry violation in strong interactions, carrying profound implications for our understanding of the matter--antimatter asymmetry in the early universe. The same environment has also stimulated extensive work on anomalous transport, including the chiral magnetic effect, chiral separation, chiral electric separation, and related real-time or kinetic descriptions~\cite{HuangLiao2013,PuWuYang2014,PuWuYang2015,JiangHuangLiao2015,Satow2014,ChenIshiiPuYamamoto2016,EbiharaFukushimaPu2017,StephanovYin2012,SonYamamoto2013,HidakaPuYang2019,HuangShiJiangLiaoZhuang2018,GaoLiangWangWang2018,LiuGaoMamedaHuang2019,LinShukla2019,LinYang2020,FukushimaKharzeevWarringa2010Realtime,CopingerFukushimaPu2018,ShengFangWangRischke2019,FengHouLiuRenWuWu2017,WuHouRen2017,LinYang2018MassCorrection,HorvathHouLiaoRen2020,FengHouRen2019,HouLin2018,LinYanLiang2018}. These measurements have also triggered extensive theoretical work on spin polarization, local equilibrium with spin, and relativistic spin hydrodynamics, ranging from perfect-fluid formulations to dissipative and second-order frameworks~\cite{SunKoLi2016,SunKo2017CVME,SunKo2017Lambda,SunKo2018Isobar,SunKo2019LongitudinalSpin,ZhouXu2018,ZhouXu2019,LiuSunKo2020,Becattini2022,FlorkowskiFrimanJaiswalSperanza2018,MontenegroTintiTorrieri2020,BecattiniBuzzegoliPalermo2021,WeickgenanntSperanzaShengWangRischke2021,FuLiuPangSongYin2021,WeickgenanntWagnerSperanzaRischke2022,Weickgenannt2023,BiswasDaherDasFlorkowskiRyblewski2023,SheHuangHouLiao2022,SheQiuHou2025,SheQiuJiangHou2025}. Theoretically, these developments have catalyzed a series of magnetized fluid-dynamical studies, encompassing ideal and resistive relativistic magnetohydrodynamics (RMHD), anomalous MHD, {and semi-analytic Bjorken-type constructions incorporating magnetic field or magnetization effects~\cite{RoyPuRezzollaRischke2015,PuRoyRezzollaRischke2016,PuYang2016,PuYangAnalytic2017,SiddiqueWangPuWang2019,SheJiangHouYang2019,MoghaddamAlbericoSheKordAzadegan2020,InghiramiDelZannaBeraudoMoghaddamBecattiniBleicher2016,InghiramiMaceHironoDelZannaKharzeevBleicher2020,HuangSheShiHuangLiao2023,HuangWuHuang2024,ShokriSadooghi2018,NakamuraMiyoshiNonakaTakahashi2023,Peng:2022cya,Jiang:2024mts}.} Collectively, these works underscore that magnetic fields are not merely passive backgrounds; rather, they qualitatively reshape transport processes, mode propagation, and the physical interpretation of heavy-ion observables~\cite{LiuHuang2020ChiralSpin,GaoMaPuWang2020}.
	
	The applicability of such magnetohydrodynamic phenomena extends far beyond the subatomic scale, manifesting critically in the macroscopic evolution of astrophysical objects. {Magnetic fields of magnetar-level intensity are a defining element in the structure and dynamics of compact stars, where they substantially modify equilibrium configurations, transport properties, and the global evolution of strongly magnetized neutron stars~\cite{ChatterjeeNovakOertel2021,Ciolfi2020,HuangRischkeSedrakian2010,HarutyunyanSedrakian2016,VardhanGrozdanovLeutheusserLiu2024,LandryLiu2026}}. Similarly, binary neutron-star mergers provide a natural laboratory in which relativistic plasma flows, magnetic-field amplification, angular-momentum transport, and magnetically driven outflows are central to the post-merger dynamics and subsequent multi-messenger emissions~\cite{HuangSedrakianRischke2011,DionysopoulouAlicPalenzuelaRezzollaGiacomazzo2013,RezzollaZanotti2013}. In dense stellar matter, strong magnetic fields induce anisotropic transport and a directional bulk viscous response, thereby modifying the macroscopic dynamics in a manner that anticipates the essential features of relativistic magnetohydrodynamics (RMHD)~\cite{HuangRischkeSedrakian2010,HuangSedrakianRischke2011}. From this broader perspective, RMHD serves as a unified macroscopic framework for describing the long-wavelength collective behavior of matter under extreme electromagnetic environments, while maintaining a rigorous link to the underlying microscopic transport, thermodynamics, and medium response~\cite{LandauLifshitzFluid,RezzollaZanotti2013}.
	
	In response to these diverse phenomenological demands, the theoretical underpinnings of RMHD have undergone a sophisticated evolution, bridging the gap between microscopic kinetics and macroscopic constitutive laws. Covariant constitutive relations for relativistic MHD were clarified within the modern hydrodynamic framework in Ref.~\cite{HernandezKovtun2017}, {while the conservation of magnetic flux has been further illuminated through the lens of generalized global symmetries and one-form hydrodynamics~\cite{GrozdanovHofmanIqbal2017,ArmasJain2019,ArmasJain2020,GrozdanovLucasPoovuttikul2019,GrozdanovPoovuttikul2019,BenenowskiPoovuttikul2019,VardhanGrozdanovLeutheusserLiu2024,VardhanGrozdanovLeutheusserLiu2025}.} {
		An important complementary route is provided by effective-field-theory formulations based on higher-form symmetries, which organize the hydrodynamic degrees of freedom and constitutive relations directly from symmetry and effective-action principles, without taking an entropy-current construction as the starting point. Such approaches have classified non-dissipative second-order transport in strong-field MHD and have subsequently been extended to dissipative one-form fluids, with applications to magnetic-field diffusion that reproduce Ohmic decay, ambipolar diffusion, and Hall drift while also predicting additional strong-field contributions~\cite{BenenowskiPoovuttikul2019,VardhanGrozdanovLeutheusserLiu2024,VardhanGrozdanovLeutheusserLiu2025}. These developments demonstrate that the EFT approach provides an independent and systematic framework for formulating magnetohydrodynamic transport, complementary to the entropy-current method adopted in the present work.}Related anisotropic-hydrodynamic and strong-field transport constructions, including Kubo-formula analyses and anisotropic dissipative decompositions, were developed in Refs.~\cite{HuangSedrakianRischke2011,MolnarNiemiRischke2016,DashSamantaDeyGangopadhyayaGhoshRoy2020,KurianDasChandra2020,HattoriLiSatowYee2017,KurianMitraGhoshChandra2019,KurianChandra2017,Feng2017Conductivity,FukushimaHidaka2018,GreifBourasGreinerXu2014,AmatoEtAl2013,ArnoldMooreYaffe2003,ChenLiuPuSongWang2013,DasMishraMohapatra2019Transport,DasMishraMohapatra2019Hall,KerbikovAndreichikov2015,Nam2012}. On the microscopic front, kinetic-theory derivations have yielded first- and second-order dissipative RMHD equations for both ideal and resistive regimes, providing explicit expressions for transport coefficients and relaxation structures~\cite{Denicol2018,Denicol2019,PandaRoy2021,KushwahDenicol2024,deBritoKushwahDenicol2025}. Complementary approaches based on the local Gibbs ensemble and the nonequilibrium statistical operator have established a systematic bridge between microscopic quantum electrodynamics and macroscopic RMHD constitutive laws~\cite{HongoHattori2021,HattoriHongoHuang2022,Harutyunyan2018}. Concurrently, analytical and numerical investigations of magnetized Bjorken flows, anomalous MHD, dynamical electromagnetic fields, and linear waves have significantly enriched the phenomenological and conceptual landscape of RMHD in the context of heavy-ion collisions~\cite{GursoyKharzeevRajagopal2014,RoyPu2015,RoyPuRezzollaRischke2015,PuRoyRezzollaRischke2016,PuYang2016,PuYangAnalytic2017,HongoHironoHirano2017,InghiramiDelZannaBeraudoMoghaddamBecattiniBleicher2016,InghiramiMaceHironoDelZannaKharzeevBleicher2020,SiddiqueWangPuWang2019,SheJiangHouYang2019,MoghaddamAlbericoSheKordAzadegan2020,HuangSheShiHuangLiao2023,HuangWuHuang2024}. Along this line, Chern--Simons MHD was shown to exhibit an anomalous Hall instability of Alfv\'en modes in a parity-violating resistive plasma~\cite{KiamariRahbardarShokriSadooghi2021}. More recently, the issues of linear causality, stability, angular anisotropy, and short-wavelength behavior have gained increasing prominence. These studies demonstrate that magnetized relativistic fluids exhibit a far more intricate structure than their unmagnetized counterparts, {particularly when dissipation and finite relaxation effects are incorporated~\cite{Biswas2020,ArmasCamilloni2022,Fang2024PRD,HoultKovtun2025,MolnarRischke2025,Fang2025PRD}.}
	
	Before constructing a dissipative framework, it is imperative to establish the specific hierarchy of evolution equations that define the RMHD regime. In many formulations, including the one adopted in this study, the fundamental equations comprise the conservation of energy-momentum $\partial_\mu T^{\mu\nu}=0$, the conservation of particle number $\partial_\mu N^\mu=0$, and the homogeneous Maxwell equation $\partial_\mu \tilde F^{\mu\nu}=0$. Crucially, the inhomogeneous Maxwell equation, $\partial_\mu F^{\mu\nu}=J^\nu_{\rm em}$, is not treated as an independent conservation law. This treatment is physically consistent with the separation of scales inherent to the MHD regime. The equation $\partial_\mu \tilde F^{\mu\nu}=0$ constitutes the Bianchi identity and expresses magnetic flux conservation; in modern terminology, it is formally understood as a conserved magnetic one-form symmetry~\cite{GrozdanovHofmanIqbal2017,HongoHattori2021}. Conversely, the inhomogeneous equation $\partial_\mu F^{\mu\nu}=J^\nu_{\rm em}$ governs the medium's electric response rather than representing a fundamental conservation law. In a conducting plasma, electric flux is screened, and excitations in the electric sector are generically gapped, thus falling outside the hydrodynamic description. Within the derivative expansion of MHD, the magnetic field is treated as a zeroth-order hydrodynamic variable, {whereas the electric field is considered a subleading quantity, incorporated via constitutive relations, transport coefficients, and resistive corrections~\cite{HernandezKovtun2017,GrozdanovHofmanIqbal2017,ArmasJain2020,HongoHattori2021,HattoriHongoHuang2022,Denicol2019,DionysopoulouAlicPalenzuelaRezzollaGiacomazzo2013,HoultKovtun2025,PoovuttikulRajagopal2021,VardhanGrozdanovLeutheusserLiu2024}.} Consequently, the hydrodynamic framework is naturally organized around the densities of energy, momentum, particle number, and magnetic flux. Finally, we emphasize that the source term in the inhomogeneous Maxwell equation is denoted by the electromagnetic current $J^\mu_{\rm em}$, to distinguish it clearly from the particle-number current $N^\mu$ used throughout this work.
	
	While the ideal framework provides a foundational description, a physically consistent RMHD theory must navigate the long-standing challenges of causality and stability that emerge once dissipative effects are included. In first-order dissipative theories, acausality and generic instabilities are well-documented pathologies even in the unmagnetized case~\cite{IsraelStewart1979,HiscockLindblom1983,HiscockLindblom1985,PuKoideRischke2010,PuKoideWang2010AIP,DenicolKodamaKoideMota2008JPG,DenicolKodamaKoideMota2008PRC,FloerchingerGrossi2018,BemficaDisconziNoronha2019,Kovtun2019,Eckart1940}. Overcoming these difficulties has been central to the development of transient, causal relativistic hydrodynamics derived from kinetic theory~\cite{BaierRomatschkeSonStarinetsStephanov2008,BetzHenkelRischke2009PPNP,BetzHenkelRischke2009JPG,DenicolNiemiMolnarRischke2012,DenicolEtAl2012EPJA,MolnarEtAl2014}. In magnetized systems, the landscape is further complicated by the magnetic field, which decomposes transport channels into longitudinal and transverse sectors, introduces additional anisotropic dissipative coefficients, and fundamentally alters the structure of collective modes. The intrinsic link between causality and stability has also been underscored in the broader context of dissipative hydrodynamics~\cite{PuKoideRischke2010}. Although significant strides have been made in formulating second-order viscous and resistive RMHD~\cite{Denicol2018,Denicol2019,PandaRoy2021,Biswas2020,HoultKovtun2025,KushwahDenicol2024,deBritoKushwahDenicol2025}, a systematic investigation of how the interplay between anisotropic transport coefficients and relaxation times constrains the linear causal domain remains an outstanding challenge. It is also noteworthy that related questions of entropy production, causality, and stability have recently been explored in relativistic spin hydrodynamics, both from entropy-current methods and from kinetic theory~\cite{Weickgenannt2023,BiswasDaherDasFlorkowskiRyblewski2023,SheQiuHou2025,SheQiuJiangHou2025,TiwariPatra2025}.
	
	Against this backdrop of theoretical challenges and phenomenological requirements, the present work formulates a second-order RMHD framework derived from entropy-current analysis. By ensuring the non-negativity of entropy production, we derive a self-consistent set of relaxation equations for the dissipative fluxes, which naturally leads to a hyperbolic completion of the theory. We then systematically investigate the linear causality and stability of this framework in the presence of a uniform background magnetic field. By promoting the dissipative fluxes to independent dynamical variables with finite relaxation times, we construct an Israel--Stewart-type formulation for a magnetized anisotropic fluid. We linearize the minimal IS theory around a homogeneous equilibrium state and demonstrate that the excitation spectrum decomposes into magnetosonic, Alfv\'en, and charge-diffusion sectors {together with their coupling terms}. For each sector, we derive asymptotic dispersion relations in both the long-wavelength (small-$k$) and short-wavelength (large-$k$) regimes, validating them against exact numerical roots while examining their angular dependence. This analysis enables us to identify how the interplay between anisotropic transport coefficients and relaxation scales governs the admissible causal domain. Consequently, this study provides a mode-resolved and sector-dependent characterization of linear causality and stability, extending the modern understanding of second-order dissipative RMHD.
	
	Throughout this work, we adopt the metric signature $\text{diag}\left(1,-1,-1,-1\right)$ and use natural units $c=\hbar=k_B=1$. The fluid velocity $u^\mu$ is normalized as $u^\mu u_\mu = 1$, and the orthogonal projection tensor is defined as $\Delta^{\mu\nu} = g^{\mu\nu} - u^\mu u^\nu$. In this manuscript, the symmetric and antisymmetric parts of a tensor $X^{\mu\nu}$ are denoted as $X^{(\mu\nu)}=\frac{1}{2}\left(X^{\mu\nu}+X^{\nu\mu}\right)$ and $X^{[\mu\nu]}=\frac{1}{2}\left(X^{\mu\nu}-X^{\nu\mu}\right)$, respectively. The remainder of this paper is organized as follows. In Section~\ref{section2}, we establish the ideal RMHD framework using the entropy-current principle. Section~\ref{section3} and Section~\ref{section4} extend this framework to include first-order and second-order dissipative corrections, respectively, ensuring a self-consistent relativistic Israel-Stewart-type formulation. Section~\ref{section5} performs a systematic linear mode analysis to identify the magnetosonic, Alfvén, and charge-diffusion sectors. In Section~\ref{section6} and Section~\ref{section7}, we investigate the constraints on causality and stability by deriving asymptotic dispersion relations and validating them against exact numerical solutions. Finally, concluding remarks and future outlooks are presented in Section~\ref{section8}. Detailed mathematical derivations and supplementary results are provided in the Appendices. Appendix~\ref{appendixf} presents the comprehensive tensor decompositions relative to the fluid flow and the direction of magnetic anisotropy. In Appendix~\ref{appendixg}, we discuss the matching and frame conditions specifically tailored for relativistic magnetohydrodynamics. The explicit second-order forms of the entropy current and the associated transport matrices are detailed in Appendix~\ref{partial Q} and Appendix~\ref{APPENDIX 3}, respectively. Finally, Appendix~\ref{Large k expansion solution} provides the full analytical expressions for the large-$k$ expansion solutions.
	
	\section{Ideal RMHD from entropy analysis}
	\label{section2}
	
	To establish the foundational evolution laws, we follow the modern symmetry-based formulation of relativistic magnetohydrodynamics~\cite{HernandezKovtun2017,GrozdanovHofmanIqbal2017,HongoHattori2021,HattoriHongoHuang2022,VardhanGrozdanovLeutheusserLiu2024,VardhanGrozdanovLeutheusserLiu2025} and describe the plasma in terms of energy-momentum, particle number, and magnetic flux conservation. The governing equations are compactly expressed as
	\begin{equation}
		\partial_{\mu}T^{\mu\nu}=0,\quad\partial_{\mu}N^{\mu}=0,\quad\partial_{\mu}\tilde{F}^{\mu\nu}=0.
		\label{1}
	\end{equation}
	In this study, we focus on a parity-even plasma, where the macroscopic currents are constructed strictly from parity-conserving tensor structures. Consequently, anomalous effects originating from $P$-odd/$CP$-odd fluctuations, such as the chiral magnetic effect, are excluded from the present constitutive relations.
	
	Regarding the choice of hydrodynamic frames, we adopt the Landau-Lifshitz frame throughout this work, where the hydrodynamic state is defined by the energy density $\epsilon$ and the normalized four-velocity $u^\mu$. The magnetic degrees of freedom are captured by the magnetic flux vector $B^\mu \equiv \tilde{F}^{\mu\nu}u_\nu$. This four-vector is inherently transverse to the fluid flow ($u_\mu B^\mu = 0$) and, in our parity-even setup, serves as the primary dynamical variable for the magnetic sector without coupling to axial currents.
	
	The thermodynamic consistency of the theory is anchored in the local thermodynamic structure, which is governed by the entropy density $s\left(\epsilon,n,B^{\mu}\right)$. Invoking the first law of thermodynamics in the local rest frame, 
	\begin{equation}
		\beta=\frac{\partial s\left(\epsilon,n,B^{\mu}\right)}{\partial\epsilon},\,\beta\mu=-\frac{\partial s\left(\epsilon,n,B^{\mu}\right)}{\partial n},\,\beta H_{\mu}=-\frac{\partial s\left(\epsilon,n,B^{\mu}\right)}{\partial B^{\mu}}\Leftrightarrow Tds=d\epsilon-\mu dn-H_{\mu}dB^{\mu}.
		\label{2}
	\end{equation}
	To facilitate the entropy-current analysis, it is convenient to resolve $T^{\mu\nu}$, $N^\mu$, and $\tilde{F}^{\mu\nu}$ into independent tensor structures governed by symmetry arguments and the available tensors in the problem. This type of covariant decomposition is standard in relativistic MHD and closely parallels the treatments in Refs.~\cite{HernandezKovtun2017,HongoHattori2021,HattoriHongoHuang2022}. In particular, the presence of a background magnetic field introduces a preferred spatial direction $b^\mu = B^\mu/B$, breaking the rotational symmetry of the fluid. To resolve the resulting anisotropy, we utilize the projection tensor $\Xi^{\mu\nu} \equiv g^{\mu\nu} - u^\mu u^\nu + b^\mu b^\nu$, which is orthogonal to both $u^\mu$ and $b^\mu$. This allows us to parameterize the zeroth-order constitutive relations in terms of longitudinal pressure $p_\parallel$ and transverse pressure $p_\perp$
	\begin{align}
		T_{(0)}^{\mu\nu}	=&\epsilon u^{\mu}u^{\nu}-p_{\perp}\Xi^{\mu\nu}+p_{\parallel}b^{\mu}b^{\nu},\label{3}\\
		N_{(0)}^{\mu}	=&nu^{\mu},\label{4}\\
		\tilde{F}_{(0)}^{\mu\nu}	=&B^{\mu}u^{\nu}-B^{\nu}u^{\mu},\label{5}
	\end{align}
	where 
	\begin{eqnarray}
		\begin{aligned}
			n=&N_{(0)}^{\mu}u_{\mu},\quad \epsilon=T_{(0)}^{\mu\nu}u_{\mu}u_{\nu},\quad p_{\parallel}=T_{(0)}^{\mu\nu}b_{\mu}b_{\nu},\quad
			p_{\perp}=  -\frac{1}{2}T_{(0)}^{\mu\nu}\Xi_{\mu\nu},\quad B^{\mu}=\tilde{F}_{(0)}^{\mu\nu}u_{\nu},
		\end{aligned}
		\label{6}
	\end{eqnarray}
	The detailed tensor decompositions of $T^{\mu\nu}$, $\tilde{F}_{(0)}^{\mu\nu}$, and $N^{\mu}$ are discussed in Appendix~\ref{appendixf}.
	The dynamical evolution of these variables is constrained by projecting the conservation laws \eqref{1} along $u_\nu$ and $H_\nu$, yielding
	\begin{align}
		u_{\nu}\partial_{\mu}T_{(0)}^{\mu\nu}=&D\epsilon+\left(\epsilon+p_{\perp}\right)\theta+\left(p_{\perp}-p_{\parallel}\right)b^{\mu}b^{\nu}\partial_{\mu}u_{\nu}=0,\label{7}\\
		\partial_{\mu}N_{(0)}^{\mu}=&Dn+n\theta=0,\label{8}\\
		H_{\nu}\partial_{\mu}\tilde{F}_{(0)}^{\mu\nu}=&-H_{\mu}DB^{\mu}-B^{\mu}H_{\mu}\theta+B^{\mu}H_{\nu}\partial_{\mu}u^{\nu}=0.\label{9}
	\end{align}
	We now turn to the central requirement of the second law of thermodynamics by computing the divergence of the entropy current $\mathcal{S}_{(0)}^{\mu}=su^{\mu}$. Substituting the resulting evolution for $D\epsilon$, $Dn$, and $-H_{\mu}DB^{\mu}$ into the entropy divergence, we obtain the adiabaticity condition
	
	\begin{eqnarray}
		\begin{aligned}
			\partial_{\mu}\mathcal{S}_{(0)}^{\mu} =&\beta\left(Ts-\epsilon-p_{\perp}+\mu n+B^{\mu}H_{\mu}\right)\theta\\
			& -\beta b^{\mu}\left[\left(p_{\perp}-p_{\parallel}\right)b^{\nu}+BH^{\nu}\right]\partial_{\mu}u_{\nu}.
		\end{aligned}
		\label{10}
	\end{eqnarray}
	
	At the ideal level (zeroth order in the derivative expansion), the local second law of thermodynamics demands that the entropy production vanish, which in turn imposes a set of algebraic constraints. This entropy-current route to ideal RMHD is by now well established and may also be viewed as the macroscopic counterpart of the magnetic one-form symmetry formulation~\cite{GrozdanovHofmanIqbal2017,HongoHattori2021,HattoriHongoHuang2022}. We thus find the following constraints
	\begin{align}
		Ts+\mu n+B^{\mu}H_{\mu}= & \epsilon+p_{\perp},\label{11}\\
		\left(p_{\perp}-p_{\parallel}\right)b^{\nu}+BH^{\nu}= & 0.\label{12}
	\end{align}
	Physically, the first relation, Eq.~\eqref{11}, generalizes the fundamental thermodynamic identity to a medium supporting a dynamical magnetic flux. The second relation, Eq.~\eqref{12}, explicitly demonstrates that a finite magnetic flux induces a mismatch between the transverse and longitudinal pressures, expressed as $p_{\perp}-p_{\parallel}=B^{\mu}H_{\mu}>0$. Furthermore, noting that $H^\mu$ is aligned with $B^\mu$ and adopting the standard constitutive parameterization $H^\mu =- \mu_m^{-1} B^\mu$ (where $\mu_m$ denotes the magnetic permeability), we identify the pressure anisotropy as a direct measure of the magnetic permeability: $\mu_{m}^{-1} = (p_{\perp} - p_{\parallel}) / B^2$, with $B^2 \equiv -B^\mu B_\mu$.
	
	Consistent with these thermodynamic constraints, the zeroth order constitutive relations can be formulated as
	\begin{align}
		T_{(0)}^{\mu\nu}= & \epsilon u^{\mu}u^{\nu}-p_{\perp}\Xi^{\mu\nu}+\left(p_{\perp}-B^{\alpha}H_{\alpha}\right)b^{\mu}b^{\nu},\label{13}\\
		N_{(0)}^{\mu}= & nu^{\mu},\quad\tilde{F}_{(0)}^{\mu\nu}=B^{\mu}u^{\nu}-B^{\nu}u^{\mu}.\label{14}
	\end{align}
	The associated generalized thermodynamic identities for this magnetized medium are then given by
	\begin{align}
		Tds	=&d\epsilon-\mu dn-H_{\mu}dB^{\mu},\label{15}\\
		\epsilon+p_{\perp}	=&Ts+\mu n+B^{\mu}H_{\mu},\label{16}\\
		dp_{\perp}	=&sdT+nd\mu+B^{\mu}dH_{\mu}.\label{17}
	\end{align}
	To prepare for the extension to dissipative regimes, we can rewrite the entropy current in a potential-like form
	\begin{eqnarray}
		\begin{aligned}
			\mathcal{S}_{(0)}^{\mu}= p_{\perp}\beta^{\mu}+\beta_{\nu}T_{(0)}^{\mu\nu}-\alpha N_{(0)}^{\mu}+\mathcal{H}_{\nu}\tilde{F}_{(0)}^{\mu\nu},
		\end{aligned}
		\label{18}
	\end{eqnarray}
	where we defined 
	\begin{eqnarray}
		\beta=T^{-1},\beta^{\mu}=\beta u^{\mu},\alpha=\beta\mu,\mathcal{H}_{\nu}=\beta H_{\nu}.  
		\label{19}
	\end{eqnarray}
	The expression \eqref{18} is the covariant form of the relation \eqref{16}. To proceed further, it is convenient to modify Eq.\eqref{15} and \eqref{17} using the definitions \eqref{19}. We obtain for Eq.~\eqref{17}
	\begin{eqnarray}
		\begin{aligned}
			dp_{\perp}=-\beta^{-1}\left(\epsilon+p_{\perp}\right)d\beta+\beta^{-1}nd\alpha+\beta^{-1}B^{\mu}d\mathcal{H}_{\mu},
		\end{aligned}
		\label{20}
	\end{eqnarray}
	where we used Eq.~\eqref{16} in the last step. Now the first law of thermodynamics and the Gibbs-Duhem relation can be written in an alternative form
	\begin{equation}
		\begin{aligned}
			ds=&\beta d\epsilon-\alpha dn-\mathcal{H}_{\mu}dB^{\mu},\\
			\beta dp_{\perp}=&-\left(\epsilon+p_{\perp}\right)d\beta+nd\alpha+B^{\mu}d\mathcal{H}_{\mu}. 
		\end{aligned}
		\label{21}
	\end{equation}
	With the aid of Eqs.\eqref{20} and\eqref{21} , these relations can be cast into a covariant form
	\begin{align}
		d\left(p_{\perp}\beta^{\mu}\right)=&-T_{(0)}^{\mu\nu}d\beta_{\nu}+N_{(0)}^{\mu}d\alpha-\tilde{F}_{(0)}^{\mu\alpha}d\mathcal{H}_{\alpha},\label{22}\\
		d\mathcal{S}_{(0)}^{\mu}=&\beta_{\nu}dT_{(0)}^{\mu\nu}-\alpha dN_{(0)}^{\mu}+\mathcal{H}_{\nu}d\tilde{F}_{(0)}^{\mu\nu},
	\end{align}
	where we used the second relation in Eq.~\eqref{21}. One should note that, despite their vector form, these equations do not contain more information than the thermodynamic relations.
	
	\section{First-Order Derivative Corrections of RMHD}
	\label{section3}
	
	To account for dissipative processes in a magnetized medium, we extend the ideal RMHD framework by decomposing the energy-momentum tensor, particle-number current, and dual field-strength tensor into their respective ideal and first-order dissipative components
	\begin{align}
		T^{\mu\nu}	=&T_{(0)}^{\mu\nu}+T_{(1)}^{\mu\nu}=\epsilon u^{\mu}u^{\nu}-\left(p_{\perp}+\Pi_{\perp}\right)\Xi^{\mu\nu}\nonumber\\
		&+\left(p_{\parallel}+\Pi_{\parallel}\right)b^{\mu}b^{\nu}+f^{\mu}b^{\nu}+f^{\nu}b^{\mu}+\pi_{\perp}^{\mu\nu},\label{24}\\
		N^{\mu}	=&N_{(0)}^{\mu}+N_{(1)}^{\mu}=nu^{\mu}+\nu_{\perp}^{\mu},\label{25}\\
		\tilde{F}^{\mu\nu}	=&\tilde{F}_{(0)}^{\mu\nu}+\tilde{F}_{(1)}^{\mu\nu}=B^{\mu}u^{\nu}-B^{\nu}u^{\mu}+b^{\mu}\ell^{\nu}-b^{\nu}\ell^{\mu}+m_{\perp}^{\mu\nu}.\label{26}   
	\end{align}
	In this decomposition, the hydrodynamic variables are defined via the following orthogonal projections
	\begin{eqnarray}
		\begin{aligned}
			\epsilon	=&T^{\mu\nu}u_{\mu}u_{\nu},\,p_{\perp}+\Pi_{\perp}=-\frac{1}{2}T^{\mu\nu}\Xi_{\mu\nu},\,p_{\parallel}+\Pi_{\parallel}=T^{\mu\nu}b_{\mu}b_{\nu},\\
			f^{\mu}=&-\Xi_{\alpha}^{\mu}T^{\alpha\beta}b_{\beta},\,
			\pi_{\perp}^{\mu\nu}	=\Xi_{\alpha\beta}^{\mu\nu}T^{\alpha\beta},\,n=N^{\mu}u_{\mu},\,\nu_{\perp}^{\mu}=\Xi_{\nu}^{\mu}N^{\nu},\\
			B^{\mu}=&\tilde{F}^{\mu\nu}u_{\nu},\quad\ell^{\nu}=-\Xi_{\alpha}^{\nu}b_{\mu}\tilde{F}^{\mu\alpha},\quad m_{\perp}^{\mu\nu}=\Xi^{\mu[\alpha}\Xi^{\beta]\nu}\tilde{F}_{\alpha\beta}.
		\end{aligned}
		\label{27}
	\end{eqnarray}
	Here, $\Pi_\perp$ and $\Pi_\parallel$ represent the anisotropic bulk viscous pressures, $f^\mu$ denotes the transverse vector associated with the mixed shear channel, and $\pi_{\perp}^{\mu\nu}$ is the transverse shear stress tensor. The vectors $\nu_\perp^\mu$ and $\ell^\mu$ correspond to charge diffusion and magnetic induction effects, respectively. The match conditions are discussed in Appendix~\ref{appendixg}. This framework is consistent with kinetic-theory derivations of dissipative magnetohydrodynamics~\cite{HernandezKovtun2017,Denicol2018,Denicol2019,PandaRoy2021,HattoriHongoHuang2022}.
	
	The macroscopic evolution of these dissipative degrees of freedom is governed by the conservation laws $\partial_\mu T^{\mu\nu}=0$, $\partial_\mu N^\mu=0$, and $\partial_\mu \tilde{F}^{\mu\nu}=0$, which yield the following modified evolution equations
	\begin{align}
		0	=&D\epsilon+\epsilon\theta+\left(p_{\perp}+\Pi_{\perp}\right)\theta_{\perp}+\left(p_{\parallel}+\Pi_{\parallel}\right)\theta_{\parallel}\nonumber\\
		&-\left(b^{\mu}f^{\nu}+b^{\nu}f^{\mu}\right)\partial_{\mu}u_{\nu}-\pi_{\perp}^{\mu\nu}\partial_{\mu}u_{\nu},\label{28}\\
		0	=&Dn+n\theta+\partial_{\mu}\nu_{\perp}^{\mu},\label{29}\\
		0	=&H_{\nu}B^{\mu}\partial_{\mu}u^{\nu}-H_{\nu}DB^{\nu}-H_{\nu}B^{\nu}\theta+H_{\nu}\ell^{\nu}\partial_{\mu}b^{\mu}\nonumber\\
		&+H_{\nu}b^{\mu}\partial_{\mu}\ell^{\nu}-H_{\nu}\ell^{\mu}\partial_{\mu}b^{\nu}-H_{\nu}b^{\nu}\partial_{\mu}\ell^{\mu}+H_{\nu}\partial_{\mu}m_{\perp}^{\mu\nu}.\label{30}    
	\end{align}
	For an interacting fluid, one generalizes the ideal entropy current by including the dissipative corrections to the constitutive relations. This is the standard starting point for deriving first-order RMHD transport constraints from the local second law \cite{HernandezKovtun2017,Denicol2018,HongoHattori2021}. To derive the constitutive relations, we evaluate the divergence of the non-equilibrium entropy current in the Navier-Stokes (NS) limit. Starting from the ansatz
	\begin{eqnarray}
		\begin{aligned}
			\mathcal{S}_{\text{NS}}^{\mu}= & p_{\perp}\beta^{\mu}+\beta_{\nu}T^{\mu\nu}-\alpha N^{\mu}+\mathcal{H}_{\nu}\tilde{F}^{\mu\nu}\\
			= & p_{\perp}\beta^{\mu}+\beta_{\nu}T_{(0)}^{\mu\nu}+\beta_{\nu}T_{(1)}^{\mu\nu}-\alpha N_{(0)}^{\mu}\\
			& -\alpha N_{(1)}^{\mu}+\mathcal{H}_{\nu}\tilde{F}_{(0)}^{\mu\nu}+\mathcal{H}_{\nu}\tilde{F}_{(1)}^{\mu\nu}+\mathcal{O}\left(\partial^{2}\right)\\
			= & \mathcal{S}_{(0)}^{\mu}+\beta_{\nu}T_{(1)}^{\mu\nu}-\alpha N_{(1)}^{\mu}+\mathcal{H}_{\nu}\tilde{F}_{(1)}^{\mu\nu}+\mathcal{O}\left(\partial^{2}\right),
		\end{aligned}  
		\label{31}
	\end{eqnarray}
	where we make use of the equilibrium entropy current $\mathcal{S}_{(0)}^{\mu}$ defined in Eq.~\eqref{18}. The divergence of the entropy current \eqref{31} can now be expressed as
	\begin{eqnarray}
		\begin{aligned}
			\partial_{\mu}\mathcal{S}_{\text{NS}}^{\mu}= & \partial_{\mu}\left(p_{\perp}\beta^{\mu}+\beta_{\nu}T^{\mu\nu}-\alpha N^{\mu}+\mathcal{H}_{\nu}\tilde{F}^{\mu\nu}+\mathcal{O}\left(\partial^{2}\right)\right)\\
			=&T_{(1)}^{\mu\nu}\partial_{\mu}\beta_{\nu}-N_{(1)}^{\mu}\partial_{\mu}\alpha+\tilde{F}_{(1)}^{\mu\nu}\partial_{\mu}\mathcal{H}_{\nu}+\mathcal{O}\left(\partial^{3}\right).
		\end{aligned}   
		\label{32}
	\end{eqnarray}  
	Substituting the specific tensor structures defined in Eq.~\eqref{24}-\eqref{26} into the entropy production rate, we obtain
	\begin{equation}
		\begin{aligned}
			\partial_{\mu}\mathcal{S}_{\text{NS}}^{\mu}= & -\beta\Pi_{\perp}\theta_{\perp}-\beta\Pi_{\parallel}\theta_{\parallel}+2f^{\mu}b^{(\alpha}\Xi_{\mu}^{\beta)}\partial_{\alpha}\beta_{\beta}+\beta\pi_{\perp}^{\mu\nu}\Xi_{\mu\nu}^{\alpha\beta}\partial_{\alpha}u_{\beta}\\
			&-\nu_{\perp}^{\mu}\Xi_{\mu}^{\rho}\partial_{\rho}\alpha+2\ell^{\mu}b^{[\alpha}\Xi_{\mu}^{\beta]}\partial_{\alpha}\mathcal{H}_{\beta}+m_{\perp}^{\mu\nu}\Xi_{\mu}^{[\alpha}\Xi_{\nu}^{\beta]}\partial_{\alpha}\mathcal{H}_{\beta}.
		\end{aligned}
		\label{33}
	\end{equation}  
	The construction of linear constitutive laws is further constrained by symmetry principles, most notably Curie’s Principle, which dictates that dissipative fluxes only couple to thermodynamic forces of the same tensorial rank and parity. In our magnetized parity-even plasma, the anisotropy vector $b^\mu$ behaves as a pseudovector, causing the thermodynamic forces to bifurcate into two distinct sectors: parity-even $\{\Pi_{\parallel},\Pi_{\perp},f^{\mu},\pi_{\perp}^{\mu\nu}\}$ and parity-odd $\{\nu_{\perp}^{\mu},\ell^{\mu},m_{\perp}^{\mu\nu}\}$.
	
	At first order, the resulting constitutive relations take the same general form as in the kinetic-theory derivations of non-resistive and resistive dissipative RMHD, though here they are obtained from the entropy-current analysis rather than from the Boltzmann or Boltzmann--Vlasov equations \cite{Denicol2018,Denicol2019,PandaRoy2021}.
	By adopting linear relations between the fluxes and their conjugate forces to enforce $\partial_{\mu}\mathcal{S}_{\text{NS}}^{\mu}\geq0$, we arrive at the first-order constitutive relations
	\begin{align}
		\Pi_{\perp}= & -\zeta_{\perp}\theta_{\perp}-\zeta_{\times}\theta_{\parallel},\label{34}\\
		\Pi_{\parallel}= & -\zeta_{\parallel}\theta_{\parallel}-\zeta_{\times}^{\prime}\theta_{\perp},\label{35}\\
		f^{\mu}= & -2\eta_{\parallel}b^{(\alpha}\Xi^{\beta)\mu}\partial_{\alpha}\beta_{\beta},\label{36}\\
		\pi_{\perp}^{\mu\nu}= & 2\eta_{\perp}\Xi^{\mu\nu\alpha\beta}\partial_{\alpha}u_{\beta},\label{37}\\
		\nu_{\perp}^{\mu}= & \kappa\Xi^{\mu\rho}\partial_{\rho}\alpha+2\kappa_{\times}b^{[\alpha}\Xi^{\beta]\mu}\partial_{\alpha}\mathcal{H}_{\beta},\label{38}\\
		\ell^{\mu}= & -2\rho_{\parallel}b^{[\alpha}\Xi^{\beta]\mu}\partial_{\alpha}\mathcal{H}_{\beta}-\kappa_{\times}^{\prime}\Xi^{\mu\rho}\partial_{\rho}\alpha,\label{39}\\
		m_{\perp}^{\mu\nu}= & 2\rho_{\perp}\Xi^{\mu[\alpha}\Xi^{\beta]\nu}\partial_{\alpha}\mathcal{H}_{\beta}.\label{40}
	\end{align}
	To satisfy the second law of thermodynamics, the associated transport matrices must be positive semi-definite. Such positivity conditions are the anisotropic RMHD analogue of the usual thermodynamic constraints on first-order transport matrices and are also consistent with the structure found in kinetic-theory approaches \cite{Denicol2018,Biswas2020}. For the anisotropic bulk-viscosity sector, substituting Eqs.~\eqref{34}--\eqref{35} into Eq.~\eqref{33} yields the quadratic form
	\begin{eqnarray*}
		-\beta\Pi_{\perp}\theta_{\perp}-\beta\Pi_{\parallel}\theta_{\parallel}=\beta\left(\begin{array}{cc}
			\theta_{\perp} & \theta_{\parallel}\end{array}\right)\left(\begin{array}{cc}
			\zeta_{\perp} & \frac{\zeta_{\times}+\zeta_{\times}^{\prime}}{2}\\
			\frac{\zeta_{\times}+\zeta_{\times}^{\prime}}{2} & \zeta_{\parallel}
		\end{array}\right)\left(\begin{array}{c}
			\theta_{\perp}\\
			\theta_{\parallel}
		\end{array}\right)
	\end{eqnarray*}
	According to Sylvester's criterion, this requires $\zeta_\perp\geq 0,\zeta_\parallel\geq 0$, and $\zeta_{\perp}\zeta_{\parallel}-\frac{1}{4}\left(\zeta_{\times}+\zeta_{\times}^{\prime}\right)^{2}\geq0$. In the charge-conjugate symmetric limit where $\zeta_\times=\zeta_\times^\prime$, this simplifies to $\zeta_{\perp}\zeta_{\parallel}-\zeta_{\times}^{2}\geq0$.
	
	A similar requirement for semi-positivity applies to the coupled diffusion-induction sector, which characterizes the interplay between charge transport and magnetic field evolution. Identifying the thermodynamic forces $X_{\mu}=\Xi_{\mu}^{\rho}\partial_{\rho}\alpha$ and $Y_{\mu}=2b^{[\alpha}\Xi_{\mu}^{\beta]}\partial_{\alpha}\mathcal{H}_{\beta}$, the entropy production rate from vector dissipation becomes
	\begin{eqnarray*}
		\begin{aligned} 
			-\nu_{\perp}^{\mu}\Xi_{\mu}^{\rho}\partial_{\rho}\alpha+2\ell^{\mu}b^{[\alpha}\Xi_{\mu}^{\beta]}\partial_{\alpha}\mathcal{H}_{\beta}=-\left(\begin{array}{cc}
				X^{\mu} & Y^{\mu}\end{array}\right)\left(\begin{array}{cc}
				\kappa & \frac{\kappa_{\times}+\kappa_{\times}^{\prime}}{2}\\
				\frac{\kappa_{\times}+\kappa_{\times}^{\prime}}{2} & \rho_{\parallel}
			\end{array}\right)\left(\begin{array}{c}
				X_{\mu}\\
				Y_{\mu}
			\end{array}\right)
		\end{aligned}
	\end{eqnarray*}
	In a relativistic framework with signature $(+,-,-,-)$, the spatial projection ensures $X^{\mu}X_{\mu}\leq0$ and $Y^\mu Y_\mu\leq0$, requiring the symmetric transport matrix to be positive semi-definite.
	
	In summary, these stability conditions ensure that the intensity of cross-coupling effects does not overwhelm the direct dissipation channels. Specifically, we require $\kappa \ge 0, \rho_\parallel \ge 0$, and $\kappa\rho_{\parallel} - \frac{1}{4}(\kappa_{\times}+\kappa_{\times}^{\prime})^{2} \ge 0$. While these coefficients are related via Onsager reciprocity $\kappa_{\times}\left(\boldsymbol{B}\right)=\kappa_{\times}^{\prime}\left(-\boldsymbol{B}\right)$, they vanish identically in a charge-conjugate symmetric plasma. Collectively, the non-negativity of $\{\zeta_{\perp}, \zeta_{\parallel}, \eta_{\parallel}, \eta_{\perp}, \kappa, \rho_{\parallel}, \rho_{\perp}\}$ ensures a thermodynamically consistent Navier–Stokes limit for RMHD, preserving the thermodynamic arrow of time under magnetic anisotropy.
	
	\section{Second-Order Derivative Corrections of RMHD}
	\label{section4}
	
	The inherent pathologies of relativistic Navier-Stokes (NS) theory, particularly regarding stability and causality, necessitate a transition to a second-order transient framework. Historically, it has been established that the first-order NS formulation is ill-posed for relativistic fluids. When subjected to linear perturbations around a global equilibrium, the system often exhibits non-physical instabilities where disturbances grow exponentially rather than decaying \cite{IsraelStewart1979, HiscockLindblom1983, HiscockLindblom1985}. Crucially, while these issues may remain hidden in the local rest frame, they become manifest and catastrophic in Lorentz-boosted frames. These instabilities are deeply rooted in the parabolic nature of the NS equations, which permit infinite signal propagation speeds, thereby violating the fundamental principle of relativistic causality.
	
	To resolve these conceptual conflicts, we adopt the Israel-Stewart (IS) approach, which restores hyperbolicity by promoting dissipative fluxes to dynamical variables. By introducing finite relaxation times for the dissipative processes,
	the IS framework replaces the parabolic Navier--Stokes dynamics by a
	transient structure that allows for finite characteristic propagation
	speeds, subject to additional constraints on the transport coefficients
	and relaxation times \cite{DenicolNiemiMolnarRischke2012}. In the context of magnetohydrodynamics, this transient sector is further enriched by the breakdown of rotational symmetry, leading to anisotropic transport channels and intricate couplings between the fluid and the magnetic field \cite{Denicol2018, Denicol2019, Biswas2020, HoultKovtun2025, MolnarRischke2025}.
	
	The construction of this second-order theory begins with an expansion of the entropy current $\mathcal{S}^\mu$ for states near local equilibrium. We define the IS entropy current as
	\begin{eqnarray}
		\begin{aligned}
			\mathcal{S}_{\text{IS}}^{\mu}= & p_{\perp}\beta^{\mu}+\beta_{\nu}T^{\mu\nu}-\alpha N^{\mu}+\mathcal{H}_{\nu}\tilde{F}^{\mu\nu}+Q^{\mu}\\
			= & p_{\perp}\beta^{\mu}+\beta_{\nu}T_{(0)}^{\mu\nu}+\beta_{\nu}T_{(1)}^{\mu\nu}-\alpha N_{(0)}^{\mu}-\alpha N_{(1)}^{\mu}+\mathcal{H}_{\nu}\tilde{F}_{(0)}^{\mu\nu}+\mathcal{H}_{\nu}\tilde{F}_{(1)}^{\mu\nu}+Q^{\mu}\\
			= & \mathcal{S}_{\text{NS}}^{\mu}+Q^{\mu}.
		\end{aligned}
		\label{41}
	\end{eqnarray}
	where $\mathcal{S}_{\text{NS}}^{\mu}$ represents the first-order corrections, and $Q^\mu$ is a general four-vector containing quadratic terms in the dissipative fluxes: $\Pi_{\perp},\Pi_{\parallel},f^{\mu},\pi_{\perp}^{\mu\nu},\nu_{\perp}^{\mu},\ell^{\mu}$, and $m_{\perp}^{\mu\nu}$. 
	
	The functional form of $Q^\mu$ is constrained by the requirement that entropy must be maximized at equilibrium. To analyze the thermodynamic density in the fluid frame, we contract the second-order entropy current \eqref{41} with the four-velocity $u_\mu$
	\begin{eqnarray}
		u_{\mu}\mathcal{S}_{\text{IS}}^{\mu}=u_{\mu}\mathcal{S}_{\text{NS}}^{\mu}+u_{\mu}Q^{\mu}.
		\label{42}
	\end{eqnarray}
	Utilizing the first-order expression for $\mathcal{S}_{\text{NS}}^{\mu}$ from Eq.~\eqref{31}, we find that several dissipative contributions vanish due to the orthogonality conditions $u_\mu T_{(1)}^{\mu\nu} = 0$, $u_\mu N_{(1)}^\mu = 0$, and $u_\mu \tilde{F}_{(1)}^{\mu\nu} = 0$ in the Landau-Lifshitz frame
	\begin{eqnarray}
		\begin{aligned}
			u_{\mu}\mathcal{S}_{\text{IS}}^{\mu}	
			=u_{\mu}\mathcal{S}_{(0)}^{\mu}+u_{\mu}Q^{\mu}.
		\end{aligned}
		\label{43}
	\end{eqnarray}
	By substituting the equilibrium entropy current $\mathcal{S}_{(0)}^{\mu}$ from Eq.~\eqref{18} and expanding the tensor structures, we obtain the projected IS entropy density
	\begin{equation}
		\begin{aligned}
			u_{\mu}\mathcal{S}_{\text{IS}}^{\mu}	=\beta\left[\left(\epsilon+p_{\perp}-n\mu-H_{\nu}B^{\nu}\right)\right]+u_{\mu}Q^{\mu}.
		\end{aligned}
		\label{44}
	\end{equation}
	Using the generalized first law of thermodynamics provided in Eq.~\eqref{11}, we derive the following relation
	\begin{eqnarray}
		u_{\mu}\mathcal{S}_{\text{IS}}^{\mu}&=s+u_{\mu}Q^{\mu}.
		\label{45}
	\end{eqnarray}
	Given that the entropy density has a local maximum at equilibrium, the second-order correction $Q^\mu$ is subject to the fundamental thermodynamic constraint
	\begin{eqnarray}
		u_{\mu}Q^{\mu}\leq0.
		\label{46}
	\end{eqnarray}
	To satisfy this stability condition, we construct $Q^\mu$ from the symmetry-allowed second-order structures retained in the present truncation
	
	\begin{eqnarray}
		\begin{aligned}
			Q^{\mu}= & u^{\mu}\left(a_{1}\Pi_{\perp}^{2}+a_{2}\Pi_{\parallel}^{2}+a_{3}f^{\lambda}f_{\lambda}+a_{4}\nu_{\perp}^{\lambda}\nu_{\perp\lambda}+a_{5}\ell^{\lambda}\ell_{\lambda}+a_{6}\pi_{\perp}^{\lambda\nu}\pi_{\perp\lambda\nu}+a_{7}m_{\perp}^{\lambda\nu}m_{\perp\lambda\nu}\right)\\
			& +\left(b_{1}\Pi_{\perp}f^{\mu}+b_{2}\Pi_{\perp}\nu_{\perp}^{\mu}+b_{3}\Pi_{\perp}\ell^{\mu}+b_{4}\Pi_{\parallel}f^{\mu}+b_{5}\Pi_{\parallel}\nu_{\perp}^{\mu}+b_{6}\Pi_{\parallel}\ell^{\mu}\right)\\
			& +\left(c_{1}\pi_{\perp}^{\mu\nu}f_{\nu}+c_{2}\pi_{\perp}^{\mu\nu}\nu_{\perp\nu}+c_{3}\pi_{\perp}^{\mu\nu}\ell_{\nu}+c_{4}m_{\perp}^{\mu\nu}f_{\nu}+c_{5}m_{\perp}^{\mu\nu}\nu_{\perp\nu}+c_{6}m_{\perp}^{\mu\nu}\ell_{\nu}\right).
		\end{aligned}
		\label{47}
	\end{eqnarray}    
	
	The formulation of $Q^\mu$ in Eq.~\eqref{47} is guided by both thermodynamic consistency and phenomenological simplicity. Specifically, we exclude non-linear cross-coupling and axial scalar-vector terms based on the following considerations. First, the entropy-density correction is restricted to a diagonalized quadratic form, a standard approach in transient hydrodynamic theories that ensures a transparent derivation of relaxation equations and coefficient constraints \cite{IsraelStewart1979, DenicolNiemiMolnarRischke2012, Harutyunyan2018}. By omitting cross-coupling terms between different dissipative sectors (e.g., $f^{\mu}\nu_{\perp\mu}$), the stability requirement $u_\mu Q^\mu \le 0$ remains analytically tractable. This diagonalization ensures that the system’s stability is guaranteed through straightforward sign constraints on the individual coefficients $a_1$ to $a_7$, circumventing the complexities of enforcing semi-definiteness on a high-dimensional coefficient matrix. Second, axial terms such as $b^\mu \Pi^2_\perp$ are neglected. Since the magnetic field is transverse to the fluid flow ($u_\mu b^\mu = 0$), these terms do not contribute to the second-order entropy density correction. Furthermore, there is no strong phenomenological justification for a scenario where pure scalar fluctuations drive an independent entropy flux exclusively along the direction of the magnetic field. The dimensionful coefficients $a_i, b_i$, and $c_i$ represent the coupling strengths of these dissipative interactions. While the stability condition $u_\mu Q^\mu \leq 0$ imposes definite signatures on the diagonal coefficients—namely $a_1, a_2, a_6, a_7 \leq 0$ and $a_3, a_4, a_5 \geq 0$ (noting that $f_\lambda, \nu_\perp^\lambda, \ell^\lambda$ are space-like)—the cross-coupling coefficients $b_i$ and $c_i$ remain unconstrained by the second law alone. Determining the precise signs and values of these off-diagonal parameters typically requires a more detailed microscopic treatment, such as the kinetic theory approach.
	
	In the first-order Navier-Stokes (NS) limit, the dissipative currents $\{\Pi_\perp, \Pi_\parallel, f^\mu, \pi^{\mu\nu}_\perp, \nu^\mu_\perp, \ell^\mu, m^{\mu\nu}_\perp\}$ are algebraic functions of the primary hydrodynamic gradients. This linear mapping is a direct consequence of the local entropy production constraint $\partial_{\mu}\mathcal{S}_{\text{NS}}^{\mu} \ge 0$. In contrast, a self-consistent second-order theory treats these dissipative fluxes as independent dynamical variables. This transition is formally represented by the inclusion of second-order corrections in the entropy current $\mathcal{S}_{\text{IS}}^\mu$. Consequently, to close the system of hydrodynamic equations, one must derive relaxation-type evolution equations for these fluxes by enforcing the non-negativity of the second-order entropy divergence, $\partial_{\mu}\mathcal{S}_{\text{IS}}^{\mu} \ge 0$.
	
	We begin our derivation by considering the generalized entropy current proposed in Eq.~\eqref{41}. Taking the four-divergence and using the fundamental conservation laws $\partial_\mu T^{\mu\nu}=0$, $\partial_\mu N^\mu=0$, and $\partial_\mu \tilde{F}^{\mu\nu}=0$, we obtain the entropy production rate as
	
	\begin{eqnarray}
		\begin{aligned}
			\partial_{\mu}\mathcal{S}_{\text{IS}}^{\mu}= &\partial_{\mu}\left(p_{\perp}\beta^{\mu}\right)+T_{(0)}^{\mu\nu}\partial_{\mu}\beta_{\nu}+T_{(1)}^{\mu\nu}\partial_{\mu}\beta_{\nu}-N_{(0)}^{\mu}\partial_{\mu}\alpha\\
			&-N_{(1)}^{\mu}\partial_{\mu}\alpha+\tilde{F}_{(0)}^{\mu\nu}\partial_{\mu}\mathcal{H}_{\nu}+\tilde{F}_{(1)}^{\mu\nu}\partial_{\mu}\mathcal{H}_{\nu}+\partial_{\mu}Q^{\mu}.
		\end{aligned}
		\label{48}
	\end{eqnarray}    
	Moreover, using the thermodynamic relations, it can be easily shown that
	\begin{eqnarray}
		\partial_{\mu}\left(p_{\perp}\beta^{\mu}\right)+T_{(0)}^{\mu\nu}\partial_{\mu}\beta_{\nu}-N_{(0)}^{\mu}\partial_{\mu}\alpha+\tilde{F}_{(0)}^{\mu\nu}\partial_{\mu}\mathcal{H}_{\nu}=0,
		\label{49}
	\end{eqnarray}
	the entropy divergence simplifies to a form driven solely by dissipative fluxes and the second-order vector $Q^\mu$
	\begin{eqnarray}
		\partial_{\mu}\mathcal{S}_{\text{IS}}^{\mu}=T_{(1)}^{\mu\nu}\partial_{\mu}\beta_{\nu}-N_{(1)}^{\mu}\partial_{\mu}\alpha+\tilde{F}_{(1)}^{\mu\nu}\partial_{\mu}\mathcal{H}_{\nu}+\partial_{\mu}Q^{\mu}.     
		\label{50}
	\end{eqnarray}
	Substituting the explicit tensor structures for the dissipative components into Eq.~\eqref{50}, we obtain the final expression for the second-order entropy production rate
	
	\begin{eqnarray}
		\begin{aligned}
			\partial_{\mu}\mathcal{S}_{\text{IS}}^{\mu}= & -\beta\Pi_{\perp}\theta_{\perp}-\beta\Pi_{\parallel}\theta_{\parallel}+2f^{\mu}b^{(\alpha}\Xi_{\mu}^{\beta)}\partial_{\alpha}\beta_{\beta}+\beta\pi_{\perp}^{\mu\nu}\Xi_{\mu\nu}^{\alpha\beta}\partial_{\alpha}u_{\beta}-\nu_{\perp}^{\mu}\Xi_{\mu}^{\rho}\partial_{\rho}\alpha\\
			&+2\ell^{\mu}b^{[\alpha}\Xi_{\mu}^{\beta]}\partial_{\alpha}\mathcal{H}_{\beta}+m_{\perp}^{\mu\nu}\Xi_{\mu}^{[\alpha}\Xi_{\nu}^{\beta]}\partial_{\alpha}\mathcal{H}_{\beta}+\partial_{\mu}Q^{\mu}.
		\end{aligned}
		\label{51}
	\end{eqnarray}
	As a final step in our theoretical derivation, we analyze the divergence term $\partial_\mu Q^\mu$. This is achieved by substituting the explicit definition of the second-order entropy current $Q^\mu$ from Eq.~\eqref{47} into the divergence expression. To begin the derivation of the constituent scalars, vectors, and tensors, we first compute the partial derivative of $Q^\mu$. It is important to account for the spatial and temporal gradients of the parameters $a_i, b_i$, and $c_i$, which are generally non-vanishing. Subsequently, we reorganize it by grouping terms associated with common dissipative currents. During this grouping, certain cross-terms arise, such as $\Pi_{\perp}f^{\mu}\tilde{\nabla}_{\mu}b_{1}$, which require careful partitioning. To handle these systematically, we introduce a set of partitioning constants, $l$ and $\tilde{l}$, which satisfy the following decomposition identity
	\begin{equation}
		\Pi_{\perp}f^{\mu}\tilde{\nabla}_{\mu}b_{1}=l_{\Pi_{\perp}f}\Pi_{\perp}f^{\mu}\tilde{\nabla}_{\mu}b_{1}+\left(1-l_{\Pi_{\perp}f}\right)\Pi_{\perp}f^{\mu}\tilde{\nabla}_{\mu}b_{1}.
		\label{52}
	\end{equation}
	To derive the causal evolution equations for dissipative fluxes, we perform an orthogonal decomposition of the partial derivative operator: $\partial_{\mu} = u_{\mu}D + b_{\mu}\tilde{D} + \tilde{\nabla}_{\mu}$. Here, $D = u^{\mu}\partial_{\mu}$ denotes the material derivative along the fluid four-velocity, $\tilde{D} = -b^{\mu}\partial_{\mu}$ represents the spatial derivative along the magnetic field line, and $\tilde{\nabla}_{\mu} = \Xi_{\mu}^{\nu}\partial_{\nu}$ is the transverse gradient operator. Applying this decomposition to the general form of the entropy current, we obtain the structured divergence
	
	\begin{eqnarray}
		\begin{aligned}\partial_{\mu}Q^{\mu}= \Pi_{\perp}\mathcal{A}+\Pi_{\parallel}\mathcal{B}+f^{\lambda}\mathcal{C}_{\lambda}+\nu_{\perp}^{\lambda}\mathcal{D}_{\lambda}+\ell^{\lambda}\mathcal{E}_{\lambda}+\pi_{\perp}^{\lambda\nu}\mathcal{F}_{\lambda\nu}+m_{\perp}^{\lambda\nu}\mathcal{G}_{\lambda\nu}.
		\end{aligned}
		\label{53}
	\end{eqnarray}
	In this expression, the scalars $\{\mathcal{A}, \mathcal{B}\}$, vectors $\{\mathcal{C}_\lambda, \mathcal{D}_\lambda, \mathcal{E}_\lambda\}$, and tensors $\{\mathcal{F}_{\lambda\nu}, \mathcal{G}_{\lambda\nu}\}$, which encompass the complex interplay between hydrodynamic gradients and second-order coupling terms, are defined in Appendix~\ref{partial Q}.
	
	Note that the dissipative fluxes multiplying these quantities satisfy the following properties: $f^{\lambda},\nu_{\perp}^{\lambda}$, $\ell^{\lambda}$ are orthogonal to both $u^\mu$ and $b^\mu$, $\pi_{\perp}^{\lambda\nu}$ is also orthogonal to both $u^\mu$ and $b^\mu$ as well as symmetric and traceless, $m_{\perp}^{\lambda\nu}$ is also orthogonal to both $u^\mu$ and $b^\mu$ as well as antisymmetric. The full form of the divergence of entropy current in the second-order theory can be written as
	\begin{eqnarray}
		\begin{aligned}
			\partial_{\mu}\mathcal{S}_{\text{IS}}^{\mu}= & -\beta\Pi_{\perp}\left(\theta_{\perp}-T\mathcal{A}\right)-\beta\Pi_{\parallel}\left(\theta_{\parallel}-T\mathcal{B}\right)+f^{\mu}\left(2b^{(\alpha}\Xi_{\mu}^{\beta)}\partial_{\alpha}\beta_{\beta}+\mathcal{C}_{\mu}\right)\\
			&+\beta\pi_{\perp}^{\mu\nu}\left(\Xi_{\mu\nu}^{\alpha\beta}\partial_{\alpha}u_{\beta}+T\mathcal{F}_{\mu\nu}\right)-\nu_{\perp}^{\mu}\left(\Xi_{\mu}^{\rho}\partial_{\rho}\alpha-\mathcal{D}_{\mu}\right)\\
			&+\ell^{\mu}\left(2b^{[\alpha}\Xi_{\mu}^{\beta]}\partial_{\alpha}\mathcal{H}_{\beta}+\mathcal{E}_{\mu}\right)+m_{\perp}^{\mu\nu}\left(\Xi_{\mu}^{[\alpha}\Xi_{\nu}^{\beta]}\partial_{\alpha}\mathcal{H}_{\beta}+\mathcal{G}_{\mu\nu}\right).
		\end{aligned}
		\label{61}
	\end{eqnarray}    
	To ensure local thermodynamic consistency ($\partial_{\mu}\mathcal{S}_{\text{IS}}^{\mu} \ge 0$), the dissipative currents must satisfy the following relaxation-type constitutive relations
	\begin{align}
		\Pi_{\perp}= & -\zeta_{\perp}\left(\theta_{\perp}-T\mathcal{A}\right)-\zeta_{\times}\theta_{\parallel},\label{62}\\
		\Pi_{\parallel}= & -\zeta_{\parallel}\left(\theta_{\parallel}-T\mathcal{B}\right)-\zeta_{\times}^{\prime}\theta_{\perp},\label{63}\\
		f^{\mu}= & -\eta_{\parallel}\left(2b^{(\alpha}\Xi^{\beta)\mu}\partial_{\alpha}\beta_{\beta}+\mathcal{C}^{\mu}\right),\label{64}\\
		\pi_{\perp}^{\mu\nu}= & 2\eta_{\perp}\left(\Xi^{\mu\nu\alpha\beta}\partial_{\alpha}u_{\beta}+T\mathcal{F}^{\mu\nu}\right),\label{65}\\
		\nu_{\perp}^{\mu}= & \kappa\left(\Xi^{\mu\rho}\partial_{\rho}\alpha-\mathcal{D}^{\mu}\right)+2\kappa_{\times}b^{[\alpha}\Xi^{\beta]\mu}\partial_{\alpha}\mathcal{H}_{\beta},\label{66}\\
		\ell^{\mu}= & -\rho_{\parallel}\left(2b^{[\alpha}\Xi^{\beta]\mu}\partial_{\alpha}\mathcal{H}_{\beta}+\mathcal{E}^{\mu}\right)-\kappa_{\times}^{\prime}\Xi^{\mu\rho}\partial_{\rho}\alpha,\label{67}\\
		m_{\perp}^{\mu\nu}= & 2\rho_{\perp}\left(\Xi^{\mu[\alpha}\Xi^{\beta]\nu}\partial_{\alpha}\mathcal{H}_{\beta}+\mathcal{G}^{\mu\nu}\right).\label{68}
	\end{align}
	Substituting the explicit forms of the thermodynamic forces into Eqs.~\eqref{61} yields the final evolution equations for the dissipative fluxes. These equations represent the magnetized analogue of the Israel-Stewart equations, characterized by additional anisotropic channels absent in unmagnetized systems \cite{Denicol2018, Denicol2019, PandaRoy2021, HoultKovtun2025, MolnarRischke2025}. For instance, the evolution of the transverse bulk pressure $\Pi_{\perp}$ is given by

	\begin{eqnarray}
		\begin{aligned}
			\tau_{\Pi_{\perp}}D\Pi_{\perp}+\Pi_{\perp}= & -\zeta_{\perp}\theta_{\perp}-\zeta_{\times}\theta_{\parallel}+\zeta_{\perp}T\biggl[a_{1}\Pi_{\perp}\theta_{\perp}+a_{1}\Pi_{\perp}\theta_{\parallel}+\Pi_{\perp}Da_{1}+l_{\Pi_{\perp}f}f^{\mu}\tilde{\nabla}_{\mu}b_{1}\\
			&-\tilde{l}_{\Pi_{\perp}f}b_{1}f^{\mu}Du_{\mu} -\left(1-l_{f\Pi_{\perp}}\right)b_{1}f^{\mu}\tilde{D}b_{\mu}+b_{1}\tilde{\nabla}_{\mu}f^{\mu}+l_{\Pi_{\perp}\nu}\nu_{\perp}^{\mu}\tilde{\nabla}_{\mu}b_{2}\\
			&-\tilde{l}_{\Pi_{\perp}\nu}b_{2}\nu_{\perp}^{\mu}Du_{\mu}-\left(1-l_{\nu\Pi_{\perp}}\right)b_{2}\nu_{\perp}^{\mu}\tilde{D}b_{\mu}+b_{2}\tilde{\nabla}_{\mu}\nu_{\perp}^{\mu}+l_{\Pi_{\perp}\ell}\ell^{\mu}\tilde{\nabla}_{\mu}b_{3}\\
			&-\tilde{l}_{\Pi_{\perp}\ell}b_{3}\ell^{\mu}Du_{\mu}-\left(1-l_{\ell\Pi_{\perp}}\right)b_{3}\ell^{\mu}\tilde{D}b_{\mu}+b_{3}\tilde{\nabla}_{\mu}\ell^{\mu}\biggr],
		\end{aligned}
		\label{70}
	\end{eqnarray}    
	Similar dynamical closures are obtained for $\Pi_{\parallel}, f^\mu, \nu_{\perp}^\mu, \ell^\mu, \pi_{\perp}^{\mu\nu},$ and $m_{\perp}^{\mu\nu}$
	
	\begin{eqnarray}
		\begin{aligned}
			\tau_{\Pi_{\parallel}}D\Pi_{\parallel}+\Pi_{\parallel}	=&-\zeta_{\parallel}\theta_{\parallel}-\zeta_{\times}^{\prime}\theta_{\perp}+\zeta_{\parallel}T\biggl[a_{2}\Pi_{\parallel}\theta_{\perp}+a_{2}\Pi_{\parallel}\theta_{\parallel}+\Pi_{\parallel}Da_{2}+l_{\Pi_{\parallel}f}f^{\mu}\tilde{\nabla}_{\mu}b_{4}\\
			&-\tilde{l}_{\Pi_{\parallel}f}b_{4}f^{\mu}Du_{\mu}-\left(1-l_{f\Pi_{\parallel}}\right)b_{4}f^{\mu}\tilde{D}b_{\mu}+b_{4}\tilde{\nabla}_{\mu}f^{\mu}+l_{\Pi_{\parallel}\nu}\nu_{\perp}^{\mu}\tilde{\nabla}_{\mu}b_{5}\\
			&-\tilde{l}_{\Pi_{\parallel}\nu}b_{5}\nu_{\perp}^{\mu}Du_{\mu}-\left(1-l_{\nu\Pi_{\parallel}}\right)b_{5}\nu_{\perp}^{\mu}\tilde{D}b_{\mu}+b_{5}\tilde{\nabla}_{\mu}\nu_{\perp}^{\mu}+l_{\Pi_{\parallel}\ell}\ell^{\mu}\tilde{\nabla}_{\mu}b_{6}\\
			&-\tilde{l}_{\Pi_{\parallel}\ell}b_{6}\ell^{\mu}Du_{\mu}-\left(1-l_{\ell\Pi_{\parallel}}\right)b_{6}\ell^{\mu}\tilde{D}b_{\mu}+b_{6}\tilde{\nabla}_{\mu}\ell^{\mu}\biggr],
		\end{aligned}
		\label{71}
	\end{eqnarray}

	\begin{eqnarray}
		\begin{aligned}
			\tau_{f}\Xi_{\lambda}^{\nu}Df_{\nu}+f_{\lambda}	=&-2\eta_{\parallel}b^{(\alpha}\Xi_{\lambda}^{\beta)}\partial_{\alpha}\beta_{\beta}-\eta_{\parallel}\biggl[a_{3}f_{\lambda}\theta_{\perp}+a_{3}f_{\lambda}\theta_{\parallel}+f_{\lambda}Da_{3}+\left(1-l_{\Pi_{\perp}f}\right)\Pi_{\perp}\tilde{\nabla}_{\lambda}b_{1}\\
			&+b_{1}\tilde{\nabla}_{\lambda}\Pi_{\perp}-\left(1-\tilde{l}_{\Pi_{\perp}f}\right)b_{1}\Pi_{\perp}\Xi_{\lambda}^{\nu}Du_{\nu}
			-l_{f\Pi_{\perp}}b_{1}\Pi_{\perp}\Xi_{\lambda}^{\nu}\tilde{D}b_{\nu}\\
			&+\left(1-l_{\Pi_{\parallel}f}\right)\Pi_{\parallel}\tilde{\nabla}_{\lambda}b_{4}+b_{4}\tilde{\nabla}_{\lambda}\Pi_{\parallel}-\left(1-\tilde{l}_{\Pi_{\parallel}f}\right)b_{4}\Pi_{\parallel}\Xi_{\lambda}^{\nu}Du_{\nu}\\
			&-l_{f\Pi_{\parallel}}b_{4}\Pi_{\parallel}\Xi_{\lambda}^{\nu}\tilde{D}b_{\nu}
			+l_{f\pi}\pi_{\perp\mu\lambda}\tilde{\nabla}^{\mu}c_{1}-\tilde{l}_{f\pi}c_{1}\pi_{\perp\mu\lambda}Du^{\mu}\\
			&-\left(1-l_{\pi f}\right)c_{1}\pi_{\perp\mu\lambda}\tilde{D}b^{\mu}+c_{1}\Xi_{\lambda}^{\nu}\tilde{\nabla}^{\mu}\pi_{\perp\mu\nu}+l_{fm}m_{\perp\mu\lambda}\tilde{\nabla}^{\mu}c_{4}
			\\
			&-\tilde{l}_{fm}c_{4}m_{\perp\mu\lambda}Du^{\mu}-\left(1-l_{mf}\right)c_{4}m_{\perp\mu\lambda}\tilde{D}b^{\mu}+c_{4}\Xi_{\lambda}^{\nu}\tilde{\nabla}^{\mu}m_{\perp\mu\nu}\biggr],
		\end{aligned}
		\label{72}
	\end{eqnarray}
	
	\begin{eqnarray}
		\begin{aligned}
			\tau_{\nu}\Xi_{\lambda}^{\nu}D\nu_{\perp\nu}+\nu_{\perp\lambda}	=&\kappa\Xi_{\lambda}^{\rho}\partial_{\rho}\alpha+2\kappa_{\times}b^{[\alpha}\Xi_{\lambda}^{\beta]}\partial_{\alpha}\mathcal{H}_{\beta}-\kappa\biggl[a_{4}\nu_{\perp\lambda}\theta_{\perp}+a_{4}\nu_{\perp\lambda}\theta_{\parallel}+\nu_{\perp\lambda}Da_{4}\\
			&+\left(1-l_{\Pi_{\perp}\nu}\right)\Pi_{\perp}\tilde{\nabla}_{\lambda}b_{2}+b_{2}\tilde{\nabla}_{\lambda}\Pi_{\perp}-\left(1-\tilde{l}_{\Pi_{\perp}\nu}\right)b_{2}\Pi_{\perp}\Xi_{\lambda}^{\nu}Du_{\nu}
			\\
			&-l_{\nu\Pi_{\perp}}b_{2}\Pi_{\perp}\Xi_{\lambda}^{\nu}\tilde{D}b_{\nu}+\left(1-l_{\Pi_{\parallel}\nu}\right)\Pi_{\parallel}\tilde{\nabla}_{\lambda}b_{5}+b_{5}\tilde{\nabla}_{\lambda}\Pi_{\parallel}\\
			&-\left(1-\tilde{l}_{\Pi_{\parallel}\nu}\right)b_{5}\Pi_{\parallel}\Xi_{\lambda}^{\nu}Du_{\nu}-l_{\nu\Pi_{\parallel}}b_{5}\Pi_{\parallel}\Xi_{\lambda}^{\nu}\tilde{D}b_{\nu}
			+l_{\nu\pi}\pi_{\perp\mu\lambda}\tilde{\nabla}^{\mu}c_{2}\\
			&-\tilde{l}_{\nu\pi}c_{2}\pi_{\perp\mu\lambda}Du^{\mu}-\left(1-l_{\pi\nu}\right)c_{2}\pi_{\perp\mu\lambda}\tilde{D}b^{\mu}+c_{2}\Xi_{\lambda}^{\nu}\tilde{\nabla}^{\mu}\pi_{\perp\mu\nu}\\
			&+l_{\nu m}m_{\perp\mu\lambda}\tilde{\nabla}^{\mu}c_{5}
			-\tilde{l}_{\nu m}c_{5}m_{\perp\mu\lambda}Du^{\mu}-\left(1-l_{m\nu}\right)c_{5}m_{\perp\mu\lambda}\tilde{D}b^{\mu}\\
			&+c_{5}\Xi_{\lambda}^{\nu}\tilde{\nabla}^{\mu}m_{\perp\mu\nu}\biggr],
		\end{aligned}
		\label{73}
	\end{eqnarray}
	
	\begin{eqnarray}
		\begin{aligned}
			\tau_{\ell}\Xi_{\lambda}^{\nu}D\ell_{\nu}+\ell_{\lambda}	=&-2\rho_{\parallel}b^{[\alpha}\Xi_{\lambda}^{\beta]}\partial_{\alpha}\mathcal{H}_{\beta}-\kappa_{\times}^{\prime}\Xi_{\lambda}^{\rho}\partial_{\rho}\alpha-\rho_{\parallel}\biggl[a_{5}\ell_{\lambda}\theta_{\perp}+a_{5}\ell_{\lambda}\theta_{\parallel}+\ell_{\lambda}Da_{5}\\
			&+\left(1-l_{\Pi_{\perp}\ell}\right)\Pi_{\perp}\tilde{\nabla}_{\lambda}b_{3}+b_{3}\tilde{\nabla}_{\lambda}\Pi_{\perp}-\left(1-\tilde{l}_{\Pi_{\perp}\ell}\right)b_{3}\Pi_{\perp}\Xi_{\lambda}^{\nu}Du_{\nu}
			\\
			&-l_{\ell\Pi_{\perp}}b_{3}\Pi_{\perp}\Xi_{\lambda}^{\nu}\tilde{D}b_{\nu}+\left(1-l_{\Pi_{\parallel}\ell}\right)\Pi_{\parallel}\tilde{\nabla}_{\lambda}b_{6}+b_{6}\tilde{\nabla}_{\lambda}\Pi_{\parallel}\\
			&-\left(1-\tilde{l}_{\Pi_{\parallel}\ell}\right)b_{6}\Pi_{\parallel}\Xi_{\lambda}^{\nu}Du_{\nu}-l_{\ell\Pi_{\parallel}}b_{6}\Pi_{\parallel}\Xi_{\lambda}^{\nu}\tilde{D}b_{\nu}
			+l_{\ell\pi}\pi_{\perp\mu\lambda}\tilde{\nabla}^{\mu}c_{3}\\
			&-\tilde{l}_{\ell\pi}c_{3}\pi_{\perp\mu\lambda}Du^{\mu}-\left(1-l_{\pi\ell}\right)c_{3}\pi_{\perp\mu\lambda}\tilde{D}b^{\mu}+c_{3}\Xi_{\lambda}^{\nu}\tilde{\nabla}^{\mu}\pi_{\perp\mu\nu}+l_{\ell m}m_{\perp\mu\lambda}\tilde{\nabla}^{\mu}c_{6}
			\\
			&-\tilde{l}_{\ell m}c_{6}m_{\perp\mu\lambda}Du^{\mu}-\left(1-l_{m\ell}\right)c_{6}m_{\perp\mu\lambda}\tilde{D}b^{\mu}+c_{6}\Xi_{\lambda}^{\nu}\tilde{\nabla}^{\mu}m_{\perp\mu\nu}\biggr],
		\end{aligned}
		\label{74}
	\end{eqnarray}

	\begin{eqnarray}
		\begin{aligned}
			\tau_{\pi}\Xi_{\lambda\nu}^{\rho\sigma}D\pi_{\perp\rho\sigma}+\pi_{\perp\lambda\nu}	=&2\eta_{\perp}\Xi_{\lambda\nu}^{\alpha\beta}\partial_{\alpha}u_{\beta}+2\eta_{\perp}T\biggl[a_{6}\pi_{\perp\lambda\nu}\theta+\pi_{\perp\lambda\nu}Da_{6}+\left(1-l_{f\pi}\right)f_{(\nu}\tilde{\nabla}_{\lambda)}c_{1}\\
			&-\left(1-\tilde{l}_{f\pi}\right)c_{1}\Xi_{\lambda\nu}^{\rho\sigma}f_{\sigma}Du_{\rho}-l_{\pi f}c_{1}\Xi_{\lambda\nu}^{\rho\sigma}f_{\sigma}\tilde{D}b_{\rho}
			+c_{1}\Xi_{\lambda\nu}^{\rho\sigma}\tilde{\nabla}_{\rho}f_{\sigma}\\
			&+\left(1-l_{\nu\pi}\right)\nu_{\perp(\nu}\tilde{\nabla}_{\lambda)}c_{2}-\left(1-\tilde{l}_{\nu\pi}\right)c_{2}\Xi_{\lambda\nu}^{\rho\sigma}\nu_{\perp\sigma}Du_{\rho}\\
			&-l_{\pi\nu}c_{2}\Xi_{\lambda\nu}^{\rho\sigma}\nu_{\perp\sigma}\tilde{D}b_{\rho}+c_{2}\Xi_{\lambda\nu}^{\rho\sigma}\tilde{\nabla}_{\rho}\nu_{\perp\sigma}
			+\left(1-l_{\ell\pi}\right)\ell_{(\nu}\tilde{\nabla}_{\lambda)}c_{3}\\
			&-\left(1-\tilde{l}_{\ell\pi}\right)c_{3}\Xi_{\lambda\nu}^{\rho\sigma}\ell_{\sigma}Du_{\rho}-l_{\pi\ell}c_{3}\Xi_{\lambda\nu}^{\rho\sigma}\ell_{\sigma}\tilde{D}b_{\rho}+c_{3}\Xi_{\lambda\nu}^{\rho\sigma}\tilde{\nabla}_{\rho}\ell_{\sigma}\biggr],
		\end{aligned}
		\label{75}
	\end{eqnarray}

	\begin{eqnarray}
		\begin{aligned}
			\tau_{m}\Xi_{\lambda}^{[\rho}\Xi_{\nu}^{\sigma]}Dm_{\perp\rho\sigma}+m_{\perp\lambda\nu}=&2\rho_{\perp}\Xi_{\lambda}^{[\alpha}\Xi_{\nu}^{\beta]}\partial_{\alpha}\mathcal{H}_{\beta}+2\rho_{\perp}\biggl[a_{7}m_{\perp\lambda\nu}\theta+m_{\perp\lambda\nu}Da_{7}\\
			&+\left(1-l_{fm}\right)f_{[\nu}\tilde{\nabla}_{\lambda]}c_{4}-\left(1-\tilde{l}_{fm}\right)c_{4}\Xi_{\lambda}^{[\rho}\Xi_{\nu}^{\sigma]}f_{\sigma}Du_{\rho}
			\\
			&-l_{mf}c_{4}\Xi_{\lambda}^{[\rho}\Xi_{\nu}^{\sigma]}f_{\sigma}\tilde{D}b_{\rho}+c_{4}\Xi_{\lambda}^{[\rho}\Xi_{\nu}^{\sigma]}\tilde{\nabla}_{\rho}f_{\sigma}+\left(1-l_{\nu m}\right)\nu_{\perp[\nu}\tilde{\nabla}_{\lambda]}c_{5}\\
			&-\left(1-\tilde{l}_{\nu m}\right)c_{5}\Xi_{\lambda}^{[\rho}\Xi_{\nu}^{\sigma]}\nu_{\perp\sigma}Du_{\rho}
			-l_{m\nu}c_{5}\Xi_{\lambda}^{[\rho}\Xi_{\nu}^{\sigma]}\nu_{\perp\sigma}\tilde{D}b_{\rho}\\
			&+c_{5}\Xi_{\lambda}^{[\rho}\Xi_{\nu}^{\sigma]}\tilde{\nabla}_{\rho}\nu_{\perp\sigma}+\left(1-l_{\ell m}\right)\ell_{[\nu}\tilde{\nabla}_{\lambda]}c_{6}\\
			&-\left(1-\tilde{l}_{\ell m}\right)c_{6}\Xi_{\lambda}^{[\rho}\Xi_{\nu}^{\sigma]}\ell_{\sigma}Du_{\rho}
			-l_{m\ell}c_{6}\Xi_{\lambda}^{[\rho}\Xi_{\nu}^{\sigma]}\ell_{\sigma}\tilde{D}b_{\rho}\\
			&+c_{6}\Xi_{\lambda}^{[\rho}\Xi_{\nu}^{\sigma]}\tilde{\nabla}_{\rho}\ell_{\sigma}\biggr].    
		\end{aligned}
		\label{76}
	\end{eqnarray}
	The characteristic relaxation times are defined as
	\begin{eqnarray}
		\begin{aligned}
			\tau_{\Pi_{\perp}}=&-2a_{1}\zeta_{\perp}T\geq 0,\quad \tau_{\Pi_{\parallel}}=-2a_{2}\zeta_{\parallel}T\geq 0,\quad \tau_{f}=2\eta_{\parallel}a_{3}\geq0,\quad \tau_{\nu}=2a_{4}\kappa\geq 0,\\
			\tau_{\ell}=&2a_{5}\rho_{\parallel}\geq 0,\quad\tau_{\pi}=-4a_{6}\eta_{\perp}T\geq 0,\quad\tau_{m}=-4a_{7}\rho_{\perp}\geq0.
		\end{aligned}
		\label{77}
	\end{eqnarray}
	To ensure causality and linear stability, these relaxation times must be non-negative. Collectively, these equations describe how dissipative currents relax toward their respective Navier-Stokes limits, maintaining a thermodynamically consistent arrow of time under magnetic anisotropy.

	\section{Linear mode analysis}\label{S5}
	\label{section5}
	To evaluate the causality and stability of the second-order RMHD framework, we investigate the dynamics of infinitesimal fluctuations around a state of global equilibrium. Each physical variable is decomposed into a uniform background value and a first-order perturbation
	\begin{eqnarray}
		\begin{aligned}
			\epsilon	&\to\epsilon+\delta\epsilon,\quad n\to n_{0}+\delta n,\quad u^{\mu}\to u^{\mu}+\delta u^{\mu},\quad B^{\mu}\to B^{\mu}+\delta B^{\mu},\\
			\Pi_{\perp}	&\to0+\delta\Pi_{\perp},\quad\Pi_{\parallel}\to0+\delta\Pi_{\parallel},\quad f^{\mu}\to0+\delta f^{\mu},\quad\nu_{\perp}^{\mu}\to0+\delta\nu_{\perp}^{\mu},\\
			\ell^{\mu}	&\to0+\delta\ell^{\mu},\quad\pi_{\perp}^{\mu\nu}\to0+\delta\pi_{\perp}^{\mu\nu},\quad m_{\perp}^{\mu\nu}\to0+\delta m_{\perp}^{\mu\nu}.
		\end{aligned}
		\label{perurbation}
	\end{eqnarray}
	The analysis is conducted in the background rest frame $u^{\mu}=\left(1,0,0,0\right)$ with a constant magnetic field oriented along the $z$-axis, $B^{\mu}=\left(0,0,0,B\right)$. In this linearized regime, the background quantities $\{\epsilon, n_0, B\}$ are space-time independent, while all dissipative fluxes vanish at zeroth order.
	
	We adopt a plane-wave ansatz for the fluctuations to derive the characteristic dispersion relations
	\begin{eqnarray}
		\delta f(t,\boldsymbol{x})=\delta f\left(\omega,\boldsymbol{k}\right)e^{-i\omega t+i\boldsymbol{k}\cdot\boldsymbol{x}}.
	\end{eqnarray}

	Applying the Fourier transform maps the differential operators to the algebraic domain via $\partial_{0} \to -i\omega$ and $\partial_{j} \to ik^{j}$. This transformation reduces the system of governing partial differential equations to a secular matrix equation, where the existence of non-trivial solutions requires the vanishing of the characteristic determinant. This approach allows for a rigorous determination of the mode spectrum, ensuring the theory adheres to the constraints of relativistic causality and stability.
	
	In the global equilibrium state, the total energy density and pressures are decomposed into additive contributions from the matter and magnetic sectors
	\begin{eqnarray}
		\epsilon=\varepsilon+\frac{B^{2}}{2\mu_{m}},\quad p_{\perp}=p+\frac{B^{2}}{2\mu_{m}},\quad p_{\parallel}=p-\frac{B^{2}}{2\mu_{m}},
	\end{eqnarray}
	and the enthalpy density $h$ can be defined as 
	\begin{eqnarray}
		h=\epsilon+p_\perp=\varepsilon+p_\perp+\frac{B^{2}}{2\mu_{m}}.
	\end{eqnarray}
	This decomposition of the total energy density and pressure is motivated by the need to distinguish the fluid contributions from the electromagnetic ones, thereby providing a more transparent physical picture of how the magnetic field modifies various collective modes. 
	
	We now collect all the projected conservation laws of energy, momentum, current, and field strength, which can be represented in the following form:
	\be\label{conservation laws before linearize}
	u_{\nu}\partial_{\mu}T^{\mu\nu}&=&0,\quad b_{\nu}\partial_{\mu}T^{\mu\nu}=0,\quad 
	\Xi^{\rho}{}_{\nu}\partial_{\mu}T^{\mu\nu}=0,\quad\partial_{\mu}N^{\mu}=0,\nonumber\\
{	u_\nu\partial_{\mu}\tilde{F}^{\mu\nu}}&{=}&{0},\quad b_{\nu}\partial_{\mu}\tilde{F}^{\mu\nu}=0,\quad
	\Xi_{\nu}^{\rho}\partial_{\mu}\tilde{F}^{\mu\nu}=0.
	\ee
	By linearizing the above conservation equations and the IS evolution equation with respect to the perturbation specified, we can obtain the conservation laws as follows:
	
	\be
	0&=&-i\omega\delta\varepsilon+i\omega\frac{B}{\mu_{m}}\delta B_{z}-\frac{B^{2}}{\mu_{m}}\left(-\partial_{z}\delta u_{z}\right)+h\left(-\partial_{x}\delta u_{x}-\partial_{y}\delta u_{y}-\partial_{z}\delta u_{z}\right),\label{linearized conservation laws}\\
	0&=&-i\omega h(1-v_A^2)\delta u_{z}-ik_{z}\bigg[c_{s}^{2}\delta\varepsilon+\left(\frac{\partial p}{\partial n}\right)\delta n+\delta\Pi_{\|}\bigg]+ik_{x}\delta f_{x}+ik_{y}\delta f_{y},\\
	0&=&i\omega h\delta u_{x}+ik_{x}\bigg[c_{s}^{2}\delta\varepsilon+\left(\frac{\partial p}{\partial n}\right)\delta n-\frac{B}{\mu_{m}}\delta B_{z}+\delta\Pi_{\perp}\bigg]
	+\frac{B}{\mu_{m}}ik_{z}\delta B_{x}\nonumber\\
	&&-ik_{z}\delta f_{x}+ik_{x}\delta\pi_{\perp xx}+ik_{y}\delta\pi_{\perp xy},\\
	0&=&i\omega h\delta u_{y}+ik_{y}\bigg[c_{s}^{2}\delta\varepsilon+\left(\frac{\partial p}{\partial n}\right)\delta n-\frac{B}{\mu_{m}}\delta B_{z}+\delta\Pi_{\perp}\bigg]
	+\frac{B}{\mu_{m}}ik_{z}\delta B_{y}\nonumber\\
	&&-ik_{z}\delta f_{y}+ik_{x}\delta\pi_{\perp xy}-ik_{y}\delta\pi_{\perp xx},\\
	0&=&-i\omega\delta n-in(k_{x}\delta u_{x}+k_{y}\delta u_{y}+k_{z}\delta u_{z})-ik_{x}\delta \nu_{\perp x}-ik_{y}\delta \nu_{\perp y},\\
	{0}&{=}&{ik_x \delta B_x+ik_y \delta B_y+ik_z \delta B_z,}\label{magnetic constraint}\\
	0&=&i\omega\delta B_{z}-iB\left(k_{x}\delta u_{x}+k_{y}\delta u_{y}\right)-ik_{x}\delta\ell_x-ik_{y}\delta\ell_y,\\
	0&=&-i\omega\delta B_{x}-iBk_{z}\delta u_{x}-ik_{z}\delta\ell_x-ik_{y}\delta m_{\perp xy},\\
	0&=&-i\omega\delta B_{y}-iBk_{z}\delta u_{y}-ik_{z}\delta\ell_y+ik_{x}\delta m_{\perp xy}.
	\ee
	where $c_s^2=(\partial p/\partial\varepsilon)_{n,B}$ defines the squared speed of sound. {In particular, the homogeneous Maxwell equation $\partial_\mu\tilde F^{\mu\nu}=0$ contains both evolution equations and a constraint equation. Its projection along the fluid velocity $u_\nu\partial_\mu\tilde F^{\mu\nu}=0$ does not provide an additional independent evolution equation for the magnetic field. Instead, in the equilibrium rest frame it reduces to the magnetic Gauss constraint  $\nabla\cdot\boldsymbol B=0$. The projections along $b^\mu$ and the transverse subspace $\Xi^\rho{}\nu$, on the other hand, determine the induction dynamics of the magnetic-field perturbations.} The linearized IS equation is in the following form 
	\be
	0&=&\left(1-i\omega\tau_{\Pi_{\perp}}\right)\delta\Pi_{\perp}-i\zeta_{\perp}\left(k_{x}\delta u_{x}+k_{y}\delta u_{y}\right)-i\zeta_{\times}k_{z}\delta u_{z},\\
	0&=&\left(1-i\omega\tau_{\Pi_{\|}}\right)\delta\Pi_{\|}-i\zeta_{\|}k_{z}\delta u_{z}-i\zeta_{\times}^{\prime}\left(k_{x}\delta u_{x}+k_{y}\delta u_{y}\right),\\
	0&=&\left(1-i\omega\tau_{f}\right)\delta f_{x}+i\eta_{f}\left(k_{x}\delta u_{z}+k_{z}\delta u_{x}\right),\\
	0&=&\left(1-i\omega\tau_{f}\right)\delta f_{y}+i\eta_{f}\left(k_{y}\delta u_{z}+k_{z}\delta u_{y}\right),\\
	0&=&\left(1-i\omega\tau_{\nu}\right)\delta\nu_{\perp x}-i\kappa k_{x}\delta\alpha+i\frac{\kappa_{\times}}{\mu_{m}}\left(k_{z}\delta B_{x}-k_{x}\delta B_{z}\right),\\
	0&=&\left(1-i\omega\tau_{\nu}\right)\delta\nu_{\perp y}-i\kappa k_{y}\delta\alpha+i\frac{\kappa_{\times}}{\mu_{m}}\left(k_{z}\delta B_{y}-k_{y}\delta B_{z}\right),\\
	0&=&\left(1-i\omega\tau_{\ell}\right)\delta\ell_{x}-i\frac{\rho_{\|}}{\mu_{m}}\left(k_{z}\delta B_{x}-k_{x}\delta B_{z}\right)+i\kappa_{\times}^{\prime}k_{x}\delta\alpha,\\
	0&=&\left(1-i\omega\tau_{\ell}\right)\delta\ell_{y}-i\frac{\rho_{\|}}{\mu_{m}}\left(k_{z}\delta B_{y}-k_{y}\delta B_{z}\right)+i\kappa_{\times}^{\prime}k_{y}\delta\alpha,\\
	0&=&\left(1-i\omega\tau_{m}\right)\delta m_{\perp xy}+i\frac{\rho_{\perp}}{\mu_{m}}\left(k_{x}\delta B_{y}-k_{y}\delta B_{x}\right),\\
	0&=&\left(1-i\omega\tau_{\pi}\right)\delta\pi_{\perp xx}-i\eta_{\perp}\left(k_{x}\delta u_{x}-k_{y}\delta u_{y}\right),\\
	0&=&\left(1-i\omega\tau_{\pi}\right)\delta\pi_{\perp xy}-i\eta_{\perp}\left(k_{x}\delta u_{y}+k_{y}\delta u_{x}\right),\label{linearized IS equation}
	\ee
	where we use the shear tensor property $\pi_{\perp xx}=-\pi_{\perp yy}$ and $\pi_{\perp xy}=\pi_{\perp yx}$. Further, we define $\eta_f=\eta_\parallel/T_0$, where $T_0$ denotes the equilibrium temperature. Since we retain only terms linear in the perturbations, the higher-order coupling terms vanish in the above equations. 
	
	Collecting all the independent linearized evolution equations,  Eqs.\eqref{linearized conservation laws}-\eqref{linearized IS equation} can be rewritten in matrix form $M_{19\times19}X_{19\times1}=0$, where
	\be\label{linear eq}
	M_{19\times19}&=&\left(\begin{array}{ccc}
		A_{10\times10} & 0 & M_{AC}\\
		0 & B_{6\times6} & M_{BC}\\
		M_{CA} & M_{CB} & C_{3\times3}
	\end{array}\right),\\
	X_{19\times1}&=&(X_A,\quad X_B,\quad X_C)^T
	\ee
	{with the perturbation vector decomposed as}\be
	X_{A}^{T}&=&\Bigl(\delta\varepsilon,\delta u_{x},\delta u_{z},\delta B_{x},\delta B_{z},\delta\Pi_{\|},\delta\Pi_{\perp},\delta f_{x},\delta\pi_{xx},\delta\ell_x\Bigr),\nonumber\\
	X_{B}^{T}&=&\Bigl(\delta u_{y},\delta B_{y},\delta f_{y},\delta\pi_{xy},\delta m_{\perp xy},\delta\ell_y\Bigr),\nonumber\\
	X_{C}^{T}&=&\Bigl(\delta n,\delta \nu_{\perp x},\delta \nu_{\perp y}\Bigr),
	\ee
	where the matrices $A_{10\times10}$, $B_{6\times6}$, and $C_{3\times3}$, which form the diagonal blocks of $M_{19\times19}$, describe the "magneto-sonic", "Alfv\'en" and "charge diffusion" modes, respectively. The coupling terms $M_{AC}$, $M_{BC}$, $M_{CA}$ and $M_{CB}$ arise from the dissipative coefficient $\kappa_\times$, the number density $n_0$ and the thermodynamic quantity $\partial p/\partial n$. The explicit form of matrix $M_{19\times19}$ and its decomposition are presented in Appendix~\ref{APPENDIX 3}. 
	
	In the above steps, to decouple and simplify the $M_{19\times19}$ matrix, we choose the wave vector to lie in the $x-z$ plane by setting $k_y=0$ , $k_x=k\sin\theta$ and $k_z=k\cos\theta$ so that we can also directly parameterize the momentum in the two-dimensional plane in terms of $k$ and $\theta$, where $\theta$ is the angle between momentum and the $z$-direction of the magnetic field. 
{	Under this choice, the magnetic constraint in Eq.~\eqref{magnetic constraint} reduces to}
	\begin{equation}
		{\sin\theta\,\delta B_x+\cos\theta\,\delta B_z=0. }\label{magnetic constraint1}
	\end{equation}
{	Therefore, the two magnetic perturbations $\delta B_x$ and $\delta B_z$ in the propagation plane are not independent. This constraint removes the longitudinal magnetic perturbation along the wave vector and restricts the magnetic fluctuations to the divergence-free subspace.}
{
	Although the Fourier-space evolution vector contains 19 amplitudes,
	the magnetic Gauss constraint removes one independent combination of
	$\delta B_x$ and $\delta B_z$. The physical perturbation space therefore
	contains 18 independent degrees of freedom: 9 associated with the
	$A$ block, 6 with the $B$ block, and 3 with the $C$ block.
	Before the additional assumptions introduced in the next section,
	these three blocks are generally coupled through the off-diagonal
	terms in Eq.~\eqref{linear eq}.
}
	We introduce the Alfv\'en-wave velocity by utilizing
	\be
	v_A=\frac{B}{\sqrt{\mu_m h}}
	\ee
	to extract independent eigenvalues (intrinsic velocity) corresponding to the Alfv\'en branch. The Alfv\'en mode mainly excites transverse shear degrees of freedom in the direction perpendicular to both the wave vector and the background magnetic field. This leads to a dispersion structure of the form $\omega^2-v_A^2 k_{\|}^2$.
	
	The stability and causality of the linearized magnetohydrodynamic system can be investigated through the solutions $\omega$ of the above secular equation. In the following section, we analyze these solutions using theoretical approximations and numerical calculations based on Eq.\eqref{linear eq}.

	\section{APPROXIMATE SOLUTIONS ANALYSIS}
	\label{section6}
	In this section, we derive approximate solutions of the secular equation in Section~\ref{S5}. Using the method of series expansion to match the coefficients, one can  obtain the approximate expression of the solution $\omega$. In the following, we conduct a detailed study of approximate solutions in these two scenarios: small momentum and large momentum, respectively.
	
	In the following calculation, we adopt a conformal equation of state appropriate for a gas of massless particles and antiparticles. The squared speed of sound, the derivative of the pressure with respect to the number density, and the perturbation $\delta\alpha$ are then given by:
	\be\label{ass1}
	c_s^2=\frac{1}{3},\quad
	\frac{\partial p}{\partial n}=0,\quad
	\delta\alpha=\frac{1}{\bar{n}}\delta n.
	\ee
	It should be noted that the particle number density $\bar{n}\ne0$ and should be distinguished from the net-particle number density $n_0$ in Eq.\eqref{perurbation}. For simplicity, we evaluate Eq.\eqref{linear eq} in a charge-conjugation-symmetric background with vanishing vector couplings:
	\be\label{ass2}
	n_0=0, \qquad \kappa_{\times}=\kappa_{\times}'=0.
	\ee
	Therefore, the decoupled channels will form the secular equation as 
	\be\label{Det}
	\det\left(M_{19\times19}\right)=\det A\det B\det C=0
	\ee
	by utilizing the properties of diagonal blocks for the matrix $M_{19\times19}$, its determinant becomes the product of $\det A$, $\det B$ and $\det C$.
	{
		Under the assumptions in Eqs.~\eqref{ass1}--\eqref{ass2}, the
		off-diagonal couplings vanish and the 18 physical modes separate into
		the magneto-sonic, Alfv\'en, and charge-diffusion sectors. In the
		small-$k$ limit, these sectors contain, respectively, four propagating
		hydrodynamic branches and five relaxation modes, two propagating
		branches and four relaxation modes, and one diffusion mode and two
		relaxation modes. In the large-$k$ limit, the same 18 modes reorganize
		into six propagating and three finite-frequency branches, four
		propagating and two finite-frequency branches, and two propagating and
		one finite-frequency branch, respectively. Thus, the total number of
		physical modes remains unchanged, while their asymptotic character
		depends on the momentum regime.
	}
	\subsection{Small-$k$ expansion}
	First, we investigate the solutions of the magneto-sonic channel $A_{10\times10}$, the Alfv\'en channel  $B_{6\times6}$ and the charge diffusion channel $C_{3\times3}$ by analyzing the determinant of Eqs.\eqref{Det} in the small-$k$ limit. One can expand each secular equation by writing $\omega$ as a power series in $k$ as $k\to0$:
	\be\label{Small k expansion DET}
	0&=&\det A_{10\times10}/.\{\omega^A\to\sum_{i=0}^\infty\omega_i k^i\},\\
	0&=&\det B_{6\times6}/.\{\omega^B\to\sum_{i=0}^\infty\omega_i k^i\},\\
	0&=&\det C_{3\times3}/.\{\omega^C\to\sum_{i=0}^\infty\omega_i k^i\}.
	\ee 
	Setting $k=0$ yields undisturbed fluid solutions $\omega_0$, the so-called leading order solutions; further, one can substitute the leading order solutions in Eqs.\eqref{Small k expansion DET} to obtain the higher order solutions of $\omega_i$.
	
	\textbf{1. The leading order solution of $k^0$ terms:}
	
	Magneto-sonic sector $A_{10 \times 10}$:  fourfold zero mode + five relaxation modes for $\omega^A$:
	\be
	&&\omega_0=0\ {\text{(Quadruple Root)}},\quad\omega_0=-\frac{i}{\tau_\ell},\quad\omega_0=-\frac{i}{\tau_{\Pi_\parallel}},\nonumber\\
	&&\omega_0=-\frac{i}{\tau_{\Pi_\perp}},\qquad\omega_0=-\frac{i}{\tau_f},\qquad\omega_0=-\frac{i}{\tau_\pi}.
	\ee
	
{For nonzero $k$, the magnetic constraint in Eq.~\eqref{magnetic constraint1} removes the longitudinal magnetic perturbation; the fourfold zero-frequency physical manifold is understood as the $k\to0$ limit at fixed propagation direction.}
	
{	Fourfold} $\omega_0=0$: "no-gradient, no-restoring-force" hydrodynamic manifold.
	In this sector, homogeneous perturbations corresponding to energy density, longitudinal/transverse velocity components in the propagation plane, and magnetic-field components in the same plane do not experience any restoring force at $k=0$. 
	
	Gapped modes $\omega_0=-i / \tau$: transient relaxation of dissipative and electromagnetic-response variables. Each nonzero root corresponds to a purely decaying mode associated with one independent relaxation equation in the extended theory. These modes are non-hydrodynamic: when taking the Navier-Stokes limit, i.e., when the relaxation times are sent to zero, $\tau\to0$ the non-hydrodynamic modes go to infinity.

	Alfv\'en sector $B_{6 \times 6}$: double zero mode + four relaxation modes for $\omega^B$:
	\be\label{B root in k=0}
	&&\omega_0=0\ \text{(Double Root)},\qquad\quad\omega_0=-\frac{i}{\tau_f},\nonumber\\
	&&\omega_0=-\frac{i}{\tau_\ell},\qquad\omega_0=-\frac{i}{\tau_m},\qquad \omega_0=-\frac{i}{\tau_\pi}.
	\ee
	
The double zero root $\omega_0=0$ corresponds to the pair of
counter-propagating Alfv\'en branches, which collapse to $\omega=0$
in the homogeneous limit, with the associated gapless subspace
spanned primarily by $(\delta u_y,\delta B_y)$.
The four gapped roots correspond to the relaxation of the dissipative
variables in this sector: $\tau_f$ and $\tau_\pi$ govern
$\delta f_y$ and $\delta\pi_{xy}$, respectively, whereas
$\tau_m$ and $\tau_\ell$ are associated with
$\delta m_{\perp xy}$ and $\delta\ell_y$.
	
	Charge diffusion sector $C_{3\times3}$:
	\be
	&&\omega_0=-\frac{i}{\tau_\nu}\quad \text{(Double Root)},\qquad\omega_0=0.\qquad 
	\ee
	
	In the homogeneous limit $k=0$, the charge conservation law contains no spatial divergence and therefore admits a neutral mode $\omega=0$. This mode corresponds to a spatially uniform perturbation of the conserved charge density $\delta n$, which cannot relax in time without gradients. The double root $-i / \tau_\nu$ reflects the independent relaxation of the transverse diffusion currents $\nu_{\perp x}$ and $\nu_{\perp y}$.
	
	Hence, in the small $k$ regime the hydrodynamic branches originate from the $k=0$ zero modes,  while the non-hydrodynamic branches originate from the gapped roots $-i/\tau_i$ and receive only perturbative $k$ corrections.
	
	\textbf{2. Higher-order terms in the solutions:}
	
	The complete physical roots $\omega^A$ for the magneto-sonic sector
	$A_{10\times10}$ are
	\be\label{small k A}
	\omega_{{r_{1,2}}}
	&=&
	\pm \frac{k}{\sqrt{2}}
	\sqrt{
		\mathcal V+
		\sqrt{
			\mathcal V^2
			-4c_s^2v_A^2\cos^2\theta
		}
	}
	+\mathcal{O}(k^2),
	\nonumber\\
	\omega_{{r_{3,4}}}
	&=&
	\pm \frac{k}{\sqrt{2}}
	\sqrt{
		\mathcal V-
		\sqrt{
			\mathcal V^2
			-4c_s^2v_A^2\cos^2\theta
		}
	}
	+\mathcal{O}(k^2),
	\nonumber\\
	\omega_{{r_5}}
	&=&
	-\frac{i}{\tau_\ell}
	+i\frac{h v_A^2\rho_\parallel}{B^2}k^2
	+\mathcal{O}(k^4),
	\nonumber\\
	\omega_{{r_6}}
	&=&
	-\frac{i}{\tau_{\Pi_\parallel}}
	+i\frac{\zeta_\parallel\cos^2\theta}{h(1-v_A^2)}k^2
	+\mathcal{O}(k^4),
	\\
	\omega_{{r_7}}
	&=&
	-\frac{i}{\tau_{\Pi_\perp}}
	+i\frac{\zeta_\perp\sin^2\theta}{h}k^2
	+\mathcal{O}(k^4),
	\nonumber\\
	\omega_{{r_8}}
	&=&
	-\frac{i}{\tau_f}
	+i\frac{\eta_f}{h}\left(\cos^2\theta+\frac{\sin^2\theta}{1-v_A^2}\right)k^2
	+\mathcal{O}(k^4),
	\nonumber\\
	\omega_{{r_9}}
	&=&
	-\frac{i}{\tau_\pi}
	+i\frac{\eta_\perp\sin^2\theta}{h}k^2
	+\mathcal{O}(k^4).
	\nonumber
	\ee
	where we define
	$\mathcal{V}=c_s^2+v_A^2-c_s^2v_A^2\sin^2\theta$.
	The {fourfold} zero-frequency mode at $k=0$ splits into two
	counter-propagating pairs at finite momentum, corresponding to the fast
	and slow magneto-sonic branches. At leading order in the long-wavelength
	expansion, their dispersions can be written as
	\be
	\omega=\pm v_{\rm fast}(\theta)k,
	\qquad
	\omega=\pm v_{\rm slow}(\theta)k.
	\ee
	
	The angular dependence of the corresponding phase velocities originates
	from the anisotropy induced by the background magnetic field.
	Physically, the fast and slow magneto-sonic waves arise from the coupling
	between compressional fluid motion and magnetic pressure and tension. The remaining five solutions are non-hydrodynamic relaxation modes.
	At $k=0$, their frequencies are determined by the relaxation scales
	$\tau_\ell$, $\tau_{\Pi_\parallel}$, $\tau_{\Pi_\perp}$, $\tau_f$, and
	$\tau_\pi$, while finite momentum produces the $\mathcal{O}(k^2)$
	corrections displayed in Eq.~\eqref{small k A}.
	
	The complete roots $\omega^B$ for the Alfvén sector $B_{6 \times 6}$:
	\be\label{small k B}
	\omega_{r_{1,2}}&=&\pm v _A\cos\theta k-\frac{i\,k^{2}}{2B^{2}h}\Big[
	\bigl(B^{2}\eta_{f}+h^{2}v_A^{2}\rho_\parallel\bigr)\cos^{2}\theta\nonumber\\&&+\bigl(B^{2}\eta_\perp+ h^2 v_A^2\rho_\perp\bigr)\sin^{2}\theta
	\Big]+\mathcal{O}(k^{3}),\nonumber\\
	\omega_{r_3}&=&-\frac{i}{\tau_\ell}+i\frac{h v_A^2 \rho_\parallel\cos^2\theta}{B^2}k^2+i k^4\frac{h  v_A^4  \rho_\parallel  \tau_\ell  \cos^2\theta  }{B^4 (\tau_\ell - \tau_m)}\big[(h \rho_\parallel - B^2 \tau_\ell)\nonumber\\&&* (\tau_\ell - \tau_m) \cos^2\theta + h \rho_\perp \tau_\ell \sin^2\theta\big],\nonumber\\
	\omega_{r_4}&=&-\frac{i}{\tau_m}+i\frac{h v_A^2 \rho_\perp\sin^2\theta}{B^2}k^2+i k^4\frac{h  v_A^4  \rho_\perp  \tau_m  \sin^2\theta  }{B^4 (\tau_m - \tau_\ell)}\big[(h \rho_\parallel\nonumber\\&&+ B^2(\tau_\ell-\tau_m))\tau_m \cos^2\theta + h \rho_\perp (\tau_m-\tau_\ell) \sin^2\theta\big],\\
	\omega_{r_5}&=&-\frac{i}{\tau_f}+i\frac{\eta_f\cos^2\theta}{h}k^2+ik^4\frac{\eta_f \tau_f \cos^2\theta}{h^2 (\tau_f -\tau_\pi)}\big[(\eta_f-h v_A^2 \tau_f) \nonumber\\&&*(\tau_f - \tau_\pi) \cos^2\theta + \eta_\perp \tau_f \sin^2\theta\big],\nonumber\\
	\omega_{r_6}&=&-\frac{i}{\tau_\pi}+i\frac{\eta_\perp\sin^2\theta}{h}k^2+ik^4\frac{\eta_\perp\tau_\pi \sin^2\theta}{h^2 (\tau_\pi-\tau_f)}\big[(\eta_f\nonumber\\&&+h v_A^2( \tau_f-\tau_\pi))\tau_\pi  \cos^2\theta + \eta_\perp (\tau_\pi-\tau_f) \sin^2\theta\big].\nonumber
	\ee
	The double root $\omega_0=0$ in Eqs.\eqref{B root in k=0} splits into a counter-propagating Alfvén pair with $\mathcal O(k)$ phase velocity and $\mathcal O(k^2)$ attenuation:
	\be
	\omega_{ \pm}= \pm v_A \cos \theta k-i \Gamma(\theta) k^2+\mathcal{O}\left(k^3\right),
	\ee
	where the imaginary part contribution of $\omega_2$ is represented by $\Gamma(\theta)$. {Besides the propagating Alfv\'en pair, the sector contains four
		non-hydrodynamic relaxation modes with gaps $1/\tau_\ell$, $1/\tau_m$, $1/\tau_f$, and $1/\tau_\pi$, associated with
		$\delta f_y$, $\delta\pi_{xy}$, $\delta m_{\perp xy}$, and
		$\delta\ell_y$. Their $k^2$ and higher-order corrections arise from
		gradient-induced mixing between the viscous and induction variables.}

	The complete roots $\omega^C$ for sector $C_{3 \times 3}$:
	\be\label{small k C}
	\omega_{r_1}&=&-ik^2\frac{\kappa\sin^2\theta}{\bar{n}}-ik^4\frac{\kappa^2\tau_\nu\sin^4\theta}{\bar{n}^2}-ik^6\frac{2\kappa^3\tau_\nu^2\sin^6\theta}{\bar{n}^3}+\mathcal{O}(k^{8}),\nonumber\\
	\omega_{r_2}&=&-\frac{i}{\tau_\nu},\\
	\omega_{r_3}&=&-\frac{i}{\tau_\nu}+ik^2\frac{\kappa\sin^2\theta}{\bar{n}}+ik^4\frac{\kappa^2\tau_\nu\sin^4\theta}{\bar{n}^2}+ik^6\frac{2\kappa ^3 \tau_\nu ^2\sin^6\theta}{\bar{n}^3}+\mathcal{O}(k^{8}).\nonumber
	\ee
	
	In the charge-diffusion sector, $\omega_{r_1}$ is the hydrodynamic (gapless) diffusion mode associated with the conserved density perturbation $\delta n$. 
	Its leading dispersion is purely imaginary and starts at $\mathcal O(k^2)$, as expected for diffusion. 
	The angular factor $\sin^2\theta$ reflects the fact that only the
	momentum component transverse to the background magnetic field
	enters this diffusion channel, which is controlled by $\kappa$
	and $\tau_\nu$.
	In the Navier--Stokes limit $\tau_\nu\to0$, these gapped poles decouple, and $\omega_{r_1}$ reduces to the standard diffusive dispersion with an effective diffusion constant determined by the corresponding first-order transport coefficients.
	The remaining roots $\omega_{r_2}$ and $\omega_{r_3}$ are non-hydrodynamic relaxation poles, $\omega\simeq -i/\tau_\nu$, which acquire $k^2$ (and higher) corrections due to weak gradient-induced coupling to $\delta n$. 
	
	The small-$k$ analysis thus provides a transparent physical classification of modes: 
	magneto-sonic and Alfv\'en waves emerge from the hydrodynamic manifold, whereas the remaining poles are non-hydrodynamic relaxation modes controlled by the corresponding $\tau_i$. 
	At this order, dissipation enters as $k^2$ attenuation for the propagating branches and as small shifts of the gapped poles, consistent with the Navier--Stokes limit supplemented by Israel--Stewart relaxation. 
	
	However, the long-wavelength expansion cannot address the high-$k$ behavior that is most sensitive to causality, stability, and the role of finite relaxation times. We therefore proceed to a large-$k$ expansion to probe the short-wavelength asymptotics of the secular equations and to clarify how the spectrum reorganizes when gradient terms dominate over relaxation.

	\subsection{Large-$k$ expansion}
	
	Having established the long-wavelength structure of the spectrum and identified the leading attenuation mechanisms, we now turn to the opposite regime of large momentum, where short-wavelength dynamics probes the causal completion of the theory, and the competition between propagation and relaxation becomes essential.
	
	One can assume that $\omega$ can be expanded as a power series in terms of $\epsilon=1/k$ under large momentum ($k\to\infty$) conditions:
	\be\label{Large k expansion DET}
	0&=&\det A_{10\times10}/.\{\omega^A\to\sum_{i=\alpha}^\infty\omega_i \epsilon^i\},\\
	0&=&\det B_{6\times6}/.\{\omega^B\to\sum_{i=\alpha}^\infty\omega_i \epsilon^i\},\label{Large k expansion DETB}\\
	0&=&\det C_{3\times3}/.\{\omega^C\to\sum_{i=\alpha}^\infty\omega_i \epsilon^i\},
	\ee 
	where the power of the leading order coefficient $\alpha$ is uncertain, determined by the mode of solution.

	We first deal with the situation of $\alpha=-1$ propagating mode of these determinants, respectively.{ For the coefficient $\omega_{A,-1}$, we take the leading order $\epsilon^{-9}$, then can arrive}
	\be\label{omega^a_-1 equation}
{	\omega_{A,-1}^3(\mathbf{a}_A + \mathbf{b}_A \omega_{A,-1}^2 + \mathbf{c}_A \omega_{A,-1}^4 + \mathbf{d}_A \omega_{A,-1}^6) = 0,}
	\ee
{	where $\omega_{A,-1}$ is the solution of this equation and the abbreviations $\mathbf{a}_A,\mathbf{b}_A,\mathbf{c}_A$ and $\mathbf{d}_A$ of coefficients extracted from $\det A_{10\times10}$ are very complex, the detailed form of them we list in Appendix \ref{Large k expansion solution}. Then the six leading order roots of the propagation mode can be written as the roots of a cubic equation $\omega_{A,-1}=\pm\sqrt{x_a}$ while the remaining triple zero roots $\omega_{A,-1}=0$ are non-hydrodynamic mode the leading order solution for $A_{10\times10}$.}

	Taking $\alpha=0$ to solve the non-hydrodynamic mode in $A_{10\times10}$, one can obtain the coefficient equation of leading order $\epsilon^{-6}$:
	\be\label{omega^a_0 equation}
{	\mathbf{a}_A' + \mathbf{b}_A'  \omega_0' + \mathbf{c}_A'  \omega_0'^2 + \mathbf{d}_A'  \omega_0'^3= 0,}
	\ee
	where $\omega_0'$ is the solution of the above {cubic} equation and the abbreviations $\mathbf{a}_A'$, $\mathbf{b}_A'$, $\mathbf{c}_A'$ and $\mathbf{d}_A'$ also are coefficients for different powers of $\omega_0'$.

	We will individually substitute the leading order solutions into the expansion, match the coefficients to obtain the equations for each subsequent order, and solve for subsequent solutions.
	The solutions of magneto-sonic mode $A_{10\times10}$ can be written
	\be\label{large k A}
	\omega_{r_1-r_6}&=&\pm\sqrt{x_a} k +\mathcal{O}(k^{0})+\mathcal{O}(k^{-1})+ \mathcal{O}(k^{-3}),\nonumber\\
	\omega_{r_7-{r_{9}}}&=&\omega_{A,0}'+ \mathcal{O}(k^{-2})+ \mathcal{O}(k^{-4}).
	\ee
	
	In order not to violate the stability $ \operatorname{Im} \omega_{-1} = 0$ of the propagation mode as well as $\operatorname{Im}\omega_{A,0}'<0$ of the relaxation mode, we must adhere to the necessary condition for each solution in Eq.\eqref{omega^a_-1 equation} and Eq.\eqref{omega^a_0 equation} using Routh–Hurwitz stability criterion. 
	Then one can rapidly confirm that $x_a > 0$ and $\operatorname{Im}\omega_{A,0}'<0$ are true when taking the special angle $\theta=0$ or $\theta=\pi/2$ by symbolic verification while verifying the validity of this expression when $\theta$ is uncertain is extremely challenging due to its excessively lengthy nature. The detail of its progress is shown in the Appendix \ref{Large k expansion solution}.

	Let us next investigate the Alfv\'en sector $B_{6\times 6}$. 
	As in the magneto-sonic case, the large-$k$ asymptotics naturally separates into two classes: propagating branches with $\alpha=-1$, whose frequencies scale linearly with $k$, and non-hydrodynamic relaxation branches with $\alpha=0$, which remain finite in the short-wavelength limit. 
	This distinction is particularly transparent in the Alfv\'en channel, since the transverse fluid and magnetic perturbations continue to support wave-like propagation, whereas the additional dissipative variables only contribute finite relaxation scales.
	
	For the propagation modes, the leading-order coefficient is determined from the $\epsilon^{-6}$ terms:
	\be\label{omegab-1 equation}
	\omega_{B,-1}^2(\mathbf{a}_B + \mathbf{b}_B  \omega_{B,-1}^2+\mathbf{c}_B\omega_{B,-1}^4)=0.
	\ee
	The double factor $\omega_{B,-1}^2$ indicates that not all modes in this sector scale linearly with $k$; two branches instead remain finite and must be treated separately through the $\alpha=0$ ansatz. 
{	The remaining four roots $\omega_{B,-1}^4$ correspond to the short-wavelength propagating part of the Alfv\'en sector. The leading order contribution linear in $k$ can be written as follows:
	\be
	v_B
	=
	\pm
	\sqrt{
		\frac{
			-\mathbf{b}_B\pm\sqrt{\Delta_B}
		}{
			2h\tau_\ell\tau_m\tau_f\tau_\pi
		}
	},
	\ee
	
	where the definitions of $\mathbf{b}_B$ and $\Delta_B$, together with the
	proof that the corresponding characteristic velocities are real, are
	given in Appendix~\ref{Large k expansion solution}. The Alfv\'en sector therefore admits the following necessary linear
	high-$k$ causality condition:
	\be\label{theoretical constraints of Alfven sector}
	|v_g|<1
	\quad\Longrightarrow\quad
	-\mathbf{b}_B+\sqrt{\Delta_B}
	<
	2h\tau_\ell\tau_m\tau_f\tau_\pi .
	\ee
	The equality in Eq.~\eqref{theoretical constraints of Alfven sector} defines the luminal boundary of the Alfv\'en sector. This condition demonstrates that the short-wavelength causal propagation is governed by a nontrivial interplay between anisotropic dissipative transport and the corresponding relaxation dynamics. In particular, the coefficients
	$\mathbf{b}_B$ and $\Delta_B$ depend on the electromagnetic-response coefficients $\rho_\parallel$ and $\rho_\perp$, the viscous coefficients $\eta_f$ and $\eta_\perp$, their associated relaxation times, and the propagation angle $\theta$. This analytic inequality provides the theoretical basis for the multi-parameter causality maps of the Alfv\'en sector presented later in Fig.~\ref{figcb}.}
	
	For the non-hydrodynamic branches of the Alfv\'en sector, taking $\alpha=0$ yields the leading-order equation
	\be\label{omega0 equation}
	\mathbf{a}'_B + \mathbf{b}'_B  \omega_{B,0}+\mathbf{c}'_B\omega_{B,0}^2=0,
	\ee
	from which one obtains
	\be
	\omega_{B,0,r_1}=-i
	\frac{\rho_\parallel\cos^2\theta+\rho_\perp\sin^2\theta}
	{\rho_\parallel\tau_m\cos^2\theta+\rho_\perp\tau_\ell\sin^2\theta},
	\omega_{B,0,r_2}=-i
	\frac{\eta_f\cos^2\theta+\eta_\perp\sin^2\theta}
	{\eta_f\tau_\pi\cos^2\theta+\eta_\perp\tau_f\sin^2\theta}.
	\ee
The first pole is a mixed electromagnetic-response relaxation mode.
Its damping rate depends on the propagation angle and reduces to
$1/\tau_m$ at $\theta=0$ and $1/\tau_\ell$ at $\theta=\pi/2$.
The second pole is a mixed viscous relaxation mode controlled jointly
by $\eta_f$, $\eta_\perp$, and their associated relaxation times.
Both poles remain purely damped under the assumed positivity conditions.
	
	Again, by substituting the leading-order roots into Eq.\eqref{Large k expansion DETB}, one obtains the asymptotic solutions of the Alfv\'en sector:
	\be\label{large k B}
	\omega_{r_1-r_4}&=&{\pm k
	\sqrt{
		\frac{
			-\mathbf{b}_B\pm\sqrt{\Delta_B}
		}{
			2h\tau_\ell\tau_m\tau_f\tau_\pi
		}
	}} + \mathcal{O}(k^{0})+\mathcal{O}(k^{-1})+\mathcal{O}(k^{-2}),\nonumber\\
	\omega_{r_5}&=&-i
	\frac{\rho_\parallel\cos^2\theta+\rho_\perp\sin^2\theta}
	{\rho_\parallel\tau_m\cos^2\theta+\rho_\perp\tau_\ell\sin^2\theta}+\mathcal{O}(k^{-1})+\mathcal{O}(k^{-2})+\mathcal{O}(k^{-3}),\\
	\omega_{r_6}&=&-i
	\frac{\eta_f\cos^2\theta+\eta_\perp\sin^2\theta}
	{\eta_f\tau_\pi\cos^2\theta+\eta_\perp\tau_f\sin^2\theta}+\mathcal{O}(k^{-1})+....\nonumber
	\ee
	
	Therefore, in the large-$k$ limit the Alfv\'en sector still exhibits a clear separation between causal propagating branches and finite relaxation poles: the former probe the hyperbolic structure of the theory, while the latter retain the memory of the underlying dissipative and electromagnetic response times.
	
	The large-$k$ propagating roots $\omega\simeq k\,\omega_{B,-1}$ encode the characteristic (front) velocities of the hyperbolic completion. 
	In an Israel--Stewart-type theory these characteristic speeds are not solely given by the ideal Alfv\'en velocity $v_A$, but also receive contributions from telegraph-type dynamics associated with the relaxation of dissipative variables. 
	
	We now turn to the charge-diffusion sector $C_{3\times3}$. 
	In contrast to the small-$k$ regime, where this sector contains a diffusive hydrodynamic branch, the large-$k$ limit reveals the causal completion of charge transport: the diffusion mode is converted into a pair of propagating short-wavelength branches with finite characteristic speed, while the remaining modes stay as finite relaxation poles. 
	This is the standard Maxwell--Cattaneo/Israel--Stewart mechanism by which an acausal diffusion equation is replaced by a hyperbolic telegraph-type dynamics.
	
	At leading order $\epsilon^{-3}$, one finds
	\be
	\omega_{C,-1}(\mathbf{a}_C  + \mathbf{b}_C\omega_{C,-1}^2)=0.
	\ee
	The corresponding roots are
	\be
	\omega_{C,-1,r_1}=0,\qquad
	\omega_{C,-1,r_2,r_3}=\pm \sqrt{\frac{\kappa\sin^{2}\theta}{\bar{n}\,\tau_\nu}}.
	\ee
	The zero root shows again that not all modes in this sector scale as $k$, so the remaining branches must be obtained from the $\alpha=0$ expansion. 
	By contrast, the nonzero pair already displays the essential large-$k$ feature: the charge sector supports signal propagation with a finite effective velocity
	determined by the diffusion coefficient $\kappa$ and the relaxation time $\tau_\nu$.
	
	For the finite-frequency branches, one obtains
	\be
	\omega_{C,0}=-\frac{i}{\tau_\nu}.
	\ee
	This root is purely damped under the positivity assumptions on the transport coefficients and relaxation times, and therefore does not contribute to asymptotic propagation.
	
	Collecting the above results, the large-$k$ solutions of the charge-diffusion sector can be written as
	\be\label{large k C}
	\omega_{r_1}&=&k\sqrt{\frac{\kappa\sin^{2}\theta}{\bar{n}\,\tau_\nu}}-\frac{i}{2\tau_\nu}-k^{-1}\frac{{\sqrt{\bar{n}}}\csc\theta}{8\sqrt{\kappa}\,\tau_\nu^{3/2}}+\mathcal{O}(k^{-2}),\nonumber\\
	\omega_{r_2}&=&-k\sqrt{\frac{\kappa\sin^{2}\theta}{\bar{n}\,\tau_\nu}}-\frac{i}{2\tau_\nu}+k^{-1}\frac{{\sqrt{\bar{n}}}\csc\theta}{8\sqrt{\kappa}\,\tau_\nu^{3/2}}+\mathcal{O}(k^{-2}),\nonumber\\
	\omega_{r_3}&=&-\frac{i}{\tau_\nu}.
	\ee
	
	Hence, the large-$k$ behavior of the charge sector provides a particularly clear manifestation of causality restoration: the diffusive dynamics of the long-wavelength regime reorganizes into a propagating pair with finite front velocity, while the purely relaxational poles remain finite and damped. The limit $\theta\to0$ does not commute with the large-$k$ expansion and must be treated separately because $\csc\theta$ appears at order $k^{-1}$ in the charge diffusion sector.
	
	Moreover, the charge sector yields a particularly transparent causality condition because the large-$k$ propagating pair is explicitly
	\be
	\omega \simeq \pm v_Ck+\cdots,
	\,v_C^2=\frac{\kappa\sin^{2}\theta}{\bar{n}\,\tau_\nu}.
	\ee
	Requiring a subluminal front velocity, $v_C(\theta)\le 1$, gives the necessary constraint
	\be\label{c condition}
	\kappa\sin^{2}\theta\,\le\ \bar{n}\,\tau_\nu,
	\label{C-causality-necessary}
	\ee
	This condition gives a straight causality boundary in the $\kappa/(\bar n\tau_\nu)$ plane at fixed $\theta$ and will be used as part of the numerical causality check in Sec.~\ref{s7}.

	\section{EXACT SOLUTIONS ANALYSIS}\label{s7}
	\label{section7}
	Having obtained analytical approximations in the small-$k$ and large-$k$ limits, we now proceed to a systematic analysis of the full spectrum. 
	The purpose of this section is twofold. 
	First, we numerically solve the secular equations for generic $(k,\theta)$ and benchmark the numerical roots against the analytical expressions obtained in Sec.~\ref{section6}, thus quantifying the domains of validity and the crossover between different asymptotic regimes. 
	Second, we use the numerical spectrum to assess the causal structure of the theory, focusing on the high-momentum behavior that is most sensitive to relaxation-time effects and to the hyperbolic completion of transport.
	
	\begin{table}[htbp]
		\centering
		
		\setlength{\tabcolsep}{6.5pt}
		
		\caption{Parameter set used in the numerical calculations}
		\label{table1}
		
		\begin{tabular}{cccccc}
			\toprule
			$c_s^2$ 
			& $v_A$ 
			& $B {h^{-1/2}}$ 
			& $\rho_\parallel h^{1/4}$ 
			& $\rho_\perp h^{1/4}$ 
			& $\zeta_\parallel h^{-3/4}$ \\
			\midrule
			$1/3$ 
			& $0.1$ 
			& $0.22$ 
			& $0.13$ 
			& $0.12$ 
			& $0.12$ \\
			\bottomrule
			\toprule
			$\zeta_\perp h^{-3/4}$ 
			& $\zeta_\times h^{-3/4}$ 
			& $\eta_f h^{-3/4}$ 
			& $\eta_\perp h^{-3/4}$ 
			& $\tau_\ell h^{1/4}$ 
			& $\tau_m h^{1/4}$ \\
			\midrule
			$0.132$ 
			& $0.1$ 
			& $0.165$ 
			& $0.33$ 
			& $3$ 
			& $3.2$ \\
			\bottomrule
			\toprule
			$\tau_{\Pi_\parallel} h^{1/4}$ 
			& $\tau_{\Pi_\perp} h^{1/4}$ 
			& $\tau_f h^{1/4}$ 
			& $\tau_\pi h^{1/4}$ 
			& $\kappa \bar{n}^{-{2}/{3}}$ 
			& $\tau_\nu \bar{n}^{1/{3}}$ \\
			\midrule
			$2.8$ 
			& $4$ 
			& $2.5$ 
			& $2.3$ 
			& $0.5$ 
			& $2.7$ \\
			\bottomrule
		\end{tabular}
		
	\end{table}
	
{	Unless otherwise specified, the numerical calculations in this section
	use the parameter set listed in Table~\ref{table1}.} To present the
	numerical results in a scale-invariant form, we use the equilibrium
	enthalpy density $h$ and the particle density $\bar n$ as reference
	scales to nondimensionalize the transport coefficients and relaxation
	times.
	
	\subsection{Comparison between numerical and analytical solutions}
	\label{section7.1}
	
	In this subsection, we compare the numerical roots of the three decoupled
	channels with the corresponding analytical approximations. Concretely, for each sector (magneto-sonic $A_{10\times10}$, Alfv\'en $B_{6\times6}$, and charge diffusion $C_{3\times3}$), we compute the full set of complex eigenfrequencies $\omega(k,\theta)$ by solving the secular equation and then confront them with the small-$k$ expansions (hydrodynamic and relaxation branches) and with the large-$k$
	asymptotics (propagating and finite-frequency poles). {Throughout this subsection, the propagation angle is fixed at $\theta=\pi/4$, representing a generic oblique propagation direction between the parallel and perpendicular limits. The dependence of the mode spectrum on the propagation angle $\theta$ will be investigated separately
		in Sec.~\ref{sec:theta_dependence}.} 
	\begin{figure}[htbp]
		\centering
		\includegraphics[width=1\textwidth]{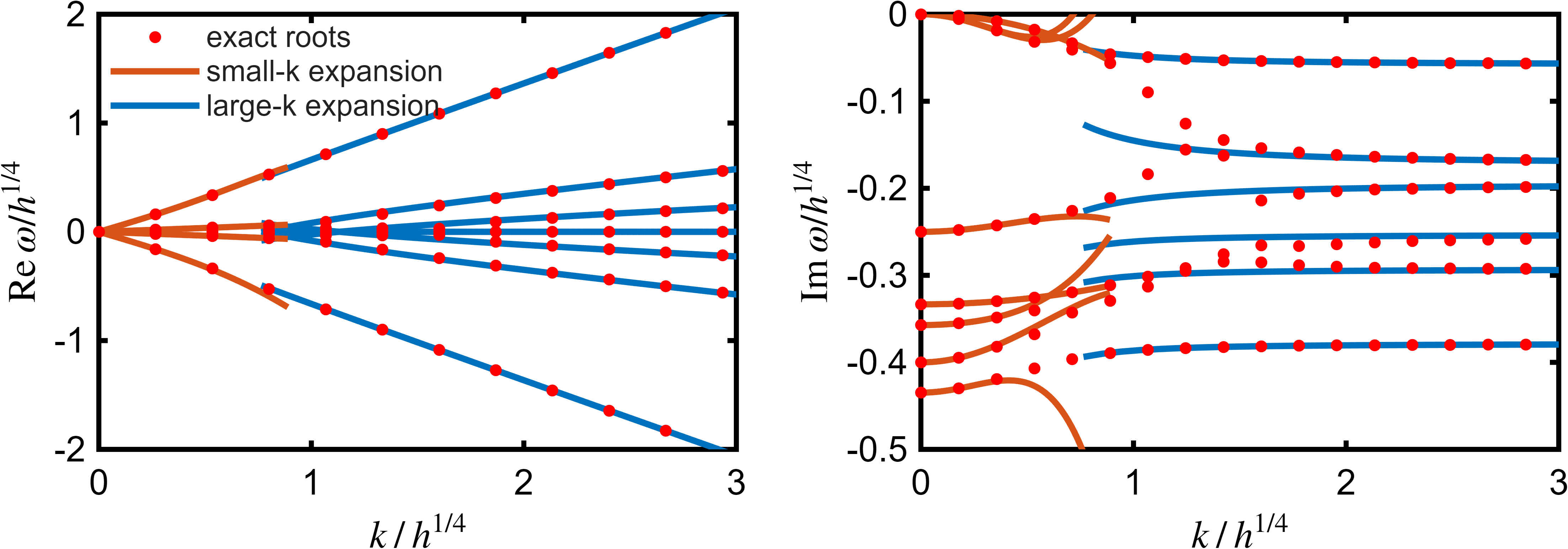}
		\caption{The comparison for magneto-sonic modes. The small k expansion curves relate to Eq.\eqref{small k A} and the large k expansion curves relate to Eq.\eqref{large k A}.}
		\label{fig1}
	\end{figure}
	\begin{figure}[htbp]
		\centering
		\includegraphics[width=1\textwidth]{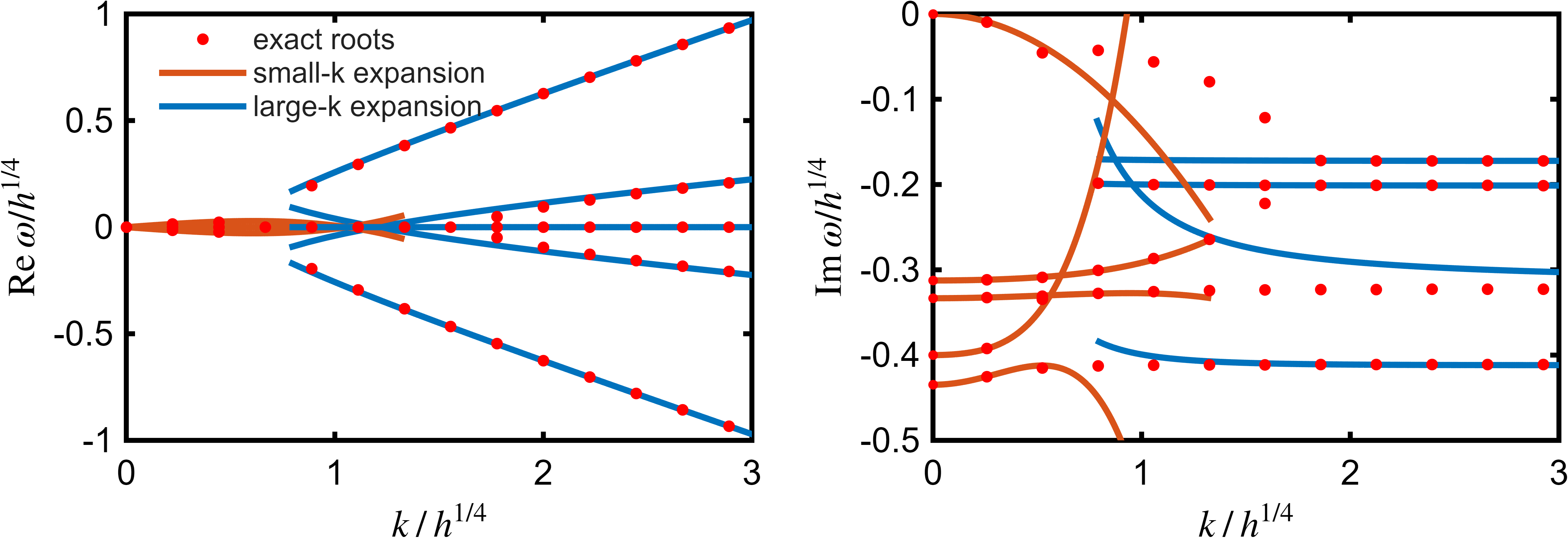}
		\caption{The comparison for Alfv\'en modes. The small k expansion curves relate to Eq.\eqref{small k B} and the large k expansion curves relate to Eq.\eqref{large k B}.}
		\label{fig2}
	\end{figure}
	\begin{figure}[htbp]
		\centering
		\includegraphics[width=1\textwidth]{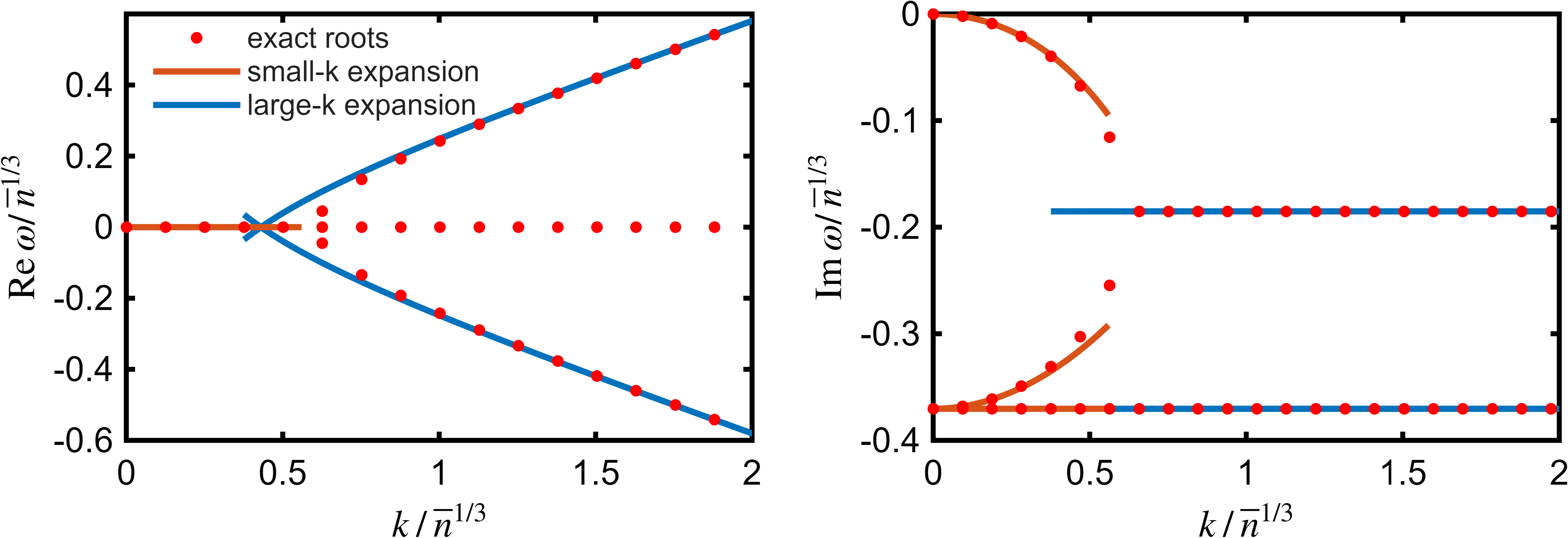}
		\caption{The comparison for charge diffusion modes. The small k expansion curves relate to Eq.\eqref{small k C} and the large k expansion curves relate to Eq.\eqref{large k C}.}
		\label{fig3}
	\end{figure}

	The comparisons of the solution $\omega$ are shown in Figs.~\ref{fig1}--\ref{fig3}, where the left panel shows the magnitude of the real part, $\mathrm{Re}\,\omega$ while the right panel shows $\mathrm{Im}\,\omega$. Because the propagating branches come in counter-propagating pairs, $\omega_\pm\simeq \pm v k+\cdots$, plotting $\mathrm{Re}\,\omega$ causes the two branches to overlap into a single curve; therefore, the damping information is contained entirely in the $\mathrm{Im}\,\omega$ panels. 
	The insets highlight the near-origin region where the long-wavelength series is expected to be most accurate.

	Taken together, Figs.~\ref{fig1}--\ref{fig3} confirm the expected two-scale structure of the spectrum: the small-$k$ expansion accurately captures the hydrodynamic manifold and the leading gradient-induced attenuation, while the large-$k$ expansion captures the short-wavelength propagation/relaxation structure that is most sensitive to causality and to the presence of finite relaxation times. 
	The numerical roots provide a continuous interpolation between the two limits and allow for an unambiguous branch identification across the crossover region.
	
	For the magneto-sonic mode in Fig.~\ref{fig1}, the small-$k$ expansion correctly reproduces the exact spectrum near $k\to0$: two propagating pairs emerge with $|\mathrm{Re}\,\omega|\propto k$ (fast/slow magneto-sonic branches), while the remaining modes start as gapped relaxation poles with $\mathrm{Im}\,\omega\simeq-1/\tau_i$. 
	As $k$ increases, the propagating branches exhibit the expected leading attenuation (quadratic in $k$), whereas the gapped poles gradually approach $k$-independent damping rates. 
	In the short-wavelength regime the large-$k$ asymptotics captures both the linear growth of $|\mathrm{Re}\,\omega|$ and the saturation of $\mathrm{Im}\,\omega$, with a crossover window around $k/h^{1/4}\sim 0.5$--$1$ where branch mixing makes neither truncated expansion uniformly accurate.

	The Alfv\'en doublet at $k\to0$ splits into a counter-propagating pair $\omega_\pm\simeq \pm v_A\cos\theta\,k-i\Gamma(\theta)k^2+\cdots$ in Fig.~\ref{fig2}, and the near-origin agreement between the orange curves and the exact roots is manifest in the inset. 
	
	{At larger $k$, the spectrum reorganizes into wave-like branches with
		$|\mathrm{Re}\,\omega|\propto k$ and two purely damped relaxation poles,
		$\omega_{r_5}$ and $\omega_{r_6}$ in Eq.~\eqref{large k B}.
		The former is a mixed electromagnetic-response pole, reducing to
		$-i/\tau_m$ at $\theta=0$ and $-i/\tau_\ell$ at
		$\theta=\pi/2$, whereas the latter is a mixed viscous--anisotropic
		relaxation pole. Both remain in the lower half-plane under the assumed
		positivity conditions.}
	
	Quantitatively, in the intermediate range $k/h^{1/4}\sim 1$--$10$ the leading large-$k$ approximation for the non-propagating branches is still visibly imperfect, and the convergence of the $\tau_m$ pole $\mathrm{Im}\,\omega_{\tau_m}\sim-1/3$ to its asymptotic value becomes accurate only at rather large momenta (roughly $k/h^{1/4}\gtrsim 10$) in the present parameter set.
	
At small $k$, the charge sector in Fig.~\ref{fig3} contains a hydrodynamic diffusion mode and two non-hydrodynamic relaxation modes, in agreement with the small-$k$ expansion. As $k$ increases, the diffusion mode and one
relaxation branch approach each other and reorganize into a pair of
counter-propagating modes with opposite real parts,
$\mathrm{Re}\,\omega\simeq\pm v_C k$, while their imaginary parts tend
toward the common large-$k$ damping rate $-1/(2\tau_\nu)$.
The remaining relaxation pole stays purely damped and approaches
$\mathrm{Im}\,\omega=-1/\tau_\nu$. This behavior illustrates the
telegraph-type crossover from diffusion at long wavelengths to finite-speed
propagation at short wavelengths.
	
	\subsection{Angle analysis}
	\label{sec:theta_dependence}
	\label{section7.2}

	Before turning to causality analysis, it is useful to examine how the spectra vary with the propagation angle $\theta$. 
	For the present second-order theory, obtaining compact analytic expressions valid for generic $\theta$ is highly nontrivial. 
	Therefore, in this section, we mainly investigate the angular dependence numerically, by comparing the exact roots with the asymptotic solutions obtained from the small-$k$ and large-$k$ expansions.
	\begin{figure*}[htbp]
		\centering
		\includegraphics[width=1\textwidth]{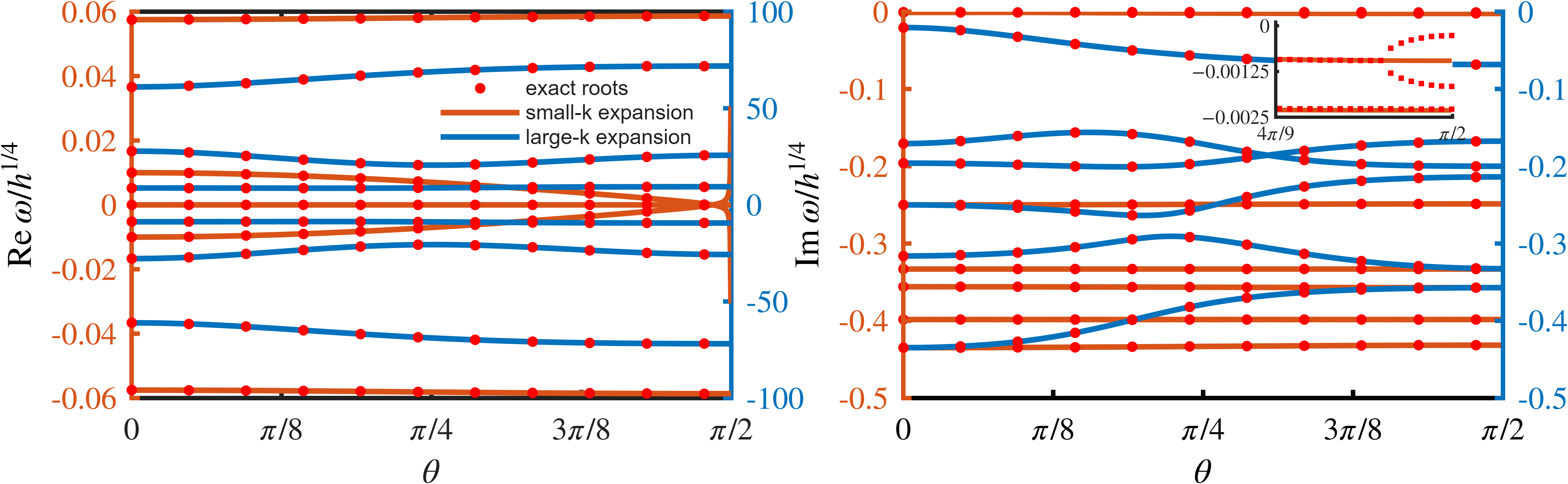}
		\caption{Angle dependence of the solution $\omega$ for magneto-sonic sector. Other parameters are taken from Table~\ref{table1}.}
		\label{figureθ1}
	\end{figure*}
	The results for the three sectors associated with the matrices $A$, $B$, and $C$ are displayed in Figs.~\ref{figureθ1}--\ref{figureθ3}. 
	For the $A$ and $B$ sectors, namely the magneto-sonic and Alfv\'en sectors, the small-$k$ expansion works well in a broad angular range, but its agreement with the exact roots deteriorates rapidly as $\theta$ approaches $\pi/2$. 
	The modes most visibly affected are precisely the propagating branches: near the right angle, the deviation between the small-$k$ approximation and the exact roots becomes substantial, while the purely damped modes are much less sensitive. 
	
This behavior is consistent with the mechanism discussed in
Refs.~\cite{Fang2024PRD,Fang2025PRD}, where the ordinary small-$k$
expansion was shown to break down near $\theta=\pi/2$ because
$\cos\theta$ becomes an additional small parameter and the corresponding
dispersion relations develop a non-analytic angular structure.
{
	Notably, the imaginary part of the numerical spectrum exhibits a splitting of the hydrodynamic modes in the small-$k$ regime as $\theta$ approaches	$\pi/2$. A qualitatively similar long-wavelength mode splitting was found by	Vardhan \textit{et al.} in an effective-field-theory formulation of strong-field MHD based on one-form symmetry~\cite{VardhanGrozdanovLeutheusserLiu2024}, where the two hydrodynamic diffusion modes become nondegenerate beyond a critical propagation angle.	This provides an independent qualitative point of comparison with the long-wavelength splitting pattern observed here.
}
	\begin{figure*}[htbp]
		\centering
		\includegraphics[width=1\textwidth]{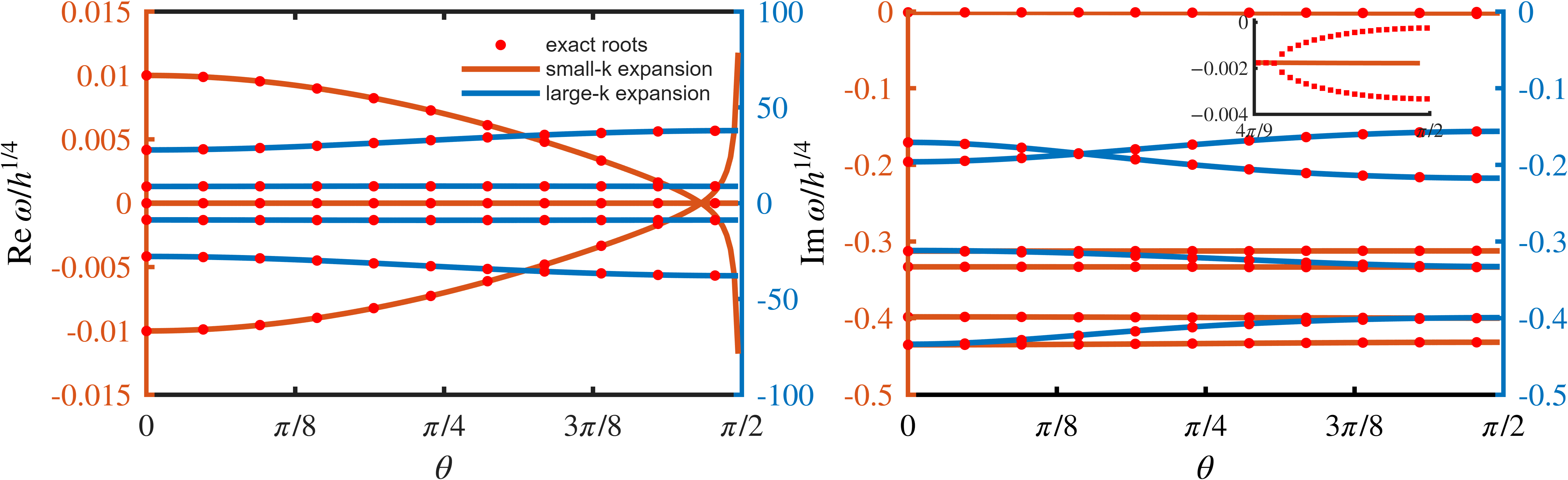}
		\caption{Angle dependence of the solution $\omega$ for Alfv\'en sector. Other parameters are taken from Table~\ref{table1}.}
		\label{figureθ2}
	\end{figure*}

	By contrast, the large-$k$ expansion is much less affected by the angular variation in the $A$ and $B$ sectors. 
	As seen in Figs.~\ref{figureθ1} and \ref{figureθ2}, the large-$k$ formulas continue to track the exact roots rather well over the entire angular interval, including the region close to $\theta=\pi/2$. 
	This indicates that, for these two sectors, the high-momentum asymptotics is considerably more robust against the angular deformation than the small-$k$ approximation.
	
	The situation is different for the charge-diffusion sector described by the matrix $C$ displayed in Fig.~\ref{figureθ3}. 
	In this case, the large-$k$ expansion remains accurate for most values of $\theta$, but becomes unreliable when $\theta\to 0$ because $\csc\theta$ terms appear at order $k^{-1}$ in \eqref{large k C}. 
	Away from this narrow region, the agreement between the exact roots and the asymptotic solution remains very good. 
	Therefore, the charge diffusion sector exhibits a complementary angular sensitivity: unlike the magneto-sonic and Alfv\'en sectors, where the main difficulty arises near $\theta=\pi/2$ in the small-$k$ regime, the charge-diffusion sector shows its visible mismatch near $\theta=0$ in the large-$k$ regime.
	
	These observations provide a useful guide for the applicability of the asymptotic formulas. 
	For the $A$ and $B$ sectors, the small-$k$ expressions should not be trusted too close to the perpendicular direction, whereas the large-$k$ expressions remain reliable in that region. 
	\begin{figure*}[htbp]
		\centering
		\includegraphics[width=1\textwidth]{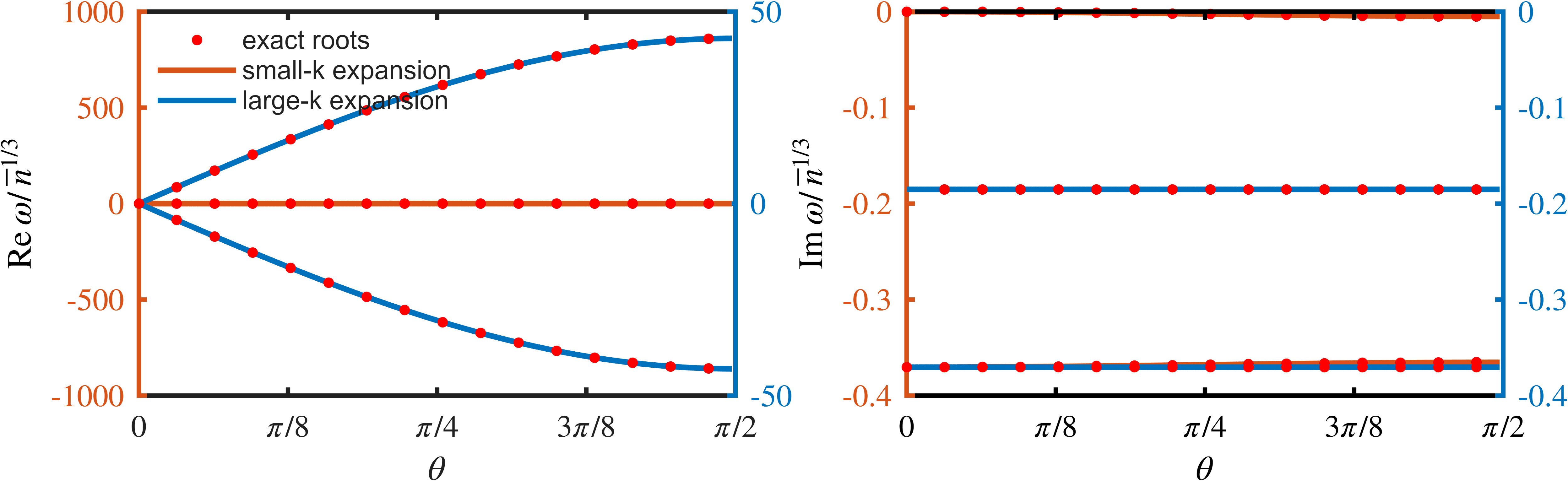}
		\caption{Angle dependence of the solution $\omega$ for charge-diffusion sector. Other parameters are taken from Table~\ref{table1}.}
		\label{figureθ3}
	\end{figure*}
	For the $C$ sector, the large-$k$ approximation is reliable for generic angles but should be used with caution near the parallel direction. 
	
 The angular dependence also motivates the causality analysis below, where the characteristic velocities and the corresponding causal constraints are examined for different propagation directions.

\subsection{Causality analysis based on numerical spectrum}
\label{section7.3}

In this subsection, we diagnose causality and stability directly from the numerical dispersion relations. 
For any propagating branch, a {necessary} causality requirement is that the asymptotic signal (group) velocity at short wavelengths does not exceed the speed of light,
\be\label{le light speed}
v_g=\lim_{k\to\infty}\,\left|\mathrm{Re}\,\frac{\partial\omega}{\partial k}\right|\le 1,
\ee
which tests whether the large-$k$ characteristic propagation remains subluminal.

In addition, a well-behaved high-momentum spectrum requires that the dispersion relations remain bounded {in the short-wavelength limit}, namely that
\be\label{boound condition}
\lim_{k\to\infty}\left|\frac{\omega}{k}\right|\ \text{is bounded},
\ee
so that the spectrum approaches finite characteristic velocities rather than developing divergent high-momentum behavior.
Finally, {linear stability requires every mode to remain in the
closed lower half of the complex-frequency plane,}
\begin{eqnarray}
	\mathrm{Im}\,\omega\leq0,\label{im part less than 0}
\end{eqnarray}
{with strict inequality for damped relaxation modes.}

It is useful to note that the numerical spectra in Figs.~\ref{fig1}--\ref{fig3} already satisfy two basic consistency requirements associated with Eqs.~\eqref{boound condition} and \eqref{im part less than 0}. 
First, at large momentum all branches remain well behaved in the sense that $|\omega|/k$ approaches finite values, corresponding to finite asymptotic velocities for the propagating branches and vanishing contributions for the finite-frequency relaxation poles. 
{ Second, all numerical branches remain in the closed lower half-plane throughout the plotted range, $\mathrm{Im}\,\omega<0$, indicating the absence of linear instabilities for the parameter set used in Figs.~\ref{fig1}--\ref{fig3}.} 
These observations motivate a more quantitative analysis of the asymptotic group velocities and their dependence on the transport parameters and propagation angle, as discussed below.

\begin{figure}[htbp]
	\centering
	\includegraphics[width=1\textwidth]{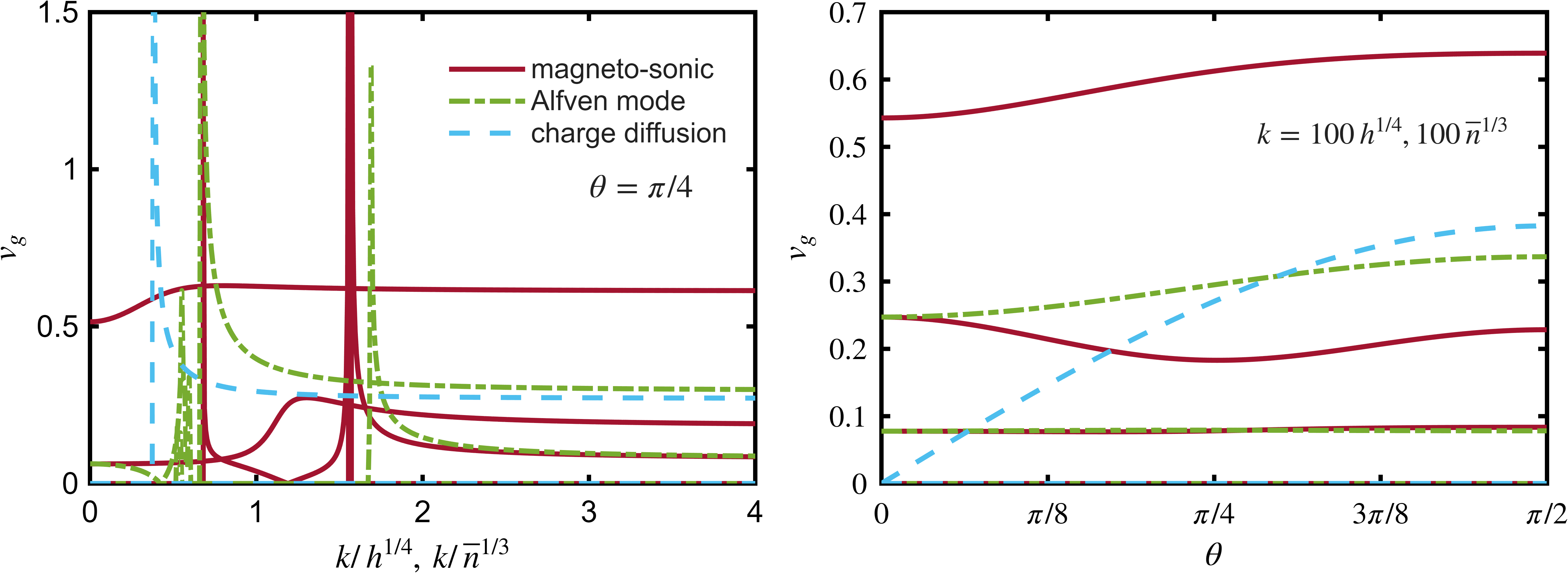}
	\caption{
		Numerically extracted group velocities in the magneto-sonic, Alfv\'en, and charge-diffusion sectors.
	{	The left panel shows their momentum dependence, while the right panel shows the angular dependence over $0\leq\theta\leq\pi/2$.}
		The reference parameters are those listed in Tab.~\ref{table1}.
	}
	\label{figvg}
\end{figure}
	{
	Figure~\ref{figvg} shows the group velocities more directly.
	The propagating branches approach finite values at large $k$, while their angular dependence varies smoothly over $0\leq\theta\leq\pi/2$.
	This behavior reflects the anisotropy induced by the background magnetic field and shows that the propagation angle plays an important role in determining the characteristic velocities.

For clarity, the two-dimensional causality maps below display the representative directions $\theta=0$, $\pi/4$, and $\pi/2$, corresponding to parallel, oblique, and perpendicular propagation.
These three directions capture the principal angular variation seen in Fig.~\ref{figvg} without overcrowding the contour plots.
In the following, we use $v_g=v_{\max}$ to denote the largest asymptotic group velocity in the corresponding sector, and the red $v_{\max}=1$ contours mark the luminal boundaries.

The two limiting directions also simplify the characteristic dynamics.
Writing $k_\parallel=k\cos\theta$ and $k_\perp=k\sin\theta$, one has $k_\perp=0$ at $\theta=0$ and $k_\parallel=0$ at $\theta=\pi/2$.
Transport terms that enter the leading large-$k$ characteristic equations only through the vanishing momentum component are therefore projected out.
This geometric structure explains the parameter-independent velocities and the vertical or horizontal contours appearing in some of the following panels.

\begin{figure*}[htbp]
	\centering
	\includegraphics[width=1\textwidth]{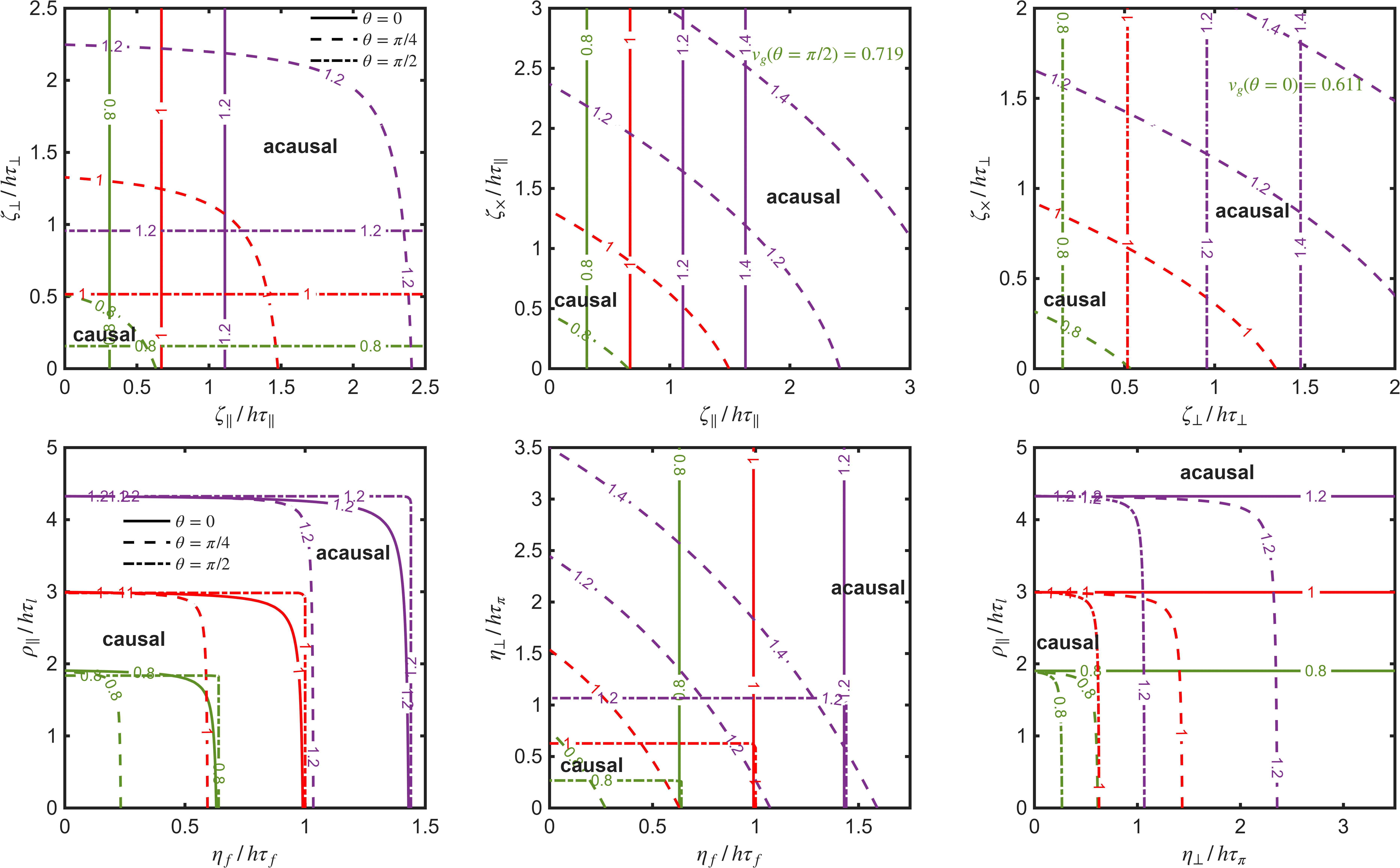}
	\caption{
		Representative causality maps of the maximal asymptotic group velocity in the magneto-sonic sector for $\theta=0$, $\pi/4$, and $\pi/2$.
		Solid, dashed, and dash-dotted curves denote the three directions, respectively.
		Green, red, and purple contours correspond to $v_{\max}<1$, $v_{\max}=1$, and $v_{\max}>1$.
		Other parameters are fixed according to Tab.~\ref{table1}.
	}
	\label{figca}
\end{figure*}

Figure~\ref{figca} shows the magneto-sonic causality maps.
The first row scans the anisotropic bulk coefficients $\zeta_\parallel$, $\zeta_\perp$, and $\zeta_\times$, while the second row illustrates their interplay with $\rho_\parallel$, $\eta_f$, and $\eta_\perp$.
The displacement of the luminal contours with $\theta$ shows that the causal boundary is intrinsically angle dependent and, for generic oblique propagation, is governed by correlated combinations of transport coefficients and relaxation scales.

The special-angle limits clearly exhibit the geometric decoupling described above.
For $\theta=0$, the characteristic velocity is independent of $\zeta_\perp$, $\zeta_\times$, and $\eta_\perp$, whereas for $\theta=\pi/2$ it is independent of $\zeta_\parallel$ and $\zeta_\times$.
The corresponding parameter-independent values $v_g(\theta=0)=0.614$ and $v_g(\theta=\pi/2)=0.719$ are both subluminal.
For oblique propagation, both $k_\parallel$ and $k_\perp$ are nonzero, so the longitudinal and transverse transport channels are simultaneously involved and the causal boundaries become genuinely multi-parameter.

\begin{figure*}[htbp]
	\centering
	\includegraphics[width=1\textwidth]{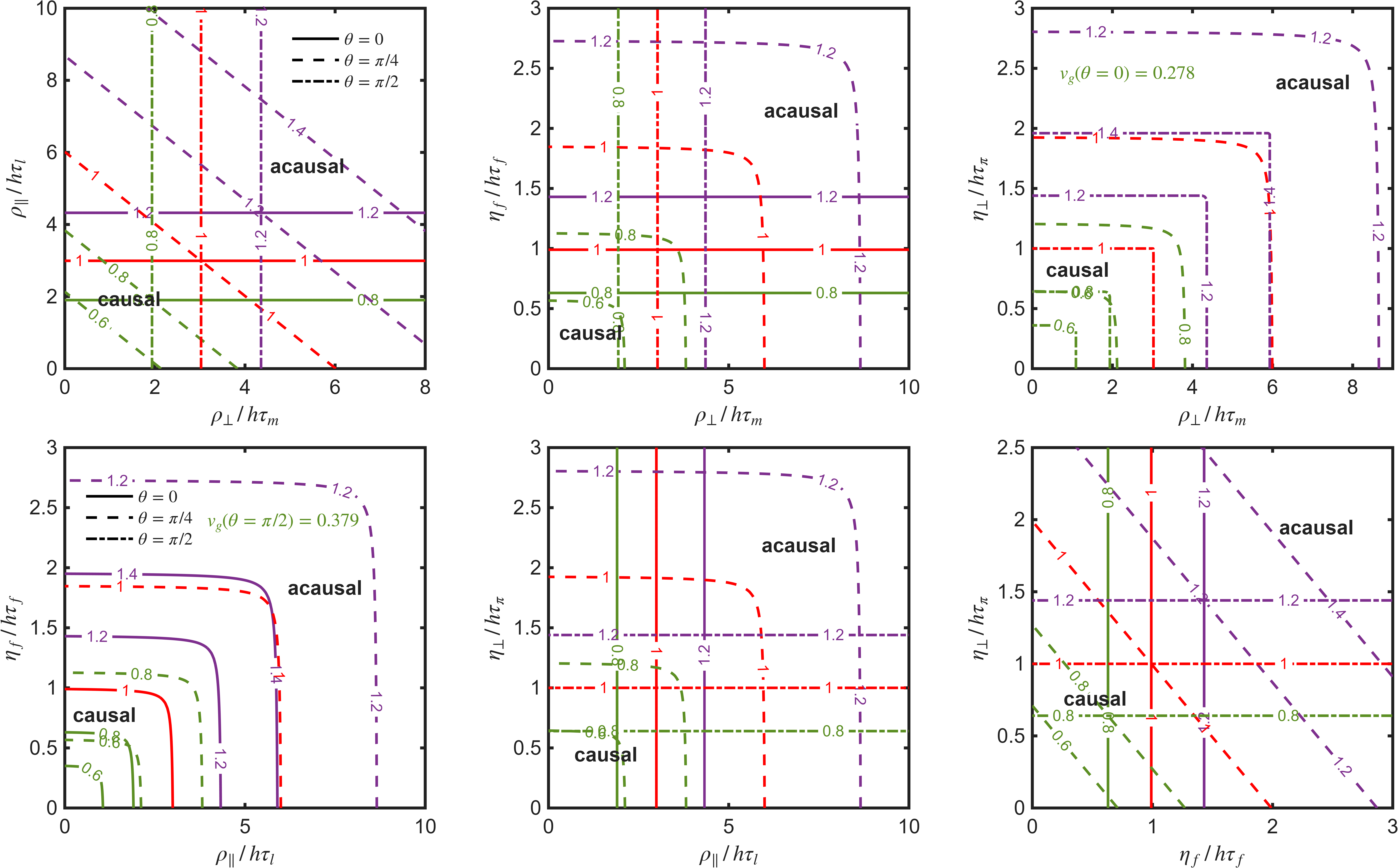}
	\caption{
		Representative causality maps of the maximal asymptotic group velocity in the Alfv\'en sector for $\theta=0$, $\pi/4$, and $\pi/2$.
		Solid, dashed, and dash-dotted curves denote the three directions, respectively.
		Green, red, and purple contours correspond to $v_{\max}<1$, $v_{\max}=1$, and $v_{\max}>1$.
		Other parameters are fixed according to Tab.~\ref{table1}.
	}
	\label{figcb}
\end{figure*}

Figure~\ref{figcb} presents the corresponding Alfv\'en-sector constraints through pairwise scans of $\rho_\perp$, $\rho_\parallel$, $\eta_f$, and $\eta_\perp$.
As in the magneto-sonic sector, the causal boundary varies with the propagation direction and generally involves correlated transport channels.

At $\theta=0$, the Alfv\'en velocity is independent of $\eta_\perp$ and $\rho_\perp$, yielding the parameter-independent value $v_g(\theta=0)=0.278$ in the corresponding scans.
At $\theta=\pi/2$, it is instead independent of $\rho_\parallel$ and $\eta_f$, with $v_g(\theta=\pi/2)=0.379$.
Thus parallel and perpendicular propagation probe different subsets of the anisotropic transport coefficients, while oblique propagation mixes the two structures.

The numerical luminal boundaries are also consistent with the analytical large-$k$ condition derived in Eq.~\eqref{theoretical constraints of Alfven sector}.
In particular, the red $v_{\max}=1$ contours closely reproduce the theoretical luminal boundary defined by
$-\mathbf{b}_B+\sqrt{\Delta_B}=2h\tau_\ell\tau_m\tau_f\tau_\pi$.
This agreement provides a direct numerical verification of the analytical Alfv\'en causality condition.

\begin{figure}[htbp]
	\centering
	\includegraphics[width=0.42\textwidth]{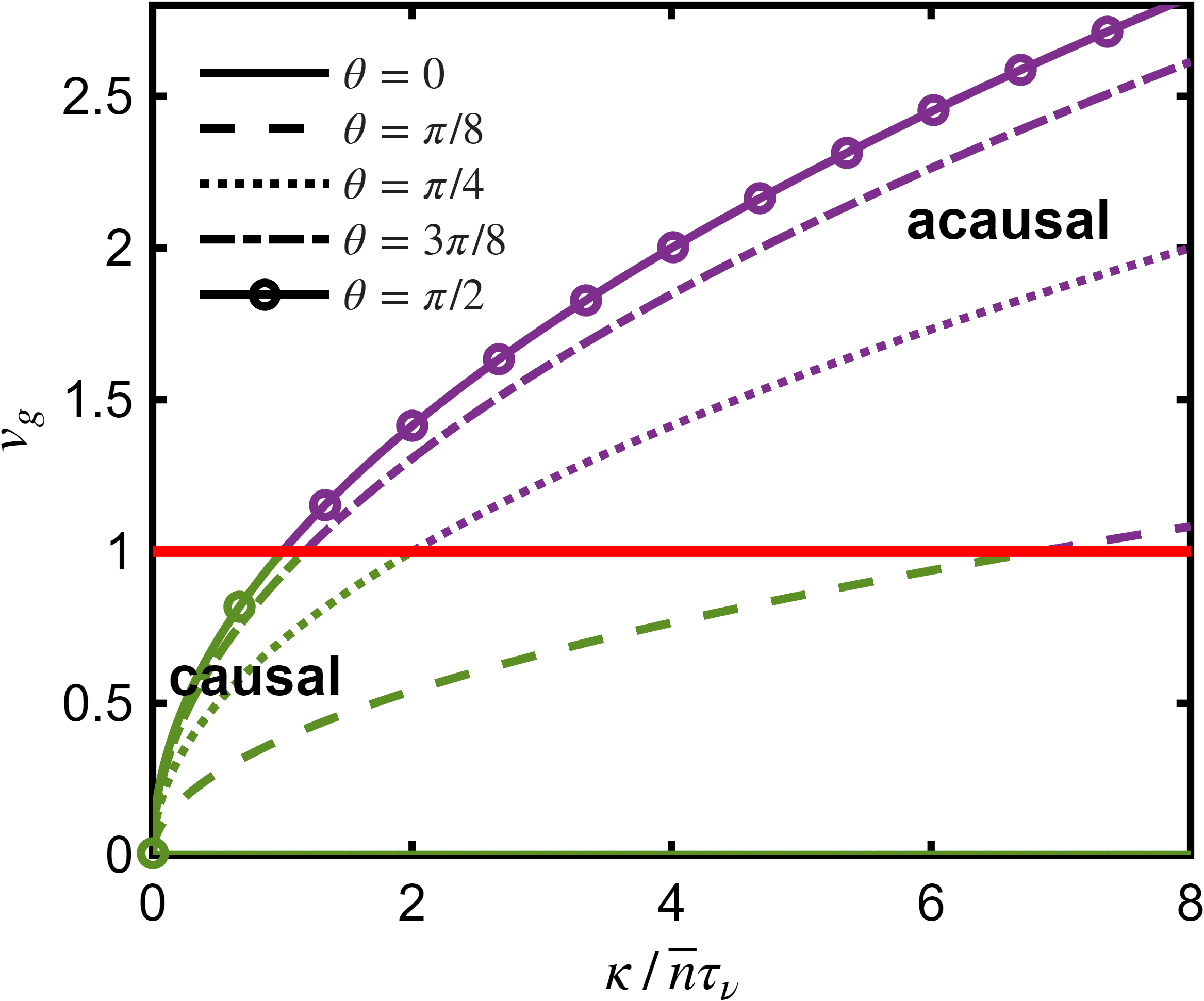}
	\caption{
		Asymptotic group velocity in the charge-diffusion sector as a function of $\kappa/(\bar n\,\tau_\nu)$ for $\theta=0$, $\pi/8$, $\pi/4$, $3\pi/8$, and $\pi/2$.
		Green and purple denote the regions $v_g<1$ and $v_g>1$, respectively, while the red line marks $v_g=1$.
		Other parameters are fixed according to Tab.~\ref{table1}.
	}
	\label{fig7}
\end{figure}

The charge-diffusion sector in Fig.~\ref{fig7} provides the simplest illustration of the angular dependence.
Its large-$k$ characteristic velocity and causality condition have already been derived in Eqs.~\eqref{large k C} and \eqref{c condition}.
For every nonzero angle, $v_g$ increases with $\kappa/(\bar n\tau_\nu)$, while the luminal crossing moves to smaller values as $\theta$ approaches $\pi/2$.
The perpendicular direction therefore gives the strongest restriction, whereas the leading characteristic velocity vanishes for parallel propagation.

The numerical $v_g=1$ crossings agree with the analytical boundary
$\kappa\sin^2\theta=\bar n\tau_\nu$ from Eq.~\eqref{c condition}.
The numerical spectrum thus reproduces both the angular dependence and the causal threshold predicted by the large-$k$ analysis.

The regions marked as ``causal'' in Figs.~\ref{figca} and \ref{figcb} should be understood as the intersection of the causal domains obtained for the three representative directions $\theta=0$, $\pi/4$, and $\pi/2$.
A strictly angle-independent causal domain, however, must satisfy the subluminal condition for every propagation direction, which in the present parity-even and axisymmetric setup is equivalently obtained by taking the intersection over $0\leq\theta\leq\pi/2$.
The present three-angle analysis therefore provides an initial determination of the location and extent of the common causal region, while a denser angular scan would systematically refine its boundary.
For the charge-diffusion sector, the situation is simpler: Eq.~\eqref{c condition} requires $\kappa\sin^2\theta\leq\bar n\tau_\nu$, and imposing this condition for all directions immediately yields the angle-independent causal bound
$\kappa\leq\bar n\tau_\nu$, with the perpendicular direction $\theta=\pi/2$ providing the most restrictive case.
	
	Taken together, Figs.~\ref{figca}--\ref{fig7} show that causality in second-order RMHD is controlled jointly by the transport coefficients, relaxation times, and propagation direction.
	Special directions can project individual dissipative channels out of the characteristic velocity, whereas oblique propagation generally produces coupled multi-parameter constraints.
	The agreement between the numerical luminal boundaries and the analytical conditions in Eqs.~\eqref{theoretical constraints of Alfven sector} and \eqref{c condition} further supports the large-$k$ causality analysis.}

\section{Conclusion}
\label{section8}

In this work, we have constructed a relativistic second-order magnetohydrodynamic framework based on entropy-current analysis and investigated its linear stability and short-wavelength causal properties around a homogeneous magnetized equilibrium. Starting from the requirement of non-negative entropy production, we derived relaxation-type evolution equations for the anisotropic dissipative fluxes and obtained a transient extension of dissipative RMHD with finite relaxation times. This construction provides a systematic framework in which the role of relaxation dynamics in the linear propagation and damping of collective modes can be analyzed, and may be viewed as a macroscopic counterpart of transient relativistic hydrodynamic approaches developed from phenomenological, kinetic-theory, and nonequilibrium statistical formulations~\cite{IsraelStewart1979,HiscockLindblom1983,HiscockLindblom1985,HernandezKovtun2017,Denicol2018,Harutyunyan2018,TiwariPatra2025}.

We then linearized the theory around a static homogeneous background with the magnetic field oriented along the $z$ direction. { Under the additional assumptions adopted in Eqs.~\eqref{ass1}-\eqref{ass2}, including the conformal equation of state, the charge-conjugation-symmetric background, and the vanishing vector cross-couplings, the secular equation factorizes into the magneto-sonic, Alfv\'en, and charge-diffusion sectors.} We derived the corresponding asymptotic dispersion relations in both the small-$k$ and large-$k$ regimes and compared them with the full numerical roots. The small-$k$ results reproduce the expected hydrodynamic and relaxation behavior, whereas the large-$k$ solutions determine the asymptotic characteristic velocities and finite relaxation poles relevant to the linear causality analysis.

The angular dependence of the spectrum was also examined explicitly. For the magneto-sonic and Alfv\'en sectors, the ordinary small-$k$ expansion becomes less accurate near the perpendicular direction, while the charge-diffusion large-$k$ expansion requires particular care close to the parallel direction. These observations illustrate that the propagation angle is an important parameter in magnetized hydrodynamics and that asymptotic momentum expansions need not remain uniformly accurate near special directions. Similar non-uniform behavior of momentum expansions near special propagation angles has also been reported in related analyses~\cite{Fang2024PRD,Fang2025PRD}. {At the same time, the pronounced angular dependence found in our long-wavelength numerical results is qualitatively consistent with effective-field-theory studies of strong-field MHD~\cite{VardhanGrozdanovLeutheusserLiu2024}. Despite the reduced accuracy of the asymptotic expansions near special directions, the exact numerical group velocities themselves vary smoothly with the propagation angle over the range considered in Fig.~\ref{figvg}.}

{Within the linear large-$k$ analysis, we obtained explicit analytical causality constraints for the Alfv\'en and charge-diffusion sectors. In the Alfv\'en sector, the condition in Eq.~\eqref{theoretical constraints of Alfven sector} determines the luminal boundary through the interplay among $\rho_\parallel$, $\rho_\perp$, $\eta_f$, $\eta_\perp$, their associated relaxation times, and the propagation angle. In the charge-diffusion sector, Eq.~\eqref{c condition} gives the corresponding analytical condition $\kappa\sin^2\theta\leq\bar n\tau_\nu$. The numerical luminal boundaries displayed in Figs.~\ref{figcb} and \ref{fig7} agree closely with these analytical results, providing an independent numerical check of the large-$k$ characteristic analysis.} The telegraph-type structure of the charge-diffusion mode is consistent with the general role of finite relaxation times in restoring finite characteristic propagation speeds in relativistic dissipative theories~\cite{PuKoideRischke2010,HiscockLindblom1983,Fang2024PRD}. 

{The angular scans in Figs.~\ref{figca} and \ref{figcb} provide a representative characterization of the causal parameter space. Requiring simultaneous subluminal propagation at $\theta=0$, $\pi/4$, and $\pi/2$ already isolates a common admissible region and exposes the underlying correlations among the transport coefficients and relaxation scales. The precise global boundary, however, is expected to be further constrained as more propagation angles are included. The present analysis should therefore be viewed as identifying the dominant structure and location of the causal domain, with a denser angular scan offering a systematic refinement of its boundary. For the charge-diffusion sector in Fig.~\ref{fig7}, imposing Eq.~\eqref{c condition} for all propagation directions further gives the angle-independent linear large-$k$ bound $\kappa\leq\bar n\tau_\nu$.}

Overall, the present study provides a systematic entropy-current construction of second-order RMHD together with a mode-resolved analysis of its linear stability and large-$k$ causal structure {under the conformal equation of state and the charge-conjugation symmetric background.} {The results indicate that the causal behavior of magnetized relativistic fluids is governed by the interplay between anisotropic transport coefficients, their corresponding relaxation times, and the propagation angle between the wave vector and the background magnetic field.} Several extensions deserve further investigation, including a more general second-order coupling structure, the analysis of non-static or nonlinear backgrounds, 
a microscopic determination of the second-order transport coefficients, and extensions to anomalous or chiral RMHD relevant to the chiral magnetic effect~\cite{BetzHenkelRischke2009JPG,BenenowskiPoovuttikul2019,MolnarRischke2025,RoyPuRezzollaRischke2015,ShokriSadooghi2018,MoghaddamAlbericoSheKordAzadegan2020,Denicol2018,Denicol2019,SiddiqueWangPuWang2019,KiamariRahbardarShokriSadooghi2021,Peng:2022cya,LandryLiu2026}. We hope that the present framework can serve as a useful basis for future studies of causality and stability in relativistic magnetohydrodynamics relevant to heavy-ion matter, compact stars, and other strongly magnetized relativistic plasmas.

	\section*{Acknowledgments}
	
	This work was supported in part by the National Key Research and Development Program of China under Contract No. 2022YFA1604900; the National Natural Science Foundation of China (NSFC) under Grant Nos. 12435009, 12275104, and 12605222; the Startup Research Fund of Henan Academy of Sciences (No. 231820058); the High-level Achievements Reward and Cultivation Projects (No. 20252320001); the Basic Research Fund of Henan Academy of Sciences (No. 20260620005); the Key Laboratory of Quark and Lepton Physics under Contract No. QLPL2025P01; and the Fundamental Research Funds for the Central Universities (No. 2026CXGR050).

	\appendix

	\section{Covariant tensor basis for magnetized RMHD}
	\label{appendixf}
	
	This appendix fixes the covariant basis used in the main text.  We use a time-like fluid velocity $u^\mu$ and a unit space-like vector $b^\mu$ specifying the local magnetic-field direction,
	\begin{equation}
		u^\mu u_\mu=1,\qquad b^\mu b_\mu=-1,\qquad u^\mu b_\mu=0 .
		\label{appA:normalization}
	\end{equation}
	The magnetic induction four-vector is written as $B^\mu=B b^\mu$, with $B\equiv\sqrt{-B^\mu B_\mu}$.  In the local rest frame one may choose
	\begin{equation}
		u^\mu_{\rm LR}=(1,0,0,0),\qquad b^\mu_{\rm LR}=(0,0,0,1),
		\label{appA:LR}
	\end{equation}
	so that the two directions left invariant by both $u^\mu$ and $b^\mu$ form the transverse plane.
	
	The two projectors used throughout the paper are
	\begin{equation}
		\Delta^{\mu\nu}=g^{\mu\nu}-u^\mu u^\nu,\qquad
		\Xi^{\mu\nu}=g^{\mu\nu}-u^\mu u^\nu+b^\mu b^\nu
		=\Delta^{\mu\nu}+b^\mu b^\nu .
		\label{appA:projectors}
	\end{equation}
	They obey
	\begin{equation}
		\Delta^{\mu\nu}u_\nu=0,\qquad 
		\Xi^{\mu\nu}u_\nu=\Xi^{\mu\nu}b_\nu=0,\qquad
		\Delta^\mu_{\ \mu}=3,\qquad
		\Xi^\mu_{\ \mu}=2 .
		\label{appA:projector-properties}
	\end{equation}
	For later use we define the transverse projections
	\begin{equation}
		A^{\langle\mu\rangle}\equiv \Delta^\mu_{\ \nu}A^\nu,\qquad
		A^{\{\mu\}}\equiv \Xi^\mu_{\ \nu}A^\nu ,
		\label{appA:vector-proj}
	\end{equation}
	and the symmetric traceless projectors
	\begin{align}
		\Delta^{\mu\nu}_{\alpha\beta}
		&\equiv
		\frac{1}{2}
		\left(
		\Delta^\mu_{\ \alpha}\Delta^\nu_{\ \beta}
		+\Delta^\mu_{\ \beta}\Delta^\nu_{\ \alpha}
		\right)
		-\frac{1}{3}\Delta^{\mu\nu}\Delta_{\alpha\beta},
		\label{appA:Delta-STT}
		\\
		\Xi^{\mu\nu}_{\alpha\beta}
		&\equiv
		\frac{1}{2}
		\left(
		\Xi^\mu_{\ \alpha}\Xi^\nu_{\ \beta}
		+\Xi^\mu_{\ \beta}\Xi^\nu_{\ \alpha}
		\right)
		-\frac{1}{2}\Xi^{\mu\nu}\Xi_{\alpha\beta}.
		\label{appA:Xi-STT}
	\end{align}
	The braces notation for a rank-two tensor is
	\begin{equation}
		A^{\{\mu\nu\}}\equiv \Xi^{\mu\nu}_{\alpha\beta}A^{\alpha\beta}.
		\label{appA:tensor-braces}
	\end{equation}
	
	Any four-vector can be split into components parallel to $u^\mu$, parallel to $b^\mu$, and transverse to both:
	\begin{equation}
		A^\mu=(A\cdot u)u^\mu-(A\cdot b)b^\mu+\Xi^\mu_{\ \nu}A^\nu .
		\label{appA:vector-decomp}
	\end{equation}
	This is the algebraic decomposition underlying the definitions of particle diffusion, magnetic induction, and transverse dissipative vectors in the main text.
	
	For a symmetric rank-two tensor $S^{\mu\nu}=S^{\nu\mu}$, the decomposition adapted to the pair $(u^\mu,b^\mu)$ is
	\begin{align}
		S^{\mu\nu}
		={}&
		\mathcal E\,u^\mu u^\nu
		+2M\,u^{(\mu}b^{\nu)}
		+\mathcal P_\parallel b^\mu b^\nu
		-\mathcal P_\perp \Xi^{\mu\nu}
		+2q^{(\mu}u^{\nu)}
		+2r^{(\mu}b^{\nu)}
		+s_\perp^{\mu\nu},
		\label{appA:symmetric-decomp}
	\end{align}
	where
	\begin{align}
		\mathcal E &= S^{\alpha\beta}u_\alpha u_\beta,
		&
		M &= -S^{\alpha\beta}u_\alpha b_\beta,
		&
		\mathcal P_\parallel &= S^{\alpha\beta}b_\alpha b_\beta,
		\label{appA:symmetric-coeff-1}
		\\
		\mathcal P_\perp &= -\frac{1}{2}S^{\alpha\beta}\Xi_{\alpha\beta},
		&
		q^\mu &= \Xi^\mu_{\ \alpha}S^{\alpha\beta}u_\beta,
		&
		r^\mu &= -\Xi^\mu_{\ \alpha}S^{\alpha\beta}b_\beta,
		\label{appA:symmetric-coeff-2}
		\\
		s_\perp^{\mu\nu} &= \Xi^{\mu\nu}_{\alpha\beta}S^{\alpha\beta}.
		\label{appA:symmetric-coeff-3}
	\end{align}
	By construction,
	\begin{equation}
		u_\mu q^\mu=b_\mu q^\mu=u_\mu r^\mu=b_\mu r^\mu=0,\qquad
		u_\mu s_\perp^{\mu\nu}=b_\mu s_\perp^{\mu\nu}=0,\qquad
		\Xi_{\mu\nu}s_\perp^{\mu\nu}=0 .
		\label{appA:symmetric-orthogonality}
	\end{equation}
	
	For an antisymmetric tensor $A^{\mu\nu}=-A^{\nu\mu}$, the corresponding split is
	\begin{equation}
		A^{\mu\nu}
		=
		2\mathcal A\,b^{[\mu}u^{\nu]}
		+2a^{[\mu}u^{\nu]}
		+2b^{[\mu}c^{\nu]}
		+a_\perp^{\mu\nu},
		\label{appA:antisymmetric-decomp}
	\end{equation}
	with
	\begin{align}
		\mathcal A &= -b_\mu u_\nu A^{\mu\nu},
		&
		a^\mu &= -\Xi^\mu_{\ \alpha}u_\beta A^{\beta\alpha},
		&
		c^\mu &= -\Xi^\mu_{\ \alpha}b_\beta A^{\beta\alpha},
		\label{appA:antisymmetric-coeff-1}
		\\
		a_\perp^{\mu\nu} &= \Xi^{\mu[\alpha}\Xi^{\beta]\nu}A_{\alpha\beta}.
		\label{appA:antisymmetric-coeff-2}
	\end{align}
	Here $a^\mu$, $c^\mu$, and both indices of $a_\perp^{\mu\nu}$ lie in the two-dimensional transverse subspace.
	
	Applying these decompositions to the hydrodynamic variables gives
	\begin{align}
		N^\mu
		&=
		\mathcal N u^\mu+n_b b^\mu+\nu_\perp^\mu,
		\label{appA:N-decomp}
		\\
		T^{\mu\nu}
		&=
		\mathcal E u^\mu u^\nu
		+2M u^{(\mu}b^{\nu)}
		+\mathcal P_\parallel b^\mu b^\nu
		-\mathcal P_\perp \Xi^{\mu\nu}
		+2h^{(\mu}u^{\nu)}
		+2f^{(\mu}b^{\nu)}
		+\pi_\perp^{\mu\nu},
		\label{appA:T-decomp}
		\\
		\tilde F^{\mu\nu}
		&=
		2\mathcal B\,b^{[\mu}u^{\nu]}
		+2g^{[\mu}u^{\nu]}
		+2b^{[\mu}\ell^{\nu]}
		+m_\perp^{\mu\nu}.
		\label{appA:F-decomp}
	\end{align}
	The scalar, vector, and tensor coefficients are defined by the projections
	\begin{align}
		\mathcal N &= u_\mu N^\mu,
		&
		n_b &= -b_\mu N^\mu,
		&
		\nu_\perp^\mu &= \Xi^\mu_{\ \nu}N^\nu,
		\label{appA:N-coeff}
		\\
		\mathcal E &= u_\mu u_\nu T^{\mu\nu},
		&
		M &= -u_\mu b_\nu T^{\mu\nu},
		&
		\mathcal P_\parallel &= b_\mu b_\nu T^{\mu\nu},
		\label{appA:T-coeff-1}
		\\
		\mathcal P_\perp &= -\frac{1}{2}\Xi_{\mu\nu}T^{\mu\nu},
		&
		h^\mu &= \Xi^\mu_{\ \alpha}T^{\alpha\beta}u_\beta,
		&
		f^\mu &= -\Xi^\mu_{\ \alpha}T^{\alpha\beta}b_\beta,
		\label{appA:T-coeff-2}
		\\
		\pi_\perp^{\mu\nu}
		&=\Xi^{\mu\nu}_{\alpha\beta}T^{\alpha\beta},
		\label{appA:T-coeff-3}
		\\
		\mathcal B &= -b_\mu u_\nu \tilde F^{\mu\nu},
		&
		g^\mu &= -\Xi^\mu_{\ \alpha}u_\beta\tilde F^{\beta\alpha},
		&
		\ell^\mu &= -\Xi^\mu_{\ \alpha}b_\beta\tilde F^{\beta\alpha},
		\label{appA:F-coeff-1}
		\\
		m_\perp^{\mu\nu}
		&=\Xi^{\mu[\alpha}\Xi^{\beta]\nu}\tilde F_{\alpha\beta}.
		\label{appA:F-coeff-2}
	\end{align}
	In the frame used in the body of the paper the variables $M$, $n_b$, $h^\mu$, and $g^\mu$ are removed by matching and frame choices, as explained in Appendix~\ref{appendixg}.
	
	The derivative operators adapted to the same split are
	\begin{equation}
		D\equiv u^\mu\partial_\mu,\qquad
		\tilde D\equiv -b^\mu\partial_\mu,\qquad
		\tilde\nabla_\mu\equiv \Xi_{\mu\nu}\partial^\nu,
		\label{appA:derivatives}
	\end{equation}
	so that
	\begin{equation}
		\partial_\mu=u_\mu D+b_\mu\tilde D+\tilde\nabla_\mu .
		\label{appA:gradient-decomp}
	\end{equation}
	The ordinary spatial gradient is therefore decomposed as
	\begin{equation}
		\nabla_\mu\equiv \Delta_{\mu\nu}\partial^\nu
		=b_\mu\tilde D+\tilde\nabla_\mu .
		\label{appA:spatial-gradient}
	\end{equation}
	
	The two scalar expansion rates used in the main text are
	\begin{equation}
		\theta_\perp\equiv \tilde\nabla_\mu u^\mu,\qquad
		\theta_\parallel\equiv b_\mu\tilde D u^\mu
		=-b^\mu b^\nu\partial_\mu u_\nu ,
		\label{appA:theta-def}
	\end{equation}
	and the usual expansion scalar satisfies
	\begin{equation}
		\theta\equiv \nabla_\mu u^\mu
		=\theta_\perp+\theta_\parallel .
		\label{appA:theta-split}
	\end{equation}
	The transverse shear and vorticity are
	\begin{align}
		\sigma_\perp^{\mu\nu}
		&\equiv
		\Xi^{\mu\nu}_{\alpha\beta}\partial^\alpha u^\beta
		=
		\tilde\nabla^{(\mu}u^{\nu)}
		-\frac{1}{2}\theta_\perp\Xi^{\mu\nu}
		+b_\alpha b^{(\mu}\tilde\nabla^{\nu)}u^\alpha,
		\label{appA:sigma-perp}
		\\
		\omega_\perp^{\mu\nu}
		&\equiv
		\Xi^\mu_{\ \alpha}\Xi^\nu_{\ \beta}\partial^{[\alpha}u^{\beta]}
		=
		\tilde\nabla^{[\mu}u^{\nu]}
		-b_\alpha b^{[\mu}\tilde\nabla^{\nu]}u^\alpha .
		\label{appA:omega-perp}
	\end{align}
	With these definitions the velocity-gradient decomposition becomes
	\begin{align}
		\partial_\mu u_\nu
		=
		u_\mu D u_\nu
		+b_\mu\tilde D u_\nu
		+\frac{1}{2}\theta_\perp\Xi_{\mu\nu}
		-b_\nu b_\alpha\tilde\nabla_\mu u^\alpha
		+\sigma_{\perp\mu\nu}
		+\omega_{\perp\mu\nu}.
		\label{appA:du-decomp}
	\end{align}
	
	For derivatives of the magnetic-field direction we define
	\begin{align}
		\vartheta_b &\equiv \tilde\nabla_\mu b^\mu,
		\label{appA:theta-b}
		\\
		\sigma_{b,\perp}^{\mu\nu}
		&\equiv
		\Xi^{\mu\nu}_{\alpha\beta}\partial^\alpha b^\beta
		=
		\tilde\nabla^{(\mu}b^{\nu)}
		-\frac{1}{2}\vartheta_b\Xi^{\mu\nu}
		-u_\alpha u^{(\mu}\tilde\nabla^{\nu)}b^\alpha,
		\label{appA:sigma-b}
		\\
		\omega_{b,\perp}^{\mu\nu}
		&\equiv
		\Xi^\mu_{\ \alpha}\Xi^\nu_{\ \beta}\partial^{[\alpha}b^{\beta]}
		=
		\tilde\nabla^{[\mu}b^{\nu]}
		+u_\alpha u^{[\mu}\tilde\nabla^{\nu]}b^\alpha .
		\label{appA:omega-b}
	\end{align}
	Then
	\begin{align}
		\partial_\mu b_\nu
		=
		u_\mu D b_\nu
		+b_\mu\tilde D b_\nu
		+\frac{1}{2}\vartheta_b\Xi_{\mu\nu}
		+u_\nu u_\alpha\tilde\nabla_\mu b^\alpha
		+\sigma_{b,\perp\mu\nu}
		+\omega_{b,\perp\mu\nu}.
		\label{appA:db-decomp}
	\end{align}
	Equations~\eqref{appA:du-decomp} and~\eqref{appA:db-decomp} are the kinematic ingredients used when reducing entropy production and second-order relaxation terms into scalar, vector, and tensor sectors.
	
	\section{Matching and frame choices in magnetized hydrodynamics}
	\label{appendixg}
	
	This appendix specifies how the local equilibrium fields and the dissipative variables are separated.  The purpose of the matching prescription is to prevent the nonequilibrium corrections from being counted twice as shifts of the thermodynamic variables.  The frame choice then fixes how transverse vector corrections are distributed among $T^{\mu\nu}$, $N^\mu$, and $\tilde F^{\mu\nu}$.
	
	For the non-anomalous parity-even sector considered in this work, the zeroth-order constitutive relations are taken to be
	\begin{align}
		T_{(0)}^{\mu\nu}
		&=
		\epsilon u^\mu u^\nu
		-p_\perp\Xi^{\mu\nu}
		+p_\parallel b^\mu b^\nu,
		\label{appB:ideal-T}
		\\
		N_{(0)}^\mu
		&=
		n u^\mu,
		\label{appB:ideal-N}
		\\
		\tilde F_{(0)}^{\mu\nu}
		&=
		2B\,b^{[\mu}u^{\nu]} .
		\label{appB:ideal-F}
	\end{align}
	Equilibrium terms such as $u^{(\mu}b^{\nu)}$ in $T^{\mu\nu}$ or $b^\mu$ in $N^\mu$ would require additional pseudoscalar equilibrium coefficients in the sector under consideration and are therefore absent.  We do not impose charge-conjugation symmetry at this stage; if it is imposed later, additional transport coefficients can be consistently set to zero.
	
	The scalar matching conditions define the local values of $\epsilon$, $n$, and $B$ directly from the full nonequilibrium fields:
	\begin{equation}
		\epsilon\equiv u_\mu u_\nu T^{\mu\nu},\qquad
		n\equiv u_\mu N^\mu,\qquad
		B\equiv -b_\mu u_\nu\tilde F^{\mu\nu}.
		\label{appB:matched-scalars}
	\end{equation}
	Equivalently, the first-order corrections satisfy
	\begin{equation}
		u_\mu u_\nu T_{(1)}^{\mu\nu}=0,\qquad
		u_\mu N_{(1)}^\mu=0,\qquad
		-b_\mu u_\nu\tilde F_{(1)}^{\mu\nu}=0 .
		\label{appB:matching-zero}
	\end{equation}
	If one starts from a pre-matched decomposition containing scalar shifts,
	\begin{align}
		T_{(1)}^{\mu\nu}\big|_{\rm scalar}
		&=
		\delta\epsilon\,u^\mu u^\nu
		-\delta p_\perp\,\Xi^{\mu\nu}
		+\delta p_\parallel\,b^\mu b^\nu,
		\label{appB:prematched-T}
		\\
		N_{(1)}^\mu\big|_{\rm scalar}
		&=
		\delta n\,u^\mu,
		\label{appB:prematched-N}
		\\
		\tilde F_{(1)}^{\mu\nu}\big|_{\rm scalar}
		&=
		2\delta B\,b^{[\mu}u^{\nu]},
		\label{appB:prematched-F}
	\end{align}
	then $\delta\epsilon$, $\delta n$, and $\delta B$ are absorbed into redefinitions of the local equilibrium variables:
	\begin{equation}
		\epsilon\rightarrow \epsilon+\delta\epsilon,\qquad
		n\rightarrow n+\delta n,\qquad
		B\rightarrow B+\delta B .
		\label{appB:scalar-redef}
	\end{equation}
	The equilibrium pressures $p_\perp(\epsilon,n,B)$ and $p_\parallel(\epsilon,n,B)$ are evaluated after this matching.  The residual scalar dissipative corrections are denoted by the anisotropic bulk pressures $\Pi_\perp$ and $\Pi_\parallel$, defined through
	\begin{equation}
		-\frac{1}{2}\Xi_{\mu\nu}T^{\mu\nu}=p_\perp+\Pi_\perp,\qquad
		b_\mu b_\nu T^{\mu\nu}=p_\parallel+\Pi_\parallel .
		\label{appB:bulk-definition}
	\end{equation}
	
	After scalar matching, but before fixing the transverse vector frame, the most general first-order corrections needed in this work can be written as
	\begin{align}
		T_{(1)}^{\mu\nu}
		&=
		-\Pi_\perp\Xi^{\mu\nu}
		+\Pi_\parallel b^\mu b^\nu
		+2h^{(\mu}u^{\nu)}
		+2f^{(\mu}b^{\nu)}
		+\pi_\perp^{\mu\nu},
		\label{appB:T1-general}
		\\
		N_{(1)}^\mu
		&=
		\nu_\perp^\mu,
		\label{appB:N1-general}
		\\
		\tilde F_{(1)}^{\mu\nu}
		&=
		2g^{[\mu}u^{\nu]}
		+2b^{[\mu}\ell^{\nu]}
		+m_\perp^{\mu\nu}.
		\label{appB:F1-general}
	\end{align}
	All vectors in Eq.~\eqref{appB:T1-general}--\eqref{appB:F1-general} are transverse to both $u^\mu$ and $b^\mu$, and $\pi_\perp^{\mu\nu}$ is symmetric, transverse, and traceless in the $\Xi^{\mu\nu}$ subspace.  The tensor $m_\perp^{\mu\nu}$ is antisymmetric and transverse in both indices.
	
	The vector frame freedom is the freedom to redefine $u^\mu$ and $b^\mu$ by first-order transverse pieces,
	\begin{equation}
		u^\mu\rightarrow u^{\prime\mu}=u^\mu-\delta u^\mu,\qquad
		b^\mu\rightarrow b^{\prime\mu}=b^\mu-\delta b^\mu,
		\label{appB:frame-shift}
	\end{equation}
	where
	\begin{equation}
		u_\mu\delta u^\mu=b_\mu\delta u^\mu
		=u_\mu\delta b^\mu=b_\mu\delta b^\mu=0 .
		\label{appB:transverse-shift}
	\end{equation}
	To first order in gradients, these transformations leave the normalization conditions $u^2=1$, $b^2=-1$, and $u\cdot b=0$ unchanged.  The scalar dissipative variables and the transverse tensors are invariant at this order, while the vector coefficients transform as
	\begin{align}
		h^{\prime\mu}
		&=
		h^\mu+w_\perp\,\delta u^\mu,
		\\
		w_\perp&\equiv \epsilon+p_\perp,
		\label{appB:h-shift}
		\\
		\nu_\perp^{\prime\mu}
		&=
		\nu_\perp^\mu+n\,\delta u^\mu,
		\label{appB:nu-shift}
		\\
		\ell^{\prime\mu}
		&=
		\ell^\mu+B\,\delta u^\mu,
		\label{appB:ell-shift}
		\\
		f^{\prime\mu}
		&=
		f^\mu-(p_\perp-p_\parallel)\delta b^\mu,
		\label{appB:f-shift}
		\\
		g^{\prime\mu}
		&=
		g^\mu+B\,\delta b^\mu .
		\label{appB:g-shift}
	\end{align}
	Thus one should not regard $h^\mu$, $\nu_\perp^\mu$, and $\ell^\mu$ as independently frame-invariant objects; they mix under a change of $u^\mu$.  Similarly, $f^\mu$ and $g^\mu$ mix under a change of the magnetic-field direction $b^\mu$.
	
	In the main text we use the energy--magnetic frame,
	\begin{equation}
		h^{\prime\mu}=0,\qquad g^{\prime\mu}=0 .
		\label{appB:main-frame-condition}
	\end{equation}
	It is obtained by choosing
	\begin{equation}
		\delta u^\mu=-\frac{h^\mu}{w_\perp},\qquad
		\delta b^\mu=-\frac{g^\mu}{B}.
		\label{appB:main-frame-shift}
	\end{equation}
	After this transformation, and after dropping primes for notational simplicity, the first-order corrections reduce to
	\begin{align}
		T_{(1)}^{\mu\nu}
		&=
		-\Pi_\perp\Xi^{\mu\nu}
		+\Pi_\parallel b^\mu b^\nu
		+2f^{(\mu}b^{\nu)}
		+\pi_\perp^{\mu\nu},
		\label{appB:T1-main-frame}
		\\
		N_{(1)}^\mu
		&=
		\nu_\perp^\mu,
		\label{appB:N1-main-frame}
		\\
		\tilde F_{(1)}^{\mu\nu}
		&=
		2b^{[\mu}\ell^{\nu]}
		+m_\perp^{\mu\nu}.
		\label{appB:F1-main-frame}
	\end{align}
	The vectors appearing in Eq.~\eqref{appB:T1-main-frame}--\eqref{appB:F1-main-frame} are the frame-transformed combinations
	\begin{align}
		f^\mu_{\rm main}
		&=
		f^\mu+\frac{p_\perp-p_\parallel}{B}g^\mu,
		\label{appB:f-main}
		\\
		\nu_{\perp,{\rm main}}^\mu
		&=
		\nu_\perp^\mu-\frac{n}{w_\perp}h^\mu,
		\label{appB:nu-main}
		\\
		\ell_{\rm main}^\mu
		&=
		\ell^\mu-\frac{B}{w_\perp}h^\mu .
		\label{appB:ell-main}
	\end{align}
	In the body of the paper we denote these transformed quantities simply by $f^\mu$, $\nu_\perp^\mu$, and $\ell^\mu$.
	
	Other frame choices are possible and lead to equivalent hydrodynamic theories.  For example, when $n\neq0$ one may eliminate the transverse particle diffusion current by taking
	\begin{equation}
		\delta u^\mu=-\frac{\nu_\perp^\mu}{n},\qquad
		\delta b^\mu=-\frac{g^\mu}{B},
		\label{appB:particle-frame}
	\end{equation}
	which gives $\nu_\perp^{\prime\mu}=0$ and $g^{\prime\mu}=0$.  In this particle frame,
	\begin{equation}
		h^{\prime\mu}=h^\mu-\frac{w_\perp}{n}\nu_\perp^\mu,\qquad
		\ell^{\prime\mu}=\ell^\mu-\frac{B}{n}\nu_\perp^\mu,\qquad
		f^{\prime\mu}=f^\mu+\frac{p_\perp-p_\parallel}{B}g^\mu .
		\label{appB:particle-frame-result}
	\end{equation}
	This frame is not convenient for charge-conjugation symmetric backgrounds with $n=0$.
	
	One may also eliminate the magnetic-induction vector by choosing, for $B\neq0$,
	\begin{equation}
		\delta u^\mu=-\frac{\ell^\mu}{B},\qquad
		\delta b^\mu=-\frac{g^\mu}{B}.
		\label{appB:magnetic-frame}
	\end{equation}
	Then $\ell^{\prime\mu}=0$ and $g^{\prime\mu}=0$, while
	\begin{equation}
		h^{\prime\mu}=h^\mu-\frac{w_\perp}{B}\ell^\mu,\qquad
		\nu_\perp^{\prime\mu}=\nu_\perp^\mu-\frac{n}{B}\ell^\mu,\qquad
		f^{\prime\mu}=f^\mu+\frac{p_\perp-p_\parallel}{B}g^\mu .
		\label{appB:magnetic-frame-result}
	\end{equation}
	Finally, if $p_\perp\neq p_\parallel$, one can use the $b^\mu$-frame freedom to remove $f^\mu$ by taking
	\begin{equation}
		\delta b^\mu=\frac{f^\mu}{p_\perp-p_\parallel}.
		\label{appB:f-frame}
	\end{equation}
	This choice produces
	\begin{equation}
		f^{\prime\mu}=0,\qquad
		g^{\prime\mu}=g^\mu+\frac{B}{p_\perp-p_\parallel}f^\mu .
		\label{appB:f-frame-result}
	\end{equation}
	Hence eliminating one vector structure generally moves the same physical information into another sector.  The frame used in Eq.~\eqref{appB:T1-main-frame}--\eqref{appB:F1-main-frame} is chosen because it keeps the energy flow and the transverse magnetic field correction absent, while retaining the particle-diffusion and induction channels needed for the linear spectrum analyzed in the main text.

	\section{Explicit form of the second-order entropy current}\label{partial Q}
	Here we repeat the Eq.~\eqref{53}:
	\begin{eqnarray}
		\begin{aligned}
			\partial_\mu Q^\mu=\Pi_{\perp}\mathcal{A}+\Pi_{\parallel}\mathcal{B}+f^{\lambda}\mathcal{C}_{\lambda}+\nu_{\perp}^{\lambda}\mathcal{D}_{\lambda}+\ell^{\lambda}\mathcal{E}_{\lambda}+\pi_{\perp}^{\lambda\nu}\mathcal{F}_{\lambda\nu}+m_{\perp}^{\lambda\nu}\mathcal{G}_{\lambda\nu}.
		\end{aligned}
		\label{pQ}
	\end{eqnarray}
	In the above equations, scalars $\mathcal{A}$ and $\mathcal{B}$, vectors $\mathcal{C}_\lambda$, $\mathcal{D}_\lambda$, and $\mathcal{E}_\lambda$, and tensors $\mathcal{F}_{\lambda\nu}$ and $\mathcal{G}_{\lambda\nu}$ are defined as 
	
	\begin{align}
		\mathcal{A}	=&a_{1}\Pi_{\perp}\theta_{\perp}+a_{1}\Pi_{\perp}\theta_{\parallel}+\Pi_{\perp}Da_{1}+2a_{1}D\Pi_{\perp}+l_{\Pi_{\perp}f}f^{\mu}\tilde{\nabla}_{\mu}b_{1}-\tilde{l}_{\Pi_{\perp}f}b_{1}f^{\mu}Du_{\mu}\nonumber\\
		&-\left(1-l_{f\Pi_{\perp}}\right)b_{1}f^{\mu}\tilde{D}b_{\mu}+b_{1}\tilde{\nabla}_{\mu}f^{\mu}+l_{\Pi_{\perp}\nu}\nu_{\perp}^{\mu}\tilde{\nabla}_{\mu}b_{2}-\tilde{l}_{\Pi_{\perp}\nu}b_{2}\nu_{\perp}^{\mu}Du_{\mu}+b_{2}\tilde{\nabla}_{\mu}\nu_{\perp}^{\mu}\nonumber\\
		&-\left(1-l_{\nu\Pi_{\perp}}\right)b_{2}\nu_{\perp}^{\mu}\tilde{D}b_{\mu}+l_{\Pi_{\perp}\ell}\ell^{\mu}\tilde{\nabla}_{\mu}b_{3}-\tilde{l}_{\Pi_{\perp}\ell}b_{3}\ell^{\mu}Du_{\mu}-\left(1-l_{\ell\Pi_{\perp}}\right)b_{3}\ell^{\mu}\tilde{D}b_{\mu}\nonumber\\
		&+b_{3}\tilde{\nabla}_{\mu}\ell^{\mu},\label{55}
	\end{align}
	\begin{align}
		\mathcal{B}	=&a_{2}\Pi_{\parallel}\theta_{\perp}+a_{2}\Pi_{\parallel}\theta_{\parallel}+\Pi_{\parallel}Da_{2}+2a_{2}D\Pi_{\parallel}+l_{\Pi_{\parallel}f}f^{\mu}\tilde{\nabla}_{\mu}b_{4}-\tilde{l}_{\Pi_{\parallel}f}b_{4}f^{\mu}Du_{\mu}\nonumber\\
		&-\left(1-l_{f\Pi_{\parallel}}\right)b_{4}f^{\mu}\tilde{D}b_{\mu}+b_{4}\tilde{\nabla}_{\mu}f^{\mu}+l_{\Pi_{\parallel}\nu}\nu_{\perp}^{\mu}\tilde{\nabla}_{\mu}b_{5}-\tilde{l}_{\Pi_{\parallel}\nu}b_{5}\nu_{\perp}^{\mu}Du_{\mu}\nonumber\\
		&-\left(1-l_{\nu\Pi_{\parallel}}\right)b_{5}\nu_{\perp}^{\mu}\tilde{D}b_{\mu}+b_{5}\tilde{\nabla}_{\mu}\nu_{\perp}^{\mu}+l_{\Pi_{\parallel}\ell}\ell^{\mu}\tilde{\nabla}_{\mu}b_{6}-\tilde{l}_{\Pi_{\parallel}\ell}b_{6}\ell^{\mu}Du_{\mu}\nonumber\\
		&-\left(1-l_{\ell\Pi_{\parallel}}\right)b_{6}\ell^{\mu}\tilde{D}b_{\mu}+b_{6}\tilde{\nabla}_{\mu}\ell^{\mu},\label{56}
	\end{align}
	\begin{align}
		\mathcal{C}_{\lambda}	=&a_{3}f_{\lambda}\theta_{\perp}+a_{3}f_{\lambda}\theta_{\parallel}+f_{\lambda}Da_{3}+2a_{3}\Xi_{\lambda}^{\nu}Df_{\nu}+\left(1-l_{\Pi_{\perp}f}\right)\Pi_{\perp}\tilde{\nabla}_{\lambda}b_{1}+b_{1}\tilde{\nabla}_{\lambda}\Pi_{\perp}\nonumber\\
		&-\left(1-\tilde{l}_{\Pi_{\perp}f}\right)b_{1}\Pi_{\perp}\Xi_{\lambda}^{\nu}Du_{\nu}-l_{f\Pi_{\perp}}b_{1}\Pi_{\perp}\Xi_{\lambda}^{\nu}\tilde{D}b_{\nu}+\left(1-l_{\Pi_{\parallel}f}\right)\Pi_{\parallel}\tilde{\nabla}_{\lambda}b_{4}+b_{4}\tilde{\nabla}_{\lambda}\Pi_{\parallel}\nonumber\\
		&-\left(1-\tilde{l}_{\Pi_{\parallel}f}\right)b_{4}\Pi_{\parallel}\Xi_{\lambda}^{\nu}Du_{\nu}-l_{f\Pi_{\parallel}}b_{4}\Pi_{\parallel}\Xi_{\lambda}^{\nu}\tilde{D}b_{\nu}+l_{f\pi}\pi_{\perp\mu\lambda}\tilde{\nabla}^{\mu}c_{1}-\tilde{l}_{f\pi}c_{1}\pi_{\perp\mu\lambda}Du^{\mu}\nonumber\\
		&-\left(1-l_{\pi f}\right)c_{1}\pi_{\perp\mu\lambda}\tilde{D}b^{\mu}+c_{1}\Xi_{\lambda}^{\nu}\tilde{\nabla}^{\mu}\pi_{\perp\mu\nu}+l_{fm}m_{\perp\mu\lambda}\tilde{\nabla}^{\mu}c_{4}-\tilde{l}_{fm}c_{4}m_{\perp\mu\lambda}Du^{\mu}\nonumber\\
		&-\left(1-l_{mf}\right)c_{4}m_{\perp\mu\lambda}\tilde{D}b^{\mu}+c_{4}\Xi_{\lambda}^{\nu}\tilde{\nabla}^{\mu}m_{\perp\mu\nu},\label{57}
	\end{align}
	\begin{align}
		\mathcal{D}_{\lambda}	=&a_{4}\nu_{\perp\lambda}\theta_{\perp}+a_{4}\nu_{\perp\lambda}\theta_{\parallel}+\nu_{\perp\lambda}Da_{4}+2a_{4}\Xi_{\lambda}^{\nu}D\nu_{\perp\nu}+\left(1-l_{\Pi_{\perp}\nu}\right)\Pi_{\perp}\tilde{\nabla}_{\lambda}b_{2}+b_{2}\tilde{\nabla}_{\lambda}\Pi_{\perp}\nonumber\\
		&-\left(1-\tilde{l}_{\Pi_{\perp}\nu}\right)b_{2}\Pi_{\perp}\Xi_{\lambda}^{\nu}Du_{\nu}-l_{\nu\Pi_{\perp}}b_{2}\Pi_{\perp}\Xi_{\lambda}^{\nu}\tilde{D}b_{\nu}+\left(1-l_{\Pi_{\parallel}\nu}\right)\Pi_{\parallel}\tilde{\nabla}_{\lambda}b_{5}+b_{5}\tilde{\nabla}_{\lambda}\Pi_{\parallel}\nonumber\\
		&-\left(1-\tilde{l}_{\Pi_{\parallel}\nu}\right)b_{5}\Pi_{\parallel}\Xi_{\lambda}^{\nu}Du_{\nu}-l_{\nu\Pi_{\parallel}}b_{5}\Pi_{\parallel}\Xi_{\lambda}^{\nu}\tilde{D}b_{\nu}+l_{\nu\pi}\pi_{\perp\mu\lambda}\tilde{\nabla}^{\mu}c_{2}-\tilde{l}_{\nu\pi}c_{2}\pi_{\perp\mu\lambda}Du^{\mu}\nonumber\\
		&-\left(1-l_{\pi\nu}\right)c_{2}\pi_{\perp\mu\lambda}\tilde{D}b^{\mu}+c_{2}\Xi_{\lambda}^{\nu}\tilde{\nabla}^{\mu}\pi_{\perp\mu\nu}+l_{\nu m}m_{\perp\mu\lambda}\tilde{\nabla}^{\mu}c_{5}-\tilde{l}_{\nu m}c_{5}m_{\perp\mu\lambda}Du^{\mu}\nonumber\\
		&-\left(1-l_{m\nu}\right)c_{5}m_{\perp\mu\lambda}\tilde{D}b^{\mu}+c_{5}\Xi_{\lambda}^{\nu}\tilde{\nabla}^{\mu}m_{\perp\mu\nu},\label{58}
	\end{align}
	\begin{align}
		\mathcal{E}_{\lambda}	=&a_{5}\ell_{\lambda}\theta_{\perp}+a_{5}\ell_{\lambda}\theta_{\parallel}+\ell_{\lambda}Da_{5}+2a_{5}\Xi_{\lambda}^{\nu}D\ell_{\nu}+\left(1-l_{\Pi_{\perp}\ell}\right)\Pi_{\perp}\tilde{\nabla}_{\lambda}b_{3}+b_{3}\tilde{\nabla}_{\lambda}\Pi_{\perp}\nonumber\\
		&-\left(1-\tilde{l}_{\Pi_{\perp}\ell}\right)b_{3}\Pi_{\perp}\Xi_{\lambda}^{\nu}Du_{\nu}-l_{\ell\Pi_{\perp}}b_{3}\Pi_{\perp}\Xi_{\lambda}^{\nu}\tilde{D}b_{\nu}
		+\left(1-l_{\Pi_{\parallel}\ell}\right)\Pi_{\parallel}\tilde{\nabla}_{\lambda}b_{6}+b_{6}\tilde{\nabla}_{\lambda}\Pi_{\parallel}\nonumber\\
		&-\left(1-\tilde{l}_{\Pi_{\parallel}\ell}\right)b_{6}\Pi_{\parallel}\Xi_{\lambda}^{\nu}Du_{\nu}-l_{\ell\Pi_{\parallel}}b_{6}\Pi_{\parallel}\Xi_{\lambda}^{\nu}\tilde{D}b_{\nu}+l_{\ell\pi}\pi_{\perp\mu\lambda}\tilde{\nabla}^{\mu}c_{3}-\tilde{l}_{\ell\pi}c_{3}\pi_{\perp\mu\lambda}Du^{\mu}
		\nonumber\\
		&-\left(1-l_{\pi\ell}\right)c_{3}\pi_{\perp\mu\lambda}\tilde{D}b^{\mu}+c_{3}\Xi_{\lambda}^{\nu}\tilde{\nabla}^{\mu}\pi_{\perp\mu\nu}+l_{\ell m}m_{\perp\mu\lambda}\tilde{\nabla}^{\mu}c_{6}-\tilde{l}_{\ell m}c_{6}m_{\perp\mu\lambda}Du^{\mu}\nonumber\\
		&-\left(1-l_{m\ell}\right)c_{6}m_{\perp\mu\lambda}\tilde{D}b^{\mu}+c_{6}\Xi_{\lambda}^{\nu}\tilde{\nabla}^{\mu}m_{\perp\mu\nu},\label{59}
	\end{align}
	\begin{align}
		\mathcal{F}_{\lambda\nu}	=&a_{6}\pi_{\perp\lambda\nu}\theta+\pi_{\perp\lambda\nu}Da_{6}+2a_{6}\Xi_{\lambda\nu}^{\rho\sigma}D\pi_{\perp\rho\sigma}+\left(1-l_{f\pi}\right)f_{(\nu}\tilde{\nabla}_{\lambda)}c_{1}-\left(1-\tilde{l}_{f\pi}\right)c_{1}\Xi_{\lambda\nu}^{\rho\sigma}f_{\sigma}Du_{\rho}\nonumber\\
		&-l_{\pi f}c_{1}\Xi_{\lambda\nu}^{\rho\sigma}f_{\sigma}\tilde{D}b_{\rho}+c_{1}\Xi_{\lambda\nu}^{\rho\sigma}\tilde{\nabla}_{\rho}f_{\sigma}+\left(1-l_{\nu\pi}\right)\nu_{\perp(\nu}\tilde{\nabla}_{\lambda)}c_{2}-\left(1-\tilde{l}_{\nu\pi}\right)c_{2}\Xi_{\lambda\nu}^{\rho\sigma}\nu_{\perp\sigma}Du_{\rho}\nonumber\\
		&-l_{\pi\nu}c_{2}\Xi_{\lambda\nu}^{\rho\sigma}\nu_{\perp\sigma}\tilde{D}b_{\rho}+c_{2}\Xi_{\lambda\nu}^{\rho\sigma}\tilde{\nabla}_{\rho}\nu_{\perp\sigma}+\left(1-l_{\ell\pi}\right)\ell_{(\nu}\tilde{\nabla}_{\lambda)}c_{3}-\left(1-\tilde{l}_{\ell\pi}\right)c_{3}\Xi_{\lambda\nu}^{\rho\sigma}\ell_{\sigma}Du_{\rho}\nonumber\\
		&-l_{\pi\ell}c_{3}\Xi_{\lambda\nu}^{\rho\sigma}\ell_{\sigma}\tilde{D}b_{\rho}+c_{3}\Xi_{\lambda\nu}^{\rho\sigma}\tilde{\nabla}_{\rho}\ell_{\sigma},\label{60}
	\end{align}
	\begin{align}
		\mathcal{G}_{\lambda\nu}	=&a_{7}m_{\perp\lambda\nu}\theta+m_{\perp\lambda\nu}Da_{7}+2a_{7}\Xi_{\lambda}^{[\rho}\Xi_{\nu}^{\sigma]}Dm_{\perp\rho\sigma}+\left(1-l_{fm}\right)f_{[\nu}\tilde{\nabla}_{\lambda]}c_{4}\nonumber\\
		&-\left(1-\tilde{l}_{fm}\right)c_{4}\Xi_{\lambda}^{[\rho}\Xi_{\nu}^{\sigma]}f_{\sigma}Du_{\rho}-l_{mf}c_{4}\Xi_{\lambda}^{[\rho}\Xi_{\nu}^{\sigma]}f_{\sigma}\tilde{D}b_{\rho}
		\nonumber\\
		&+c_{4}\Xi_{\lambda}^{[\rho}\Xi_{\nu}^{\sigma]}\tilde{\nabla}_{\rho}f_{\sigma}+\left(1-l_{\nu m}\right)\nu_{\perp[\nu}\tilde{\nabla}_{\lambda]}c_{5}-\left(1-\tilde{l}_{\nu m}\right)c_{5}\Xi_{\lambda}^{[\rho}\Xi_{\nu}^{\sigma]}\nu_{\perp\sigma}Du_{\rho}\nonumber\\
		&-l_{m\nu}c_{5}\Xi_{\lambda}^{[\rho}\Xi_{\nu}^{\sigma]}\nu_{\perp\sigma}\tilde{D}b_{\rho}+c_{5}\Xi_{\lambda}^{[\rho}\Xi_{\nu}^{\sigma]}\tilde{\nabla}_{\rho}\nu_{\perp\sigma}+\left(1-l_{\ell m}\right)\ell_{[\nu}\tilde{\nabla}_{\lambda]}c_{6}
		\nonumber\\
		&-\left(1-\tilde{l}_{\ell m}\right)c_{6}\Xi_{\lambda}^{[\rho}\Xi_{\nu}^{\sigma]}\ell_{\sigma}Du_{\rho}-l_{m\ell}c_{6}\Xi_{\lambda}^{[\rho}\Xi_{\nu}^{\sigma]}\ell_{\sigma}\tilde{D}b_{\rho}+c_{6}\Xi_{\lambda}^{[\rho}\Xi_{\nu}^{\sigma]}\tilde{\nabla}_{\rho}\ell_{\sigma}.\label{q61} 
	\end{align}
	
	\section{Explicit forms of matrices}\label{APPENDIX 3}
	In this appendix, we list the explicit forms of the matrices $A_{10\times10}$, $B_{6\times6}$, and $C_{3\times3}$ in Eq.\eqref{linear eq}. Since the full matrix is too wide to display compactly, we list only its nonzero elements. The non-zero matrix elements of $A_{10\times10}$ are displayed as follows:
	\begin{align}
		A_{11}   &= -\mathrm{i}\omega, 
		& A_{12}   &= -\mathrm{i}hk\sin\theta, 
		& A_{13}   &= -\mathrm{i}h(1-v_A^2)k\cos\theta, \nonumber\\
		A_{15}   &= \mathrm{i}\omega \frac{B}{\mu_m}, 
		& A_{21}   &= -\mathrm{i}c_s^2 k\cos\theta, 
		& A_{23}   &= -\mathrm{i}\omega h(1-v_A^2), \nonumber\\
		{}        & {}
		& A_{26}   &= -\mathrm{i}k\cos\theta, 
		& A_{28}   &= \mathrm{i}k\sin\theta, \nonumber\\
		A_{31}   &= \mathrm{i}c_s^2 k\sin\theta, 
		& A_{32}   &= \mathrm{i}\omega h, 
		& A_{34}   &= \mathrm{i}\frac{B}{\mu_m} k\cos\theta, \nonumber\\
		A_{35}   &= -\mathrm{i}\frac{B}{\mu_m} k\sin\theta, 
		& A_{37}   &= \mathrm{i}k\sin\theta, 
		& A_{38}   &= -\mathrm{i}k\cos\theta, \nonumber\\
		A_{39}   &= \mathrm{i}k\sin\theta, 
		& A_{42}   &= -\mathrm{i}Bk\sin\theta, 
		& A_{45}   &= \mathrm{i}\omega, \nonumber\\
		A_{4,10} &= -\mathrm{i}k\sin\theta, 
		& A_{52}   &= -\mathrm{i}Bk\cos\theta, 
		& A_{54}   &= -\mathrm{i}\omega, \nonumber\\
		A_{5,10} &= -\mathrm{i}k\cos\theta, 
		& A_{62}   &= -\mathrm{i}\zeta_{\times}^{\prime}k\sin\theta, 
		& A_{63}   &= -\mathrm{i}\zeta_{\parallel}k\cos\theta, \nonumber\\
		A_{66}   &= 1-\mathrm{i}\omega\tau_{\Pi_{\parallel}}, 
		& A_{72}   &= -\mathrm{i}\zeta_{\perp}k\sin\theta, 
		& A_{73}   &= -\mathrm{i}\zeta_{\times}k\cos\theta, \nonumber\\
		A_{77}   &= 1-\mathrm{i}\omega\tau_{\Pi_{\perp}}, 
		& A_{82}   &= \mathrm{i}\eta_f k\cos\theta, 
		& A_{83}   &= \mathrm{i}\eta_f k\sin\theta, \nonumber\\
		A_{88}   &= 1-\mathrm{i}\omega\tau_f, 
		& A_{92}   &= -\mathrm{i}\eta_{\perp}k\sin\theta, 
		& A_{99}   &= 1-\mathrm{i}\omega\tau_{\pi}, \nonumber\\
		A_{10,4} &= -\mathrm{i}\frac{\rho_{\parallel}}{\mu_m}k\cos\theta, 
		& A_{10,5} &= \mathrm{i}\frac{\rho_{\parallel}}{\mu_m}k\sin\theta, 
		& A_{10,10}&= 1-\mathrm{i}\omega\tau_{\ell}.
	\end{align}
	
	The nonzero matrix elements of $B_{6\times6}$ are displayed as follows:
	\begin{align}
		B_{11} &= \mathrm{i}\omega h, 
		& B_{12} &= \mathrm{i}\frac{B}{\mu_m}k\cos\theta, 
		& B_{13} &= -\mathrm{i}k\cos\theta, \nonumber\\
		B_{14} &= \mathrm{i}k\sin\theta, 
		& B_{21} &= -\mathrm{i}Bk\cos\theta, 
		& B_{22} &= -\mathrm{i}\omega, \nonumber\\
		B_{25} &= \mathrm{i}k\sin\theta, 
		& B_{26} &= -\mathrm{i}k\cos\theta, 
		& B_{31} &= \mathrm{i}\eta_f k\cos\theta, \nonumber\\
		B_{33} &= 1-\mathrm{i}\omega\tau_f, 
		& B_{41} &= -\mathrm{i}\eta_{\perp}k\sin\theta, 
		& B_{44} &= 1-\mathrm{i}\omega\tau_{\pi}, \nonumber\\
		B_{52} &= \mathrm{i}\frac{\rho_{\perp}}{\mu_m}k\sin\theta, 
		& B_{55} &= 1-\mathrm{i}\omega\tau_m, 
		& B_{62} &= -\mathrm{i}\frac{\rho_{\parallel}}{\mu_m}k\cos\theta, \nonumber\\
		B_{66} &= 1-\mathrm{i}\omega\tau_{\ell}.
	\end{align}
	
	The matrix $C_{3\times3}$ is given by
	
	\be
	C=\left(\begin{array}{ccc}
		-\mathrm{i}\omega  & -\mathrm{i}k\sin\theta & 0\\
		-\mathrm{i}\dfrac{\kappa}{\bar n}k\sin\theta  & 1-\mathrm{i}\omega\tau_{\nu} & 0\\
		0  & 0 & 1-\mathrm{i}\omega\tau_{\nu}
	\end{array}\right).
	\ee
	
	The coupling matrices $M_{AC}$, $M_{BC}$, $M_{CA}$, and $M_{CB}$ are given by
	\begin{align}
		M_{AC,21} &= -\mathrm{i}\left(\frac{\partial p}{\partial n}\right)k\cos\theta, 
		& M_{AC,31} &= \mathrm{i}\left(\frac{\partial p}{\partial n}\right)k\sin\theta, 
		& M_{AC,10,1} &= \mathrm{i}\frac{\kappa_{\times}^{\prime}}{\bar n}k\sin\theta. \nonumber\\
		M_{CA,12} &= -\mathrm{i}nk\sin\theta, 
		& M_{CA,13} &= -\mathrm{i}nk\cos\theta, 
		& M_{CA,24} &= \mathrm{i}\frac{\kappa_{\times}}{\mu_m}k\cos\theta, \nonumber\\
		M_{CA,25} &= -\mathrm{i}\frac{\kappa_{\times}}{\mu_m}k\sin\theta.&M_{CB,32} &= \mathrm{i}\frac{\kappa_{\times}}{\mu_m}k\cos\theta,&M_{BC}&=0_{6\times3},
	\end{align}
	where the coupling term $M_{BC}$ is zero since all of its elements are proportional to $k_y$ and the wave vector has been restricted to the $x-z$ plane.

	\section{Large-$k$ expansion solutions}\label{Large k expansion solution}
	In this appendix, we present the explicit coefficients in Eqs.\eqref{omega^a_-1 equation}, \eqref{omega^a_0 equation}, \eqref{omegab-1 equation} and \eqref{omega0 equation} and analyze the necessary conditions for their solutions. 
	
		{{The coefficients appearing in the large-$k$ expansion of the
	magneto-sonic sector can be recast as
	\begin{eqnarray}
		\mathbf{a}_A
		&=&
		\mathcal{N}_A
		\Big[
		\mathcal{A}_0
		-4\eta_f\mathcal{A}_2\cos 2\theta
		+\mathcal{A}_4\cos 4\theta
		\Big].
		\nonumber\\
		\mathbf{b}_A
		&=&
		-\frac{\cos\theta}{B^2}
		\Big[
		\mathcal{B}_4\cos^4\theta
		+\mathcal{B}_2\cos^2\theta\sin^2\theta
		+\mathcal{B}_0\sin^4\theta
		\Big],
		\nonumber\\
		\mathbf{c}_A
		&=&
		\frac{h\cos\theta}{2}
		\Big[
		\mathcal{C}_0-\mathcal{C}_2\cos 2\theta
		\Big],
		\nonumber\\
		\mathbf{d}_A
		&=&
		-h^2(1-v_A^2)\tau_\ell\tau_{\Pi_\parallel}\tau_f\tau_{\Pi_\perp}\tau_\pi\cos\theta .
	\end{eqnarray}
	
	For $\mathbf{a}_A$, we define
	\begin{eqnarray}
		\mathcal{A}_0
		&\equiv&
		\eta_\perp\tau_{\Pi_\perp}
		\Big(
		3\eta_f\tau_{\Pi_\parallel}+\mathcal{P}\tau_f
		\Big)
		+\tau_\pi
		\Bigg\{
		\eta_f\tau_{\Pi_\parallel}(3\zeta_\perp-\zeta_\times)
		+\tau_f
		\Big(
		\zeta_\perp\mathcal{P}
		-\zeta_\times^2
		-c_s^2h\zeta_\times\tau_{\Pi_\parallel}
		\Big)
		\nonumber\\
		&&+\eta_f\tau_{\Pi_\perp}
		\Big[
		3\zeta_\parallel-\zeta_\times
		+2c_s^2h(2-v_A^2)\tau_{\Pi_\parallel}
		\Big]
		+c_s^2h\tau_f\tau_{\Pi_\perp}
		\Big[
		\zeta_\parallel-(1-v_A^2)\zeta_\times
		\Big]
		\Bigg\},
		\nonumber\\
		\mathcal{A}_2
		&\equiv&
		\eta_\perp\tau_{\Pi_\parallel}\tau_{\Pi_\perp}
		+
		\Big(
		\zeta_\perp\tau_{\Pi_\parallel}-\zeta_\parallel\tau_{\Pi_\perp}
		+c_s^2h v_A^2\tau_{\Pi_\parallel}\tau_{\Pi_\perp}
		\Big)\tau_\pi,
		\nonumber\\
		\mathcal{A}_4
		&\equiv&
		\eta_\perp\tau_{\Pi_\perp}
		\Big(
		\eta_f\tau_{\Pi_\parallel}-\mathcal{P}\tau_f
		\Big)
		+\tau_\pi
		\Bigg\{
		\zeta_\perp
		\Big(
		\eta_f\tau_{\Pi_\parallel}-\mathcal{P}\tau_f
		\Big)
		+\zeta_\times^2\tau_f
		\nonumber\\
		&&
		+2c_s^2h(2-v_A^2)\eta_f\tau_{\Pi_\parallel}\tau_{\Pi_\perp}
		+\zeta_\parallel(\eta_f-c_s^2h\tau_f)\tau_{\Pi_\perp}
		+\zeta_\times\eta_f(\tau_{\Pi_\parallel}+\tau_{\Pi_\perp})
		\nonumber\\
		&&
		+c_s^2h\zeta_\times\tau_f
		\Big[
		\tau_{\Pi_\parallel}+(1-v_A^2)\tau_{\Pi_\perp}
		\Big]
		\Bigg\}.
	\end{eqnarray}
	
	For $\mathbf{b}_A$, the corresponding quantities are
	\begin{eqnarray}
		\mathcal{B}_4
		&\equiv&
		\Bigg[
		B^2\tau_\ell\mathcal{P}
		\Big(
		\eta_f+h v_A^2\tau_f
		\Big)
		\nonumber\\
		&&\quad
		+h^2v_A^2\rho_\parallel
		\Big(
		(1-v_A^2)\eta_f\tau_{\Pi_\parallel}+\mathcal{P}\tau_f
		\Big)
		\Bigg]\tau_{\Pi_\perp}\tau_\pi,
		\nonumber\\
		\mathcal{B}_2
		&\equiv&
		h^2v_A^2\rho_\parallel
		\Bigg\{
		(1-v_A^2)\eta_\perp\tau_{\Pi_\parallel}\tau_f\tau_{\Pi_\perp}
		\nonumber\\
		&&\quad
		+
		\Big[
		(1-v_A^2)\zeta_\perp\tau_{\Pi_\parallel}\tau_f
		+(2-v_A^2)\eta_f\tau_{\Pi_\parallel}\tau_{\Pi_\perp}
		\nonumber\\
		&&\quad
		+\big(
		\zeta_\parallel+2c_s^2h(1-v_A^2)\tau_{\Pi_\parallel}
		\big)\tau_f\tau_{\Pi_\perp}
		\Big]\tau_\pi
		\Bigg\}
		\nonumber\\
		&&
		+B^2\tau_\ell
		\Bigg\{
		\eta_\perp\mathcal{P}\tau_f\tau_{\Pi_\perp}
		+
		\Bigg[
		h v_A^2\eta_f\tau_{\Pi_\parallel}\tau_{\Pi_\perp}
		-\zeta_\times\eta_f(\tau_{\Pi_\parallel}+\tau_{\Pi_\perp})
		-\zeta_\times^2\tau_f
		+\zeta_\parallel\tau_f\mathcal{U}
		\nonumber\\
		&&
		+c_s^2h(1-v_A^2)
		\Big(
		\zeta_\perp\tau_{\Pi_\parallel}\tau_f
		+\tau_{\Pi_\parallel}(-2\eta_f+h v_A^2\tau_f)\tau_{\Pi_\perp}
		-\zeta_\times\tau_f(\tau_{\Pi_\parallel}+\tau_{\Pi_\perp})
		\Big)
		\Bigg]\tau_\pi
		\Bigg\},
		\nonumber\\
		\mathcal{B}_0
		&\equiv&
		\tau_{\Pi_\parallel}
		\Bigg\{
		h^2v_A^2\rho_\parallel
		\Big[
		(1-v_A^2)\eta_\perp\tau_f\tau_{\Pi_\perp}
		+(1-v_A^2)\zeta_\perp\tau_f\tau_\pi
		\nonumber\\
		&&\quad
		+\big(
		\eta_f+c_s^2h(1-v_A^2)\tau_f
		\big)\tau_{\Pi_\perp}\tau_\pi
		\Big]
		\nonumber\\
		&&\quad
		+B^2\eta_f\tau_\ell
		\Big[
		\eta_\perp\tau_{\Pi_\perp}+\mathcal{U}\tau_\pi
		\Big]
		\Bigg\}.
	\end{eqnarray}
	
	For $\mathbf{c}_A$, we introduce

	\begin{eqnarray}
		\mathcal{C}_0
		&\equiv&
		\tau_\ell
		\Bigg\{
		(1-v_A^2)\eta_\perp\tau_{\Pi_\parallel}\tau_f\tau_{\Pi_\perp}
		\nonumber\\
		&&\quad
		+
		\Big[
		(1-v_A^2)\zeta_\perp\tau_{\Pi_\parallel}\tau_f
		+(2-v_A^2)\eta_f\tau_{\Pi_\parallel}\tau_{\Pi_\perp}
		\nonumber\\
		&&\quad
		+\big(
		\zeta_\parallel+h(1-v_A^2)\left[2v_A^2+c_s^2(2-v_A^2)\right]\tau_{\Pi_\parallel}
		\big)\tau_f\tau_{\Pi_\perp}
		\Big]\tau_\pi
		\Bigg\}
		\nonumber\\
		&&
		+\frac{2h^2v_A^2(1-v_A^2)\rho_\parallel}{B^2}
		\tau_{\Pi_\parallel}\tau_f\tau_{\Pi_\perp}\tau_\pi,
		\nonumber\\
		\mathcal{C}_2
		&\equiv&
		\tau_\ell\tau_f
		\Big[
		(1-v_A^2)\eta_\perp\tau_{\Pi_\parallel}\tau_{\Pi_\perp}
		+
		\Big[(1-v_A^2)\zeta_\perp\tau_{\Pi_\parallel}-\zeta_\parallel\tau_{\Pi_\perp}
		\nonumber\\
		&&\quad
		-c_s^2hv_A^2(1-v_A^2)\tau_{\Pi_\parallel}\tau_{\Pi_\perp}\Big]\tau_\pi
		\Big]
		\nonumber\\
		&&
		+v_A^2\eta_f\tau_\ell\tau_{\Pi_\parallel}\tau_{\Pi_\perp}\tau_\pi.
	\end{eqnarray}}
	
	The additional abbreviations are

	\begin{align}
		\mathcal{P}&
		\equiv
		\zeta_\parallel+c_s^2h(1-v_A^2)\tau_{\Pi_\parallel},\qquad
		\mathcal{U}
		\equiv
		\zeta_\perp+h\left[v_A^2+c_s^2(1-v_A^2)\right]\tau_{\Pi_\perp},\nonumber\\
		\mathcal{N}_A&
		\equiv
		\frac{h v_A^2\rho_\parallel\cos\theta}{8B^2}.
	\end{align}
	
	Let $x_{A,i}$ ($i=1,2,3$) denote the three roots of
	Eq.~\eqref{omega^a_-1 equation}, with
	\be
	\mathbf{a}_A + \mathbf{b}_A x_{A} + \mathbf{c}_A x_{A}^2 + \mathbf{d}_A x_{A}^3=0,\qquad\omega_{A,-1}=\pm\sqrt{x_{A,i}}.
	\ee
	The reality of the leading large-$k$ characteristic velocities
	requires
	\be
	{	x_{A,i}\geq0,
	\qquad i=1,2,3.}
	\ee
	For $0\leq\theta<\pi/2$, one has $\mathbf{d}_A<0$.
	Using the Vieta relations, a necessary set of conditions for
	$x_{A,i}\geq0$ is therefore
	\be\label{omega^a_-1 solution condictions}
	\mathbf{d}_A<0,\qquad
	\mathbf{c}_A\geq0,\qquad
	\mathbf{b}_A\leq0,\qquad
	\mathbf{a}_A\geq0.
	\ee
	The inequalities become strict when all three roots are strictly
	positive.
	
	For an arbitrary propagation angle, the angular dependence of
	these sign conditions can be reduced to the following algebraic
	constraints:

	\begin{eqnarray}
		\mathbf{a}_A>0
		&\Longleftrightarrow&
	 M_A>-2 \sqrt{F_{A, \|} F_{A, \perp}}, \quad	F_{A, \|}>0, \quad F_{A, \perp}>0,
		\nonumber\\
		\mathbf{b}_A<0
		&\Longleftrightarrow&
		\mathcal{B}_2>
		-2\sqrt{\mathcal{B}_0\mathcal{B}_4},
		\nonumber\\
		\mathbf{c}_A>0
		&\Longleftrightarrow&
		\mathcal{C}_0-\mathcal{C}_2>0,
		\qquad
		\mathcal{C}_0+\mathcal{C}_2>0,
		\nonumber\\
		\mathbf{d}_A<0
		&\Longleftrightarrow&
		\tau_\ell\tau_{\Pi_\parallel}\tau_f\tau_{\Pi_\perp}\tau_\pi>0,
	\end{eqnarray}
	where $F_{A,\parallel}
	=
	\mathcal{A}_0-4\eta_f\mathcal{A}_2+\mathcal{A}_4,\,
	F_{A,\perp}
	=
	\mathcal{A}_0+4\eta_f\mathcal{A}_2+\mathcal{A}_4\,\text{and}\,
	M_A
	=
	2\mathcal{A}_0-6\mathcal{A}_4.$
	These relations show that the required signs are not guaranteed
	solely by the positivity of the individual transport coefficients
	and relaxation times, but impose additional constraints on their
	relative magnitudes.
	
	At the exactly perpendicular direction $\theta=\pi/2$, all four
	coefficients contain a common factor $\cos\theta$ and vanish
	simultaneously. The perpendicular case should therefore be
	understood as the continuous limit obtained after dividing the
	characteristic equation by this common factor.
	
	The four coefficients associated with Eq.~\ref{omega^a_0 equation} are recast as the explicit forms of the abbreviations $\mathbf{a}'_A$, $\mathbf{b}'_A$, $\mathbf{c}'_A$, and $\mathbf{d}'_A$:
	\begin{eqnarray}
		\mathbf{a}_A'
		&=&
		-c_s^2h\,\mathcal{N}_A
		\Big[
		\mathcal{Q}
		+2(2-v_A^2)\eta_f
		-4v_A^2\eta_f\cos2\theta
		\nonumber\\
		&&
		-\big(
		\mathcal{Q}-2(2-v_A^2)\eta_f
		\big)\cos4\theta
		\Big],
		\nonumber\\
		\mathbf{b}_A'
		&=&
		i\,\mathcal{N}_A
		\Big[
		\mathcal{D}_0
		+4\eta_f\mathcal{D}_2\cos2\theta
		+\mathcal{D}_4\cos4\theta
		\Big],
		\nonumber\\
		\mathbf{c}_A'
		&=&
		\mathcal{N}_A
		\Big[
		\mathcal{E}_0
		-4\eta_f\mathcal{E}_2\cos2\theta
		+\mathcal{E}_4\cos4\theta
		\Big],
		\nonumber\\
		\mathbf{d}_A'
		&=&
		-i\,\mathcal{N}_A
		\Big[
		\mathcal{A}_0
		-4\eta_f\mathcal{A}_2\cos2\theta
		+\mathcal{A}_4\cos4\theta
		\Big]
		\nonumber\\
		&=&
		-i\,\mathbf{a}_A .
	\end{eqnarray}
	
	For $\mathbf{b}_A'$, we define

	\begin{eqnarray}
		\mathcal{D}_0
		&\equiv&
		\zeta_\parallel\zeta_\perp-\zeta_\times^2+\zeta_\parallel\eta_\perp
		+\eta_f
		\big(
		3\zeta_\parallel+3\zeta_\perp-2\zeta_\times+3\eta_\perp
		\big)
		\nonumber\\
		&&
		+c_s^2h
		\Bigg\{
		\Big[
		(1-v_A^2)(\zeta_\perp+\eta_\perp)-\zeta_\times
		+2(2-v_A^2)\eta_f
		\Big]\tau_{\Pi_\parallel}
		\nonumber\\
		&&
		+\mathcal{Q}\tau_f
		+\Big[
		\zeta_\parallel+(1-v_A^2)(\eta_\perp-\zeta_\times)
		+2(2-v_A^2)\eta_f
		\Big]\tau_{\Pi_\perp}
		\nonumber\\
		&&
		+\Big[
		\zeta_\parallel+(1-v_A^2)\zeta_\perp
		-(2-v_A^2)\zeta_\times
		+2(2-v_A^2)\eta_f
		\Big]\tau_\pi
		\Bigg\},
		\nonumber\\
		\mathcal{D}_2
		&\equiv&
		\zeta_\parallel-\zeta_\perp-\eta_\perp
		-c_s^2h v_A^2
		(\tau_{\Pi_\parallel}+\tau_{\Pi_\perp}+\tau_\pi),
		\nonumber\\
		\mathcal{D}_4
		&\equiv&
		\zeta_\times^2-\zeta_\parallel(\zeta_\perp+\eta_\perp)
		+\eta_f
		\big(
		\zeta_\parallel+\zeta_\perp+2\zeta_\times+\eta_\perp
		\big)
		\nonumber\\
		&&
		+c_s^2h
		\Bigg\{
		\Big[
		\zeta_\times-(1-v_A^2)(\zeta_\perp+\eta_\perp)
		+2(2-v_A^2)\eta_f
		\Big]\tau_{\Pi_\parallel}
		-\mathcal{Q}\tau_f
		\nonumber\\
		&&
		+\Big[
		-\zeta_\parallel+(1-v_A^2)(\zeta_\times-\eta_\perp)
		+2(2-v_A^2)\eta_f
		\Big]\tau_{\Pi_\perp}
		\nonumber\\
		&&
		+\Big[
		-\zeta_\parallel-(1-v_A^2)\zeta_\perp
		+(2-v_A^2)\zeta_\times
		+2(2-v_A^2)\eta_f
		\Big]\tau_\pi
		\Bigg\}.
	\end{eqnarray}

	For $\mathbf{c}_A'$, we introduce
	\begin{eqnarray}
		\mathcal{E}_0
		&\equiv&
		\eta_f
		\big(
		3\zeta_\perp-\zeta_\times+3\eta_\perp
		\big)\tau_{\Pi_\parallel}
		+
		\big[
		\zeta_\parallel(\zeta_\perp+\eta_\perp)-\zeta_\times^2
		\big]\tau_f
		+
		\Big[
		\zeta_\parallel\eta_\perp
		+\eta_f(3\zeta_\parallel-\zeta_\times+3\eta_\perp)
		\Big]\tau_{\Pi_\perp}
		\nonumber\\
		&&
		+
		\Big[
		\zeta_\parallel\zeta_\perp-\zeta_\times^2
		+\eta_f(3\zeta_\parallel+3\zeta_\perp-2\zeta_\times)
		\Big]\tau_\pi
		+c_s^2h
		\Bigg\{
		\Big[
		(1-v_A^2)(\zeta_\perp+\eta_\perp)-\zeta_\times
		\Big]\tau_{\Pi_\parallel}\tau_f
		\nonumber\\
		&&
		+\Big[
		(1-v_A^2)\eta_\perp
		+2(2-v_A^2)\eta_f
		\Big]\tau_{\Pi_\parallel}\tau_{\Pi_\perp}
		+\Big[
		\zeta_\parallel+(1-v_A^2)(\eta_\perp-\zeta_\times)
		\Big]\tau_f\tau_{\Pi_\perp}
		\nonumber\\
		&&
		+\Big[
		(1-v_A^2)\zeta_\perp-\zeta_\times
		+2(2-v_A^2)\eta_f
		\Big]\tau_{\Pi_\parallel}\tau_\pi
		+\Big[
		\zeta_\parallel+(1-v_A^2)\zeta_\perp
		-(2-v_A^2)\zeta_\times
		\Big]\tau_f\tau_\pi
		\nonumber\\
		&&
		+\Big[
		\zeta_\parallel-(1-v_A^2)\zeta_\times
		+2(2-v_A^2)\eta_f
		\Big]\tau_{\Pi_\perp}\tau_\pi
		\Bigg\},
		\nonumber\\
		\mathcal{E}_2
		&\equiv&
		\eta_\perp(\tau_{\Pi_\parallel}+\tau_{\Pi_\perp})
		+\zeta_\perp(\tau_{\Pi_\parallel}+\tau_\pi)
		-\zeta_\parallel(\tau_{\Pi_\perp}+\tau_\pi)
		+c_s^2h v_A^2
		\big(
		\tau_{\Pi_\parallel}\tau_{\Pi_\perp}+\tau_{\Pi_\parallel}\tau_\pi+\tau_{\Pi_\perp}\tau_\pi
		\big),
		\nonumber\\
		\mathcal{E}_4
		&\equiv&
		\eta_f(\eta_\perp+\zeta_\times+\zeta_\perp)\tau_{\Pi_\parallel}
		+
		\Big[
		\zeta_\times^2-\zeta_\parallel(\eta_\perp+\zeta_\perp)
		\Big]\tau_f
		+
		\Big[
		-\zeta_\parallel\eta_\perp
		+\eta_f(\zeta_\parallel+\eta_\perp+\zeta_\times)
		\Big]\tau_{\Pi_\perp}
		\nonumber\\
		&&
		+
		\Big[
		\zeta_\times^2-\zeta_\parallel\zeta_\perp
		+\eta_f(\zeta_\parallel+\zeta_\perp+2\zeta_\times)
		\Big]\tau_\pi
		+c_s^2h
		\Bigg\{
		\Big[
		\zeta_\times-(1-v_A^2)(\eta_\perp+\zeta_\perp)
		\Big]\tau_{\Pi_\parallel}\tau_f
		\nonumber\\
		&&
		+\Big[
		-(1-v_A^2)\eta_\perp
		+2(2-v_A^2)\eta_f
		\Big]\tau_{\Pi_\parallel}\tau_{\Pi_\perp}
		+\Big[
		-\zeta_\parallel+(1-v_A^2)(\zeta_\times-\eta_\perp)
		\Big]\tau_f\tau_{\Pi_\perp}
		\nonumber\\
		&&
		+\Big[
		\zeta_\times-(1-v_A^2)\zeta_\perp
		+2(2-v_A^2)\eta_f
		\Big]\tau_{\Pi_\parallel}\tau_\pi
		+\Big[	-\zeta_\parallel-(1-v_A^2)\zeta_\perp
		+(2-v_A^2)\zeta_\times
		\Big]\tau_f\tau_\pi
		\nonumber\\
		&&
		+\Big[
		-\zeta_\parallel+(1-v_A^2)\zeta_\times
		+2(2-v_A^2)\eta_f
		\Big]\tau_{\Pi_\perp}\tau_\pi
		\Bigg\}.
	\end{eqnarray}
	
	Here the additional abbreviation is
	\begin{eqnarray}
		\mathcal{Q}
		&\equiv&
		\zeta_\parallel
		+(1-v_A^2)(\zeta_\perp+\eta_\perp)
		-(2-v_A^2)\zeta_\times .
	\end{eqnarray}
	We next consider the stability of the subleading large-$k$
	solutions $\omega_{A,0}$. Since $\mathbf{a}_A'$ and
	$\mathbf{c}_A'$ are real, whereas $\mathbf{b}_A'$ and
	$\mathbf{d}_A'$ are purely imaginary, we define $\lambda\equiv-i\omega_{A,0}$ and use $\mathbf{d}_A'=-i\mathbf{a}_A$. The Eq.~\eqref{omega^a_0 equation} can then be rewritten as the real cubic
	equation
	\begin{eqnarray}
	{		\mathbf{a}_A\lambda^3
		+\mathbf{c}_A'\lambda^2
		+\operatorname{Im}\mathbf{b}_A'\lambda
		-\mathbf{a}_A'
		=0.}
	\end{eqnarray}
	Since $\operatorname{Im}\omega_{A,0}<0$ is equivalent to
	$\operatorname{Re}\lambda<0$, the Routh--Hurwitz criterion gives
	the stability conditions
	\begin{eqnarray}
		{	\mathbf{a}_A>0,
		\qquad
		\mathbf{c}_A'>0,
		\qquad
		\operatorname{Im}\mathbf{b}_A'>0,
		\qquad
		\mathbf{a}_A'<0,
		\qquad
		\mathbf{c}_A'\operatorname{Im}\mathbf{b}_A'
		>
		-\mathbf{a}_A\mathbf{a}_A'.}
		\label{omega^a_0 RH conditions}
	\end{eqnarray}
	For the non-strict stability condition
	$\operatorname{Im}\omega_{A,0}\leq0$, the corresponding boundary
	cases should be treated separately.
	
	The Routh--Hurwitz conditions guarantee that the roots lie in the
	stable half-plane, but do not by themselves imply that
	$\omega_{A,0}$ is purely imaginary. To establish
	$\operatorname{Re}\omega_{A,0}=0$, all three roots of the real
	cubic equation for $\lambda$ must be real. This requires
	\begin{eqnarray}
		\Delta_{A,0}
		&=&
		(\mathbf{c}_A')^2
		\left(\operatorname{Im}\mathbf{b}_A'\right)^2
		-4\mathbf{a}_A
		\left(\operatorname{Im}\mathbf{b}_A'\right)^3
		+4\mathbf{a}_A'(\mathbf{c}_A')^3\nonumber\\
		&&	-27\mathbf{a}_A^2(\mathbf{a}_A')^2
		-18\mathbf{a}_A\mathbf{a}_A'
		\mathbf{c}_A'
		\operatorname{Im}\mathbf{b}_A'
		\geq0.
		\label{omega^a_0 discriminant}
	\end{eqnarray}
	Therefore, when
	Eqs.~\eqref{omega^a_0 RH conditions} and
	\eqref{omega^a_0 discriminant} are simultaneously satisfied,
	all three roots obey
	\begin{eqnarray}
	{		\operatorname{Re}\omega_{A,0}=0,
		\qquad
		\operatorname{Im}\omega_{A,0}<0,}
	\end{eqnarray}
	and are thus purely damped. For generic transport parameters, however, the condition $\Delta_{A,0}\geq0$ does not follow directly from the Routh--Hurwitz conditions. Therefore, unless the positivity of $\Delta_{A,0}$ can be established independently, linear stability can be proven without necessarily implying that all $\omega_{A,0}$ are purely imaginary.}

	The coefficients in the large-$k$ equation of the Alfv\'en sector
	can be written as
	\begin{eqnarray}
		\mathbf{a}_B
		&=&
		\frac{h v_A^2}{B^2}
		\mathcal{R}_B\mathcal{S}_B,
		\nonumber\\
		\mathbf{b}_B
		&=&
		-\tau_\ell\tau_m
		\Big(
		\mathcal{S}_B
		+h v_A^2\tau_f\tau_\pi\cos^2\theta
		\Big)
		-\frac{h^2v_A^2}{B^2}
		\tau_f\tau_\pi\mathcal{R}_B,
		\nonumber\\
		\mathbf{c}_B
		&=&
		h\tau_\ell\tau_m\tau_f\tau_\pi,
	\end{eqnarray}
	where we introduce
	\begin{eqnarray}
		\mathcal{R}_B
		&\equiv&
		\rho_\parallel\tau_m\cos^2\theta
		+\rho_\perp\tau_\ell\sin^2\theta,
		\nonumber\\
		\mathcal{S}_B
		&\equiv&
		\eta_f\tau_\pi\cos^2\theta
		+\eta_\perp\tau_f\sin^2\theta .
	\end{eqnarray}
		{
	The leading-order equation can be written as
	\begin{eqnarray}
		\mathbf{a}_B+\mathbf{b}_Bv^2+\mathbf{c}_Bv^4=0.
	\end{eqnarray}
	For convenience, we further introduce $\Delta_B
	\equiv
	\mathbf{b}_B^2
	-
	\frac{4h^2v_A^2}{B^2}
	\tau_\ell\tau_m\tau_f\tau_\pi
	\mathcal{R}_B\mathcal{S}_B
	$, the four roots are
	\begin{eqnarray}
		v_{1,2}
		&=&
		\pm
		\sqrt{
			\frac{
				-\mathbf{b}_B+\sqrt{\Delta_B}
			}{
				2h\tau_\ell\tau_m\tau_f\tau_\pi
			}
		},
		\nonumber\\
		v_{3,4}
		&=&
		\pm
		\sqrt{
			\frac{
				-\mathbf{b}_B-\sqrt{\Delta_B}
			}{
				2h\tau_\ell\tau_m\tau_f\tau_\pi
			}
		}.
	\end{eqnarray}
	
	We next verify that the above leading-order velocities are real.
	Introducing
	\begin{eqnarray}
		X
		&\equiv&
		\tau_\ell\tau_m\mathcal{S}_B,
		\nonumber\\
		Y
		&\equiv&
		h v_A^2\tau_\ell\tau_m\tau_f\tau_\pi\cos^2\theta,
		\nonumber\\
		Z
		&\equiv&
		\frac{h^2v_A^2}{B^2}
		\tau_f\tau_\pi\mathcal{R}_B,
	\end{eqnarray}
	one has $X\geq0,\, Y\geq0\, \text{and}\, Z\geq0$ provided that the transport coefficients and relaxation times are
	non-negative, with the relaxation times taken to be positive.
	Moreover,
	\begin{eqnarray}
		-\mathbf{b}_B=X+Y+Z.
	\end{eqnarray}
	The discriminant can therefore be rewritten as
	\begin{eqnarray}
		\Delta_B
		&=&
		(X+Y+Z)^2-4XZ
		\nonumber\\
		&=&
		(X-Z)^2+Y^2+2Y(X+Z)\geq0.
	\end{eqnarray}
	Thus, $\sqrt{\Delta_B}$ is real. Furthermore,
	\begin{eqnarray}
		\mathbf{b}_B^2-\Delta_B
		=
		4XZ
		\geq0.
	\end{eqnarray}
	
	Together with the formula $h\tau_\ell\tau_m\tau_f\tau_\pi>0$
	this guarantees that both quantities under the outer square roots
	are non-negative. Therefore, all four nonzero roots are real.
	Including the double non-hydrodynamic zero root, we finally obtain
	\begin{eqnarray}
		\operatorname{Im}\omega_{B,-1,r_i}=0,
		\qquad i=1,\ldots,6.
	\end{eqnarray}
	Hence, the leading large-$k$ propagating modes in the Alfv\'en
	sector do not develop an imaginary contribution proportional to
	$k$.}

	The coefficients $\mathbf{a}_B'$, $\mathbf{b}_B'$, and
	$\mathbf{c}_B'$ in Eq.~\eqref{omega0 equation} can then be written as
	\begin{eqnarray}
		\mathbf{a}_B'
		&=&
		-\mathcal{N}_B
		\widetilde{\mathcal{R}}_B
		\widetilde{\mathcal{S}}_B,
		\nonumber\\
		\mathbf{b}_B'
		&=&
		i\mathcal{N}_B
		\left(
		\mathcal{R}_B\widetilde{\mathcal{S}}_B
		+
		\widetilde{\mathcal{R}}_B\mathcal{S}_B
		\right),
		\nonumber\\
		\mathbf{c}_B'
		&=&
		\mathcal{N}_B
		\mathcal{R}_B\mathcal{S}_B
		=
		\mathbf{a}_B ,
	\end{eqnarray}
	where $\widetilde{\mathcal{R}}_B$ , $\widetilde{\mathcal{S}}_B$, and $\mathcal{N}_B$ are defined as
	\begin{eqnarray}
		\widetilde{\mathcal{R}}_B
		&\equiv&
		\rho_\parallel\cos^2\theta+\rho_\perp\sin^2\theta,
		\nonumber\\
		\widetilde{\mathcal{S}}_B
		&\equiv&
		\eta_f\cos^2\theta+\eta_\perp\sin^2\theta,
		\nonumber\\
		\mathcal{N}_B
		&\equiv&
		\frac{h v_A^2}{B^2}.
	\end{eqnarray}

	\bibliographystyle{JHEP}
	\bibliography{ref}

\end{document}